\documentclass[a4paper,12pt]{article}
\usepackage[utf8]{inputenc}
\usepackage{cancel}
\usepackage{ulem}
\usepackage{amsfonts}
\usepackage{amssymb}
\usepackage{graphicx}
\usepackage{amsmath}
\usepackage{enumerate}
\usepackage{here}
\usepackage{subfloat}
\usepackage{subfig}
\usepackage{color}
\usepackage{float}
\usepackage{here}
\usepackage{cite}
\usepackage{mathrsfs}
\usepackage{float,epsfig}
\usepackage{dcolumn}
\usepackage{widetext}

\usepackage{graphicx}
\usepackage{bm}
\usepackage{amsmath,amssymb,amsthm}
\usepackage[colorlinks=true,linkcolor=blue]{hyperref}
\title{\bf \Large The Higgs Potential in 2HDM extended with a Real Triplet Scalar: A roadmap. }

\author{B. Ait-Ouazghour, M. Chabab\footnote{mchabab@uca.ac.ma (Corresponding author)}\\
	\\ 
	{\small $^{1}$ High Energy and Astrophysics Laboratory,  FSSM, Cadi Ayyad University
	}\\
	{\small  P.O.B. 2390 Marrakech, Morocco.
	}
}
\date{}
\begin{document} 
	\maketitle
	
	\begin{abstract}
	 We perform a comprehensive study of The Higgs potential of the two Higgs doublet model extended by a real triplet scalar field $\Delta$. This model, dubbed $2\mathcal{HDM+T}$, has a rich Higgs spectrum  consisting of  three CP-even Higgs $h_{1,2,3}$, one CP-odd $A_0$ and two pairs of charged Higgs $H^\pm_{1,2}$. First, we determine the  perturbative unitarity constraints and a set of non trivial conditions for the boundedness from below (BFB). Then we derive  the Veltman conditions  by considering the quadratic divergencies of Higgs boson self energies in $2\mathcal{HDM+T}$.  We find that  the parameter space  is severely delimited by these theoretical constraints, as well as experimental exclusion limits and Higgs signal rate measurements at LEP and LHC.  Using HiggsBounds-5.3.2beta and HiggSignals-2.2.3beta public codes an exclusion test at $2\sigma$ is then performed on the physical scalars of  $2\mathcal{HDM+T}$.  Our analysis provides a clear insight on the nonstandard scalar masses, showing that the allowed ranges are strongly sensitive to the sign of mixing angle $\alpha_1$, essentially when naturalness is involved. For $\alpha_1 < 0$ scenario, our results place higher limits on the bounds of all scalar masses, and show that the pairs $(h_2, H_1^\pm)$ and $(h_3, H_2^\pm)$ are nearly mass degenerate varying within the intervals  $[130\,,\,246]$~GeV and $[160\,,\,335]$~GeV respectively.  When  $\alpha_1$ turns positive, we show that consistency with theoretical constraints and current LHC data, essentially on the diphoton decay channel, favors Higgs masses varying within wide allowed ranges: $[153\,,\,973]$~GeV for $m_{A_0}$; $[151\,,\,928]$~GeV for ($m_{h_2}$,  $m_{H_1^\pm}$) and $[186\,,\,979]$~GeV for ($m_{h_3}$, $m_{H_2^\pm}$). Finally, we find that the $\gamma \gamma$ and $Z\gamma$ Higgs decay modes are generally correlated if $\tan \beta$ lies within the reduced intervals $ 17 \le \tan \beta \le 25$ and $\lambda_b$ parameter is frozen around  $1.3$ ($ 1.1$) for $\sin \alpha_1 > 0$ ($\sin \alpha_1 < 0$). 
	
	\end{abstract}
	
	
%

\section{INTRODUCTION}
\label{intro}
 With the LHC discovery of a scalar resonance with a mass about $125$ GeV and properties compatible with the Higgs boson predicted by the Standard Model (SM), the SM has gained a status of a theory  \cite{RunI_2012_20, RunI_2012_21, RunI_2012_22, RunI_2012_23, Higgs_mass} while the Brout-Englert-Higgs mechanism has been confirmed as a fundamental mode for the mass origin of its gauge bosons and fermions \cite{Higgs_Hunter}.   However, despite its brilliant success, the SM cannot address many issues in particle physics. The mystery of (tiny) neutrino masses is one of them. Another still unsolved enigma related to dark matter and dark energy, their nature, compositions and interactions. Also the legitimate  question on existence of other nonstandard scalars which may contribute to electroweak symmetry breaking (EWSB) is not answered yet, not to mention the naturalness problem.  These major open problems indicate an urgent need for a new physics beyond the SM (BSM). Many attractive theories aiming to solve some of these issues have been proposed. Among them, a group of models assume the existence of additional fundamental triplet scalars which, through mixing with the SM Higgs boson, foresee a richer spectrum and imprints of BSM phenomena~\cite{Wess:1974tw,Salam:1974ig,Ferrara:1974pu,Martin:1997ns}. Generally, models with complex triplet scalars can serve to explain neutrino oscillations while those with real triplet scalar mainly address dark matter problem  \cite{mag80, li80,valle80, moha81, aoki08, perez09, akeroyd10, aa2011, key12, chabab14, Chen2014, mariana2015, Bonilla2015,ram1_20, ram11_20} or electroweak phase transition and Electroweak Baryogenesis (EWBG) \cite{ram2_20}. Recently we have studied several triplet extended models including type II seesaw model  \cite{HTMY2}, Higgs triplet model with null hypercharge  \cite{HTMY0}, a type II seesaw two Higgs doublet model, consisting of $2\mathcal{HDM}$ model with a complex triplet scalar  \cite{2HDMcT}.

In this context,  we consider in this work another simple framework: the Two Higgs Doublet Model augmented with a real triplet scalar, dubbed  $2\mathcal{ HDM+T}$.  Similarly to many frameworks that consider an extended Higgs sector through the addition of a real scalar field, $2\mathcal{ HDM+T}$ can also alleviate some of the $SM$ drawbacks. Indeed, this model may provide additional sources of CP violation besides the CKM matrix phase, a necessary ingredient to achieve a viable scenario based on Electroweak Baryogenesis. Generally, the addition of real scalar  results in scalar potential with a distinct thermal behaviour that could lead to a strong first order phase transition required for the realization of Baryon Asymmetry via Electroweak Baryogenesis mechanism  \cite{salam2013, ram2_20}. Also,  thanks to its more involved Higgs sector,  $2\mathcal{ HDM+T}$ can lead to a potential dark matter candidate via the neutral component of the triplet field  \cite{ram1_20, takeshi2011}.

As in previous papers, we perform a detailed study of its Higgs  potential and derive the main features of the model, namely the full set of theoretical constraints inherent to $2\mathcal{ HDM+T}$. These  include perturbative unitarity and boundedness from below (BFB), in addition to naturalness from which we determine the modified Veltman conditions. Then, to delineate the model parameter space, collider constraints originating from Higgs signals measurements are also incorporated. Both HiggsBounds-5.3.2beta \cite{HiggsBounds2011,HiggsBounds2014} and HiggSignals-2.2.3beta \cite{HiggsSignals2014} public codes are  used to test our theoretical predictions in the  $2\mathcal{  HDM+T}$ Higgs sector against exclusion experimental limits from direct Higgs searches at LEP, Tevatron and LHC. Phenomenological analysis of the Higgs decays is then performed with  the aim to highlight how the Higgs masses spectrum evolve when a specific condition is incorporated. A particular attention is given to the crucial role played by naturalness and its impact on the heavy scalars masses.  


 The rest of the paper is organized as follows. In Sec. II, we will  perform a comprehensive study of  the main features of $2\mathcal{HDM+T}$ model and present the full set of theoretical constraints on the parameters of its Higgs potential, including modified Veltman conditions resulting from naturalness problem. Sections III and IV will be devoted to delineate $2\mathcal{HDM+T}$ parameter space.  The analysis subsequently performed will take into account full set of theoretical constraints as well as the measured signal rates from  ATLAS and CMS Run I and Run II are also included in the analysis by means of HiggsBounds and  HiggsSignals codes. The obtained results are then presented with emphasis on bounds and range of variation of heavy scalars masses.  Our conclusion will be drawn in section VI. Technical details are collected in appendices.

\section{($2\mathcal{HDM+T}$) MODEL: General considerations}
In this section, we present a general overview of ($2\mathcal{HDM+T}$) model. First we discuss the salient features of its scalar potential, then we derive the Higgs spectrum and the theoretical constraints that the model must respect. The couplings of Higgs bosons to fermions are also outlined as well as the parameterization adopted in the subsequent parts of the paper.
\subsection{The Higgs Potential}
\label{sec:thehiggspot}
The $2\mathcal{ HDM+T}$ model contains two Higgs doublets $H_{i}$ (i = 1,2) in addition with one scalar field $\Delta$ transforming as a triplet under the $SU(2)_L$ gauge group with hypercharge $Y_\Delta=0$.  The most general gauge-invariant Lagrangian of the scalar sector is given by \cite{ Higgs_Hunter, branco2012, perez2009},  
%
\small{\begin{equation}
	\begin{matrix}
	\mathcal{L}=\sum_{i=1}^2(D_\mu{H_i})^\dagger(D^\mu{H_i})+Tr(D_\mu{\Delta})^\dagger(D^\mu{\Delta})\vspace*{0.12cm}\\
	\hspace{-3cm}-V(H_i, \Delta)+\mathcal{L}_{\rm Yukawa}
	\label{eq:  2HDMt-lag}
	\end{matrix}
	\end{equation}}
where the scalar potential $V(H_i,\Delta)$ can be written as:
\begin{widetext}
	\begin{equation}
	\begin{aligned}
	V(H_i,\Delta)=&\,\,m^2_{1}\, H_1^\dagger H_1 + m^2_{2}\, H_2^\dagger H_2 -\hspace*{0cm}m_{12}^2\,H_2^\dagger H_1 +\frac{\lambda_1}{2} (H_1^\dagger H_1)^2 + \frac{\lambda_2}{2}  (H_2^\dagger H_2)^2\hspace*{0cm}+\hspace*{0cm}\lambda_3\, H_1^\dagger H_1\, H_2^\dagger H_2\\
	& + \lambda_4\, H_1^\dagger H_2\, H_2^\dagger H_1+\frac{\lambda_5}{2}\hspace*{0cm}\hspace*{0cm} \left[(H_1^\dagger H_2)^2+ (H_2^\dagger H_1)^2 \right]+\hspace*{0cm}\lambda_6\,H_1^\dagger H_1 Tr\Delta^{\dagger}{\Delta} +\lambda_7\,H_2^\dagger H_2 Tr\Delta^{\dagger}{\Delta}\\
	&+\mu_1 H_1^\dagger\Delta H_1 + \mu_2H_2^\dagger\Delta H_2 + \mu_3 [H_1^\dagger\Delta H_2 \hspace*{0cm}+hc]+\hspace*{0cm}\lambda_8\,H_1^\dagger{\Delta}\Delta^{\dagger} H_1\\
	& + \lambda_9\,H_2^\dagger{\Delta}\Delta^{\dagger} H_2+\hspace*{0cm}m^2_{\Delta}\, Tr(\Delta^{\dagger}{\Delta}) +\bar{\lambda}_8(Tr\Delta^{\dagger}{\Delta})^2\hspace*{0cm}+\hspace*{0cm}\bar{\lambda}_9Tr(\Delta^{\dagger}{\Delta})^2
	\label{eq:VDelta}
	\end{aligned}
	\end{equation}
\end{widetext}

Here $Tr$ denotes the trace over 2x2 matrices. The covariant derivatives of the associated fields read as,
\begin{equation}
D_\mu{H_i}=\partial_\mu{H_i}+igT^a{W}^a_\mu{H_i}+i\frac{g'}{2}B_\mu{H_i} \label{eq:covd1}
\end{equation}
\begin{equation}
D_\mu{\Delta}=\partial_\mu{\Delta}+ig[T^a{W}^a_\mu,\Delta] \label{eq:covd2}\hspace*{1.9cm}
\end{equation}
where $B_\mu$ and ${W}^a_\mu$, stand for the SM gauge bosons,   $g'$ and $g$ are coupling constants of the $U(1)_Y$ and $SU(2)_L$ gauge symmetry respectively. The matrices $T^a$ are defined in terms of the Pauli matrices, $T^a \equiv \sigma^a/2$, with ($a=1, 2, 3$). 


 Minimization of the potential Eq.~\ref{eq:VDelta} yields the following necessary conditions,
\begin{widetext}
	\begin{equation}
	\begin{aligned}
	m_1^2=&\,\,\frac{-v_1 \left(v_t \left(2 \lambda _{a} v_t-2 \mu _1\right)+2 \lambda _1 v_1^2+2 \lambda_{345} v_2^2\right)+4 m_{12}^2 v_2+2 \mu _3 v_2 v_t}{4 v_1}\\
	m_2^2=&\,\,\frac{-v_2 \left(v_t \left(2 \lambda _{b} v_t-2 \mu _2\right)+2 \lambda_{345} v_1^2+2 \lambda _2 v_2^2\right)+4 m_{12}^2 v_1+2 \mu _3 v_1 v_t}{4 v_2}\\
	m_\Delta^2=&\,\,\frac{v_1^2 \left(\mu _1-2 \lambda _{a} v_t\right)+v_2^2 \left(\mu _2-2 \lambda_{b} v_t\right)-4 \lambda _{c} v_t^3+2 \mu _3 v_1 v_2}{4 v_t} \\
	\end{aligned}
	\label{min_con}
	\end{equation}
\end{widetext}
where we used the notation:  $\lambda_a=\lambda_6+\frac{\lambda_8}{2}$, $\lambda_b=\lambda_7+\frac{\lambda_9}{2}$, $\lambda_c=\bar{\lambda}_8+\frac{\bar{\lambda_9}}{2}$,
 and $\lambda_{345}=\lambda_3+\lambda_4+\lambda_5$.

After the electroweak symmetry breaking ($EWSB$), the triplet field $\Delta$ and Higgs doublets $H_i$ can be represented as,
 \begin{eqnarray}
 \Delta &=&\left(
 \begin{array}{cc}
 (v_t+\rho^0)/2 & \delta^+/\sqrt{2} \\
 \delta^-/\sqrt{2}  & -(v_t+\rho^0)/2\\
 \end{array}
 \right)
 \label{triplet}
 \end{eqnarray}
 \begin{eqnarray}
 H_1&=&\left(
 \begin{array}{c}
 \phi_1^+ \\
(v_1+\rho_1+i\eta_1)/\sqrt{2} \\
 \end{array}
 \right){,}~~~H_2=\left(
 \begin{array}{c}
 \phi_2^+ \\
 (v_2+\rho_2+i\eta_2)/\sqrt{2} \\
 \end{array}
 \right)
 \label{doublet}
 \end{eqnarray}

  Three of the eleven Higgs degrees of freedom corresponding to the Goldstone bosons are absorbed by the longitudinal components of vector gauge bosons, while the six remaining ones are manifested in the physical Higgs spectrum as: three CP-even scalars $h_1$, $h_2$, $h_3$ ordered according to  $m_{h_1}< m_{h_2}< m_{h_3}$, one CP-odd $A$ and two charged Higgs pair $H_1^\pm$, $H_2^\pm$ with $m_{H_1^\pm}< m_{H_2^\pm}$.

\subsection{Higgs masses and mixing angles}
\label{sec:higgsmasses}
The $11\times 11$ squared mass matrix,
\begin{equation}
{\mathcal M}^2_{ij}=\frac{1}{2} \frac{\partial^2 V}{\partial \varphi_i \partial \varphi_j} |_{H_{i} = \langle H_{i} \rangle ,
	\Delta = \langle \Delta \rangle}
\end{equation}
can be recast, using Eq.~\ref{min_con}, into a block of diagonal form composed of  two $3\times 3$ matrices, denoted ${\mathcal M}^2_\pm$ ,${\mathcal M}^2_{CP_{even}}$, and one $2\times 2$ matrix representing ${\mathcal M}^2_{CP_{odd}}$.

\subsubsection{Masses of the charged fields}
The mass matrix for the charged field is written by,
\begin{equation}
M^2_{\pm}=\left(\begin{matrix}
m^2_{G^+G^-}&m^2_{G^+H^-}&m^2_{G^+\delta^-}\\
m^2_{G^+H^+}&m^2_{H^+H^-}&m^2_{H^+\delta^-}\\
m^2_{\delta^+G^-}&m^2_{\delta^+H^-}&m^2_{\delta^+\delta^-}
\end{matrix}\right)
\end{equation}
where its elements read as,
\begin{equation}
\begin{aligned}
m^2_{G^+G^-}=&\frac{t_{\beta } \left(2 m_{12}^2+\mu _3 v_t\right)-\lambda _{345}v_d^2 s_{\beta }^2+2 \mu _1 v_t}{2}\\
\end{aligned}
\end{equation}
\begin{equation}
\begin{aligned}
m^2_{H^+H^-}=&\frac{-\lambda _{345}v_d^2 c_{\beta }^2+ct_{\beta } \left(2 m_{12}^2+\mu _3 v_t\right)+2 \mu _2 v_t}{2}\\
\end{aligned}
\end{equation}
\begin{equation}
\begin{aligned}
m^2_{\delta^+\delta^-}=&\frac{v_d^2 \left(\mu _3 s_{2\beta}+\mu _1 c_{\beta }^2+\mu _2 s_{\beta }^2\right)}{4 v_t}\\
\end{aligned}
\end{equation}
\begin{equation}
\begin{aligned}
m^2_{G^+H^-}=&\frac{\lambda _{345} v_d^2 c_{\beta } s_{\beta }-2 m_{12}^2+\mu _3 v_t}{2}\\
\end{aligned}
\end{equation}
\begin{equation}
\begin{aligned}
m^2_{G^+\delta^-}=&\frac{v_d \left(\mu _1 c_{\beta }+\mu _3 s_{\beta }\right)}{2}\,,\,\,\,m^2_{H^+\delta^-}=\frac{v_d \left(\mu _3 c_{\beta }+\mu _2 s_{\beta }\right)}{2}
\end{aligned}
\end{equation}

with  $v_d=\sqrt{v_1^2+v_2^2}$ GeV, $c_{\beta}=\cos \beta=v_1/v_d$, $s_{\beta}=\sin \beta=v_2/v_d$, $t_{\beta}=\tan \beta=v_2/v_1$ and $ct_{\beta}=1 /\tan \beta=v_1/v_2$. We show that ${\mathcal M}^2_{ij}$ can be diagonalized by the $3\times 3$ rotation  matrix ${\mathcal C}$: 
\begin{equation}
{\mathcal C}=\left(\begin{matrix}
c_{\theta^\pm_1} c_{\theta^\pm_2} & s_{\theta^\pm_1} c_{\theta^\pm_2} & s_{\theta^\pm_2}\\
-(c_{\theta^\pm_1} s_{\theta^\pm_2} s_{\theta^\pm_3} + s_{\theta^\pm_1} c_{\theta^\pm_3})
& c_{\theta^\pm_1} c_{\theta^\pm_3} - s_{\theta^\pm_1} s_{\theta^\pm_2} s_{\theta^\pm_3}
& c_{\theta^\pm_2} s_{\theta^\pm_3} \\
- c_{\theta^\pm_1} s_{\theta^\pm_2} c_{\theta^\pm_3} + s_{\theta^\pm_1} s_{\theta^\pm_3} &
-(c_{\theta^\pm_1} s_{\theta^\pm_3} + s_{\theta^\pm_1} s_{\theta^\pm_2} c_{\theta^\pm_3})
& c_{\theta^\pm_2}  c_{\theta^\pm_3}
\label{rotation_charg_matrix}
\end{matrix}\right)
\end{equation}
where $\theta_i^\pm$ (i=1,2,3) are the rotation angles,  
\begin{equation}
\cos\theta_1^\pm=\frac{v_1}{v_d},\,\,\sin\theta_1^\pm=\frac{v_2}{v_d}
\end{equation}
\begin{equation}
\cos\theta_2^\pm=-\frac{v_d}{\sqrt{4v_t^2+v_d^2}},\,\,\sin\theta_2^\pm=\frac{2 v_t}{\sqrt{4 v_t^2+v_d^2}}
\end{equation}
 with  $v=\sqrt{v_1^2+v_2^2+4v_t^2} = 246$ GeV. The $\theta_3^\pm$  mixing angle is used as input \footnote{The large analytical formula of $\theta_3^\pm$ is deferred  to appendix \ref{sec-diagonal}.}. \\ 
\\ 
The physical charged Higgs states could be regarded as combination of $\phi_1^\pm$, $\phi_2^\pm$ and $\delta^\pm$ with mixing parameterized as, 
\begin{equation}
\begin{matrix}

\left(\begin{matrix}
G_0^\pm\\
H^\pm_1\\
H^\pm_2
\end{matrix}\right)&=&
{\mathcal C}\left(\begin{matrix}
\phi^\pm_1\\
\phi^\pm_2\\
\delta^\pm
\end{matrix}\right)
\end{matrix}
\end{equation}
and mass eigenvalues given by,

\begin{widetext}
\begin{equation}
		\begin{aligned}
		m^2_{H^\pm_{1(2)}}=&\,\,\frac{\pm\sqrt{v_d^2 \left(4 s_{2 \beta } v^2 v_t \left(A B v_d c_{\beta }+A C v_d s_{\beta }-2 B C v_t\right)+\left(-2 A v_d v_t+v_d^2 \mathcal{X} c_{\beta } s_{\beta }+4 \mathcal{Y} v_t^2\right)^2\right)}}{4 c_{\beta }s_{\beta } v_t v_d^2}\\
		&+\frac{-2 A v_d^2 v_t+v_d^3 \mathcal{X} c_{\beta } s_{\beta }+4 v_d \mathcal{Y} v_t^2}{4 c_{\beta }s_{\beta } v_t v_d^2}
		\end{aligned}
		\end{equation}
\end{widetext}
with  $\mathcal{X}=B c_{\beta }+C s_{\beta }$, $\mathcal{Y}=B s_{\beta }+C c_{\beta }$, $A=m_{G^+H^-}^2$, $B=m_{G^+\delta^-}^2$, $C=m_{H^+\delta^-}^2$, $c_{2\beta}=\cos 2\beta$ and $s_{2\beta}=\sin 2\beta$.

\subsubsection{Masses of the neutral fields:}
The  squared mass matrix for the neutral scalar field reads as, 
\begin{widetext}
	\begin{equation}
	M^2_{odd}=\left(
	\begin{array}{ccc}
	\frac{v_2 \left(2 m_{12}^2+\mu _3 v_t-2 \lambda _5 v_1 v_2\right)}{2 v_1} & -m_{12}^2-\frac{\mu _3 v_t}{2}+\lambda _5 v_1 v_2\\
	-m_{12}^2-\frac{\mu _3 v_t}{2}+\lambda _5 v_1 v_2 & \frac{v_1 \left(2 m_{12}^2+\mu _3 v_t-2 \lambda _5 v_1 v_2\right)}{2 v_2} 
	\end{array}
	\right)
	\end{equation}
	\begin{equation}
	M^2_{even}=\left(
	\begin{array}{ccc}
	\frac{v_2 \left(2 m_{12}^2+\mu _3 v_t\right)}{2 v_1}+\lambda _1 v_1^2 & -m_{12}^2-\frac{\mu _3 v_t}{2}+\lambda _{345} v_1 v_2 & v_1 \left(\lambda _{a} v_t-\frac{\mu _1}{2}\right)-\frac{\mu _3 v_2}{2} \\
	-m_{12}^2-\frac{\mu _3 v_t}{2}+\left(\lambda _{345}\right) v_1 v_2 & \frac{v_1 \left(2 m_{12}^2+\mu _3 v_t\right)}{2 v_2}+\lambda _2 v_2^2 & v_2 \left(\lambda _{b} v_t-\frac{\mu _2}{2}\right)-\frac{\mu _3 v_1}{2} \\
	v_1 \left(\lambda _{a} v_t-\frac{\mu _1}{2}\right)-\frac{\mu _3 v_2}{2} & v_2 \left(\lambda _{b} v_t-\frac{\mu _2}{2}\right)-\frac{\mu _3 v_1}{2} & \frac{8 \lambda _{c} v_t^3+\mu _1 v_1^2+\mu _2 v_2^2+2 \mu _3 v_1 v_2}{4 v_t} \\
	\end{array}
	\right)
	\end{equation}
\end{widetext}
We diagonalize  $CP_{even}$ mass matrix using the formula, 

\begin{eqnarray}
{\mathcal{E}}{\mathcal{M}}_{{{\mathcal{CP}}_{even}}}^2{\mathcal{E}}^T&=&diag(m^2_{h_1},m^2_{h_2},m^2_{h_3})
\label{rota-matrix-cp-even}
\end{eqnarray}

where ${\mathcal{E}}$ stands for an orthogonal matrix given by,
\begin{eqnarray}
{\mathcal{E}} =\left( \begin{array}{ccc}
c_{\alpha_1} c_{\alpha_2} & s_{\alpha_1} c_{\alpha_2} & s_{\alpha_2}\\
-(c_{\alpha_1} s_{\alpha_2} s_{\alpha_3} + s_{\alpha_1} c_{\alpha_3})
& c_{\alpha_1} c_{\alpha_3} - s_{\alpha_1} s_{\alpha_2} s_{\alpha_3}
& c_{\alpha_2} s_{\alpha_3} \\
- c_{\alpha_1} s_{\alpha_2} c_{\alpha_3} + s_{\alpha_1} s_{\alpha_3} &
-(c_{\alpha_1} s_{\alpha_3} + s_{\alpha_1} s_{\alpha_2} c_{\alpha_3})
& c_{\alpha_2}  c_{\alpha_3}
\end{array} \right)
\label{eq:mixingmatrix}
\end{eqnarray}

The mixing angles $\alpha_1$, $\alpha_2$ and $\alpha_3$ vary in the range,
\begin{eqnarray}
- \frac{\pi}{2} \le \alpha_{1,2,3} \le \frac{\pi}{2} \;.
\end{eqnarray}
which means that $\sin {\alpha_{1,2,3}}$ can be either positive or negative, while the three mass eigenstates  being ordered as:
\begin{eqnarray}
m^2_{h_1} < m^2_{h_2} < m^2_{h_3} \;.
\end{eqnarray}

On the other hand, diagonalization of the $2\times2$ $CP_{odd}$ mass matrix  proceeds via the following matrix $\mathcal{O}$,
\begin{eqnarray}
\mathcal{O}=\left(
\begin{array}{ccc}
\cos\beta & -\sin\beta\\
\sin\beta & \cos\beta 
\end{array}\right)
\end{eqnarray}	
Among the two eigenvalues of $M^2_{odd}$ , one is zero, corresponding to the Goldstone boson $G^0$, while the other one,
\begin{eqnarray}
m^2_{A^0}=\frac{v_d^2\left(2 m_{12}^2+\mu _3 v_t-2 \lambda _5 v_1 v_2\right)}{2 v_1 v_2}
\label{masseod}
\end{eqnarray}
refers to the mass of pseudo-scalar physical state $A^0$.  
\paragraph*{}
At this stage, note that from $18$ potential parameters, only $14$  degrees of freedom are left, thanks to the minimization conditions Eq.~\ref{min_con} and to the  $VEV$s formula:   $v=\sqrt{v_1^2+v_2^2+4v_t^2} = 246$ GeV. \\

 Since many choices are possible for what to use as
input parameters, we opt in this paper for the following hybrid parameterization:
\begin{widetext}
	\begin{equation}
	\begin{aligned}
	\mathcal{P}=&\{m_{h_1},\,m_{h_2},\,m_{h_3},\,m_{H_1^\pm},\,m_{H_2^\pm},\,m_{A^0},\,\alpha_1\,\,\alpha_2,\,\alpha_3,\,\,\theta^\pm_3,\,tan\beta,\,\lambda_4,\,\mu_1,\,v_t\}
	\end{aligned}
	\end{equation}
\end{widetext}
It is also worth to stress that one can readily trade the set of Lagrangian parameters in terms of the physical Higgs masses and mixing angles as demonstrated in appendix \ref{sec-para}. 


\subsection{Yukawa Texture}
\label{sec:yukawatexture}
The Yukawa Lagrangian in our model includes all the Yukawa sector
of $2\mathcal{HDM}$:
	\begin{equation}
	\label{eq:Yukawa_Lagrangian}
	\mathcal{L}_{\rm Yukawa} = Y_{d}{\overline Q}_L\Phi_1d_R^{} 
	+ Y_{u}{\overline Q}_L\tilde{\Phi}_2u_R^{}
	+ Y_{e}{\overline L}_L\Phi_1 e_R^{}+{h.c.},
	\end{equation}
	and describes  the interactions between Higgs bosons and quarks, charged leptons.  ${\overline Q}_L$ and ${\overline L}_L$ are the left-handed quark and 
	lepton doublets, $d_R$, $u_R$ and $e_R$ are the right-handed up-type quark, 
	down-type quark and lepton singlets, respectively.
	$Y_{u}$, $Y_{d}$ and $Y_{e}$ 
	stand for the corresponding Yukawa coupling matrices
	(with $\tilde{\Phi}_2=i\sigma_2\Phi^*_2$).

\noindent
	
	It is known that an extended Higgs sector naturally induces  Flavor Changing Neutral Currents (FCNC) that have to be suppressed \cite{fcnc77}. This can be safely achieved via a $\mathbb{Z}_2$  discrete symmetry that model the Yukawa interactions. In this case, the $2HDM$ parameters $\lambda_6=\lambda_7=0$ and $\mu_{12}^2 =0$. Throughout this paper, we choose the type-II Yukawa interactions where down-type quark and charged leptons couple to $\Phi_1$ while up-type quark couples to $\Phi_2$. We also assume a softly broken $\mathbb{Z}_2$ symmetry by taking a non vanishing $\mu_{12}^2$, while the remaining parameters are real. 
	
	In this case, the quark part of Eq.~\ref{eq:Yukawa_Lagrangian} becomes,
		
		\begin{eqnarray}
		{\cal L}^{\rm 2HDMT-II}_Y  &=&
		-\frac{g}{2m_W\cos\beta}\left[\bar{q}_Dm_D\left(\mathcal{E}_{11}h_1+\mathcal{E}_{21}h_2+\mathcal{E}_{31}h_3\right)q_D-i\sin\beta\bar{q}_Dm_D\gamma_5q_DA^0\right]\nonumber\\
		&-& \frac{g}{2m_W\sin\beta}\left[\bar{q}_Um_U\left(\mathcal{E}_{12}h_1+\mathcal{E}_{22}h_2+\mathcal{E}_{32}h_3\right) q_U-i\cos\beta\bar{q}_Um_U\gamma_5q_DA^0\right]\nonumber\\ 
		&+&g\frac{V_{ud}}{\sqrt{2}m_W}\left(H^+_1\bar{q}_U\left[\frac{{\mathcal C}_{22}}{\sin\beta}m_U\frac{\left(1-\gamma_5\right)}{2}-\frac{{\mathcal C}_{21}}{\cos\beta}m_D\frac{\left(1+\gamma_5\right)}{2}\right]q_D+h.c\right)\nonumber\\
		&+&g\frac{V_{ud}}{\sqrt{2}m_W}\left(H^+_2\bar{q}_U\left[\frac{{\mathcal C}_{32}}{\sin\beta}m_U\frac{\left(1-\gamma_5\right)}{2}-\frac{{\mathcal C}_{31}}{\cos\beta}m_D\frac{\left(1+\gamma_5\right)}{2}\right]q_D+h.c\right)\,,
		\label{eq:LhY}
		\end{eqnarray}
			
where  the elements $\mathcal{C}_{ij}$ appearing in the charged Higgs  Yukawa couplings are given in Eq.~\ref{rotation_charg_matrix}.\\

On the other hand, the Higgs couplings $H_i$ to the gauge bosons $V=W,Z$  can be readily identified by expanding the covariant derivative ${\rm D}_\mu$, and performing the usual transformations on the gauge and scalar fields to generate the physical fields. A full list of these couplings as well as  those of two Higgs to a vector boson, and trilinear couplings among neutral, charged scalars and gauge bosons are also presented in the appendix \ref{scalarcoup}. 
%

\section{THEORETICAL CONSTRAINTS}
\label{constraints}
The $2\mathcal{HDM+T}$ Higgs potential parameters are not free but controlled by the theoretical and experimental constraints which delineate the parameter space of the model. Hence, no need to stress that all subsequent phenomenological analysis is performed within the parameter space scanned by potential parameters of $2\mathcal{HDM+T}$ obeying the all theoretical constraints, namely: perturbative unitarity,  boundedness form below (BFB), and naturalness.  In other words, only scan points that pass all these constraints are relevant.
\subsection{Perturbative Unitarity}
\label{unitarity}
As usual, our model has also to be confronted with unitarity constraints which require that the amplitudes $M$ of any $2 \to 2$ scalars scattering has to obey perturbative unitarity. The associated matrix $M$ is then constructed by means of all possible combination of two scalar fields both in initial as well final states. At tree level, being reals, these amplitudes lead to a condition on partial wave amplitude $a_0$, namely $|a_0|\le 1$ or $Re(a_0) \le  \frac{1}{2}$. These  can be translated, at high energies, into bounds on the eigenvalues of the scattering matrix $M$: $\lambda_i < 8 \pi$ \cite{Kanemura:1993hm, arhrib, dey14}.\\
In $2\mathcal{HDM+T}$ model, the matrix can be decomposed into several channels:  three $0-$charge channels, one $1-$charge channel and one $2-$charge channel.  Hereafter, we present the explicit formulas for all obtained eigenvalues:

\begin{equation}
\begin{aligned}
|\lambda_3+\lambda_4|\le k\pi&\;\;\;\;\;\;\;|\lambda_3+2\lambda_4\pm 3\lambda_5|\le k\pi
\end{aligned}
\end{equation}
\begin{equation}
\begin{aligned}
|\lambda_3 \pm \lambda_5|\le k\pi\;\;\;,\;\;\;|\lambda_a|\le k\pi\;\;\;,\;\;\;|\lambda_b|\le k\pi
\end{aligned}
\end{equation}
\begin{equation}
\begin{aligned}
|\frac{1}{2} \left(\lambda _1+\lambda _2\pm\sqrt{\lambda _1^2-2 \lambda _2 \lambda _1+\lambda _2^2+4 \lambda _4^2}\right)|\le k\pi
\end{aligned}
\end{equation}
\begin{equation}
\begin{aligned}
|\frac{1}{2} \left(\lambda _1+\lambda _2\pm\sqrt{\lambda _1^2-2 \lambda _2 \lambda _1+\lambda _2^2+4 \lambda _5^2}\right)|\le k\pi
\end{aligned}
\end{equation}
\begin{equation}
\begin{aligned}
|2\lambda_c|\le k\pi&\;_;\;\;\;\;\;\;\;|2\lambda_b|\le k\pi
\end{aligned}
\end{equation}

The parameter $k$ takes values $k = 8$ or $16$, depending on whether one chooses $Re(a_{0}) \le  \frac{1}{2}$ or $|a_{0}|\le 1$  as unitarity condition. \\

In addition, we have derived three other eigenvalues  by solving the cubic polynomial equation,
\begin{equation}
\begin{aligned}
x^3-x^2 \left(10 \lambda _{c}+6 \lambda _1+6 \lambda _2\right)+x \left(-12 \lambda _{a}^2-12 \lambda _{b}^2+60 \lambda _1 \lambda _{c}+60 \lambda _2 \lambda _{c}-16 \lambda _3^2-16 \lambda _4 \lambda _3-4 \lambda _4^2+36 \lambda _1 \lambda _2\right)\\+\left(-96 \lambda _3 \lambda _{a} \lambda _{b}-48 \lambda _4 \lambda _{a} \lambda _{b}+72 \lambda _2 \lambda _{a}^2+72 \lambda _1 \lambda _{b}^2+160 \lambda _3^2 \lambda _{c}+40 \lambda _4^2 \lambda _{c}-360 \lambda _1 \lambda _2 \lambda _{c}+160 \lambda _3 \lambda _4 \lambda _{c}\right)=0\;\;
\label{eq:cubic-polynom}
\end{aligned}
\end{equation}

Full technical details of this derivation are given in appendix \ref{sec-U}.

\subsection{Boundedness From Below (BFB):}
\label{bfb}
Here, we derive the relations among potential parameters that need to be respected in order to guarantee the vacuum stability. This means the potential is bounded
from below at the weak scale, and is never negative along any direction of the field space. Obviously, for large field values, the scalar potential Eq.~(\ref{eq:VDelta}) is  generally dominated by quartic terms, dubbed $V^{(4)}(H_1,H_2,\Delta) $:

\begin{widetext}
	\[
	\begin{matrix}
	V^{(4)}(H_1,H_2,\Delta) &=& \frac{\lambda_1}{2} (H_1^\dagger H_1)^2 + \frac{\lambda_2}{2} (H_2^\dagger H_2)^2 + \lambda_3\, H_1^\dagger H_1\, H_2^\dagger H_2 + \lambda_4\, H_1^\dagger H_2\, H_2^\dagger H_1\nonumber\\
	&+&\frac{\lambda_5}{2} \left[(H_1^\dagger H_2)^2+ (H_2^\dagger H_1)^2 \right] + \lambda_6\,H_1^\dagger H_1Tr\Delta^{\dagger}{\Delta} + \lambda_7\,H_2^\dagger H_2Tr\Delta^{\dagger}{\Delta}\nonumber\\
	&+& \lambda_8\,H_1^\dagger{\Delta}\Delta^{\dagger} H_1 + \lambda_9\,H_2^\dagger{\Delta}\Delta^{\dagger} H_2 + \bar{\lambda}_8 (Tr\Delta^{\dagger}{\Delta})^2+\bar{\lambda}_9 Tr(\Delta^{\dagger}{\Delta})^2
	\end{matrix}	
	\]
\end{widetext}

Hence, the application of positivity criteria to  $V^{(4)}(H_1,H_2,\Delta) $ for all directions would conduce to the full set of necessary and sufficient $BFB$ conditions. To this end,  we follow the efficient prescription used in our previous work \cite{2HDMcT} and implement the convenient parameterization, where the Higgs fields of the $2\mathcal{HDM+T}$ are defined as: 

\begin{eqnarray}
r &\equiv& \sqrt{H_1^\dagger{H_1} + H_2^\dagger{H_2} + Tr\Delta^{\dagger}{\Delta}} \label{eq:para1}\\
H_1^\dagger{H_1} &\equiv& r^2 \cos^2 \theta \sin^2 \phi  \label{eq:para2}\\
H_2^\dagger{H_2} &\equiv& r^2 \sin^2 \theta \sin^2 \phi  \label{eq:para3}\\
Tr\Delta^{\dagger}{\Delta} &\equiv& r^2 \cos^2 \phi  \label{eq:para4}\\
Tr(\Delta^{\dagger}{\Delta})^2/(Tr\Delta^{\dagger}{\Delta})^2 &\equiv& \epsilon  \label{eq:para5}\\
(H_1^\dagger{\Delta}{\Delta}^{\dagger}H_1)/(H_1^\dagger{H_1}Tr\Delta^{\dagger}{\Delta}) &\equiv&  \eta  \label{eq:para6}\\
(H_2^\dagger{\Delta}{\Delta}^{\dagger}H_2)/(H_2^\dagger{H_2}Tr\Delta^{\dagger}{\Delta}) &\equiv&  \zeta \label{eq:para7}
\end{eqnarray}
First, we show that  the $\epsilon$, $\eta$ and $\zeta$ parameters are all equal to $\frac{1}{2}$: 
\begin{equation}
\begin{aligned}
\epsilon=\eta=\zeta=\frac{1}{2}.
\end{aligned}
\end{equation}
Then, after straightforward calculations, we obtain the BFB constraints,
\begin{equation}
\begin{aligned}
\lambda_1>0,\;\;\;\lambda_2>0,\;\;\;\lambda_a>0,\;\;\;\lambda_b>0\;and\;\;\lambda_c>0
\end{aligned}
\end{equation}
\begin{equation}
\begin{aligned}
\lambda_3+\sqrt{\lambda_1\lambda_2}>0,\;\;\;\;\;\lambda_3+\lambda_4-|\lambda_5|+\sqrt{\lambda_1\lambda_2}>0
\end{aligned}
\end{equation}
\begin{equation}
\begin{aligned}
\lambda_a>-\sqrt{2\lambda_1\lambda_c},\;\;\;\;\;\;\lambda_b>-\sqrt{2\lambda_2\lambda_c}
\end{aligned}
\end{equation}
\begin{equation}
\begin{aligned}
4\left(\lambda_3+\lambda_4-|\lambda_5|\right)\lambda_c-2\lambda_a\lambda_b>\\-2\sqrt{\left(2\lambda_1\lambda_c-\lambda_a^2\right)\left(2\lambda_2\lambda_c-\lambda_b^2\right)}
\end{aligned}
\end{equation}
\begin{equation}
\begin{aligned}
4\lambda_3\lambda_c-2\lambda_a\lambda_b>-2\sqrt{\left(2\lambda_1\lambda_c-\lambda_a^2\right)\left(2\lambda_2\lambda_c-\lambda_b^2\right)}
\end{aligned}
\end{equation}

We defer the details of this derivation to Appendix \ref{bfb-appendix}.

\subsection{The modified Veltman  Conditions}
Here, our aim is to tackle the hierarchy problem in $2\mathcal{HDM+T}$ by controling the quadratic divergencies (QD) and seek how  the new degrees of freedom in this model conspire with the $2\mathcal{HDM}$ ones to modify the Veltman conditions in order to soften the divergencies \cite{Veltman}. 

\paragraph{}
To this end, we derived the quadratic divergences of the Higgs self-energies in terms of the original fields, namely the doublet $\Phi_1$, $\Phi_2$ and triplet $\Delta$, without spontaneous breaking of the SU(2)x U(1) gauge symmetry. Indeed, the calculation is straightforward when the symmetry is kept intact, compared to that performed in the broken phase where the algebra is much more involved \cite{Osland, Newton, Ma}.  To determine the modified Veltman conditions (VC), we employed dimensional regularization to collect the quadratic divergences \cite{Veltman}, since this prescription secures  gauge and Lorentz invariance. Moreover we performed this calculation in a general linear $R_{\zeta}$ gauge and checked that the obtained results are clearly free from $\zeta$-parameter as it should be.  \\

We present hereafter the essential of quadratic divergences derivation of the Higgs self-energy in the symmetry unbroken phase for $2\mathcal{HDM}+T$ model. So the calculation is performed in terms of the original scalar fields, namely the doublet $\Phi_1$, $\Phi_2$ and triplet $\Delta$ that we represent  as: \footnote{For the triplet, we used the convenient notation:
	\begin{eqnarray}
	\Delta_1=\delta^+, \hspace {0.5cm}
	\Delta_2=\delta^0, \hspace {0.5cm}
	\Delta_3=\delta^- \nonumber
	\end{eqnarray}
}
\begin{eqnarray}
\Phi_1=\left(
\begin{array}{c}
\phi_1^+ \\
\phi_1^0 \\
\end{array}
\right)~,~~~~\Phi_2=\left(
\begin{array}{c}
\phi_2^+ \\
\phi_2^0 \\
\end{array}
\right)~,~~~~~\Delta=\left(
\begin{array}{cc}
\delta^0/2 & \delta^+/\sqrt{2} \\
\delta^-/\sqrt{2}  & -\delta^0/2\\
\end{array}
\right)
\label{doublet}
\end{eqnarray}

\paragraph{}

We followed the strategy of \cite{Osland, Newton} and used the following convenient notation through our calculation: 
\begin{itemize}
	\item $(\Phi)p$ and $(\Delta)q$ denotes the $p$-component of the doublets and q-component of the triplet fields. 
	\item $I_{11}$ ($I_{22}$) as the quadratically divergent part of the two-point functions with either the upper or lower components of $\Phi_1$ ($\Phi_2$) fields, on both external lines of the relevant Feynman Diagrams. Similarly, we label $I_{33}$ for the triplet two-point functions,  with one of $\Delta$ components on external lines.
\end{itemize}

   To get the final results in symmetry unbroken phase, one has just to sum up all the possible diagrams, keeping only the coefficients of divergent parts in $I_{ij}$, to readily obtain:
\begin{eqnarray}
I_{11}&\Rightarrow&3/4g^2+1/4g^{'2}+3\lambda_1+2\lambda_3 + \lambda_4+3/2\lambda_a-\left(\frac{\sqrt{2}}{v}\right)^2 \frac{m_D^2}{ \cos^2\beta}
\label{I11} \\
I_{22}&\Rightarrow&3/4g^2+1/4g^{'2}+3\lambda_2+2\lambda_3+\lambda_4+3/2\lambda_b-\left(\frac{\sqrt{2}}{v}\right)^2 \frac{m_U^2}{sin^2\beta}
\label{I22} \\
I_{33}&\Rightarrow&g^2+\lambda_a+\lambda_b+\frac{5\lambda_c}{2}
\label{I33}
\end{eqnarray}

with $m_D^2=m_e^2+m_\mu^2+m_\tau^2+3(m_d^2+m_s^2+m_b^2)$ and $m_U^2=3(m_u^2+m_c^2+m_t^2)$. \\

At last, thanks to the relations $ \cos \beta=\frac{v_1}{v}$, $ \sin \beta=\frac{v_2}{v}$, $m_W=\frac{gv}{2}$, $e=g~ \sin\theta_W$ and $g^{'}=g~\tan \theta_W$, one can recover the tadpoles equations in the broken phase (physical Higgs fields): 
\begin{eqnarray}
\delta T_{d_1} &=&  3 \lambda _a+6 \lambda _1+4 \lambda _3+2 \lambda _4+\frac{ (2 c_W^2+1) e m_W}{c_W^2 s_W v}-\frac{ e m_D^2 Tr(I_n) v}{m_W s_W v_1^2}  
\label{Td1} \\
\delta T_{d_2} &=& 3 \lambda _b+6 \lambda _2+4 \lambda _3+2 \lambda _4+\frac{ (2 c_W^2+1) e m_W}{c_W^2 s_W v}-\frac{ e m_U^2 Tr(I_n) v}{m_W s_W v_2^2} 
\label{Td2} \\
\delta T_{t} &=& 2 \lambda _{a}+2 \lambda _{b}+5 \lambda _{c}+\frac{4 e m_W}{s_W v}  
\label{Td3}
\end{eqnarray}

Here  $Tr(I_n)$ is the trace of n-dimensional identity Dirac matrix, $Tr(I_n)=2^{\frac{n}{2}}=2$ in the subsequent calculations \footnote{The space-time dimension $n$ to pick up the quadratic divergences depends on the number of loops $L$ via the formula $n= 4-\frac{2}{L}$  \cite{jones90}}.
\paragraph{}
At this stage, several remarks are in order: First, notice that the potential parameter $\lambda_1$ is lacking in $\delta T_{d_2}$ ($\lambda_2$ in $\delta T_{d_1}$)  and $\delta T_{t}$ is obvious since $\lambda_1$  ($\lambda_2$) couples solely the $H_1$ ($H_2$) field respectively. Also, being only concerned with the triplet scalar,  the couplings  $\lambda_c$ occurs solely in $\delta T_t$. Similarly, the $\lambda_a$ is not seen in $\delta T_{d_2}$ since it is rather connected with $H_1$ potential terms. Lastly, note that the Veltman conditions for  the Higgs Triplet Model with $Y=0$ reported in \cite{HTMY0} can be readily recovered when the couplings $\lambda_2$, $\lambda_3$, $\lambda_4$, $\lambda_b$ are canceled, and $v_1$ traded for $v_d$ in Eqs.~(\ref{Td1}, \ref{Td2}, \ref{Td3}). Similarly,  we can also see that modified Veltman conditions in two Higgs doublets model \cite{Grz, Barz, Masina, kundu14, biswas15, Chowdh, maria} are reproduced if the couplings fingerprinting scalar triplet in the Lagrangian, namely $\lambda_a$, $\lambda_b$ and $\lambda_c$, are removed from Eqs.~(\ref{Td1}, \ref{Td2}, \ref{Td3}).
\paragraph{}
In order to implement the three VC's Eqs.~(\ref{I11}, \ref{I22}, \ref{I33}). in the parameter space and the subsequent scans, we generally assume  that their deviations $ \delta T$ should not exceed a magnitude of $5$.

\section{EXPERIMENTAL CONSTRAINTS:}
\subsection{Electroweak $\rho$ parameter:}
First, recall that though the $\rho$ parameter in $2\mathcal{HDM+T}$ deviates from the unity, consistency with electroweak precision measurements of $\rho=1.00039\pm 0.0019$ \cite{pdg2018} can be assured. Indeed, from a tree level calculation of $\rho$,  
\begin{eqnarray}
\rho=\frac{v_1^2+v_2^2+4v_t^2}{v_1^2+v_2^2}=1+4\frac{v_t^2}{v_d^2}  \label{eq:rho-2HDMt}
\end{eqnarray}
we see that the deviation must then satisfy the limit, $\delta \rho= (\frac{2v_t}{v_d})^2 \le 0.0006 $,  thereby setting an upper bound on the triplet $VEV$,  $v_t < 3$ GeV, when $2\sigma$ errors are assumed. 

\subsection{Constraints from Higgs data}
 Limits on heavy Higgs masses have been reported by LEP and LHC. From the LEP direct search results,  the lower bounds on neutral scalar masses, $m_{A^0, H^0} >  80-90$ GeV for models with more than one doublet, while the charged Higgs  $m_{H^{\pm}}$ below the Z boson mass has been excluded \cite{LEP2013}. The LEP II indirect limit is even higher with  $m_{H^{\pm}} \geq 125$ GeV. Indeed it is well known that searches for neutral Higgs can induce an indirect, model dependent, limits on the charged Higgs bosons. For example in $2HDM$, such correlation between the neutral and charged sectors is revealed via the relation: $$	M^2_{H^\pm} = M^2_{A} + \frac{1}{2} v^{2} (\lambda_3-\lambda_4)$$  while in $MSSM$, one considers the tree level formula  (at tree level): $$M^2_{H^\pm} = M^2_{A} +  M^2_{W}$$  \\
 This suggests an implication of the pseudo-scalar Higgs $A$ in $H^\pm$ sector. Thereby, an indirect lower bound on the charged Higgs can indeed be estimated from the LEP II limits on the neutral Higgs bosons  \cite{Roy2004, Banerjee2006}.  \\

 Many other constraints on $m_{H^\pm}$ have been established from measurements of the inclusive weak radiative B-meson decay branching ratio. Recently,  a lower bound on the mass of $H^{\pm}$ from $B \to X_s \gamma$ data has been set to around $480$ GeV In \cite{Misiak2015}.  An even higher limit, $m_{H^{\pm}} \ge 570$ GeV, has been reported in \cite{Misiak2017}.  Furthermore, ATLAS \cite{atlas_charged} and CMS  \cite{cms_charged} collaborations have searched  for the production of charged Higgs boson using several different final states.  Thus, exclusion limits were released either for  $H^{\pm}$ lighter or heavier than the top mass. Some of these limits were recently re-interpreted in the context of $BSM$ models with non minimal scalar sectors. As example, ATLAS data for search of $H^{\pm}$ produced via $VBF$ and decaying into  $W^{\pm}Z$, excluded the charged Higgs with a mass in the range $240 \le m_{H^{\pm}} \leq 700$ GeV within the Georgi-Machacek Model \cite{atlas_htm}. \\

In order to confront ATLAS and CMS measurements to $2\mathcal{ HDM+T}$ model, the signal strengths, a directly observable quantity, is generally employed. In our calculation, we  rather  use the ratio adopted in \cite{arhrib2012} for the Higgs decay to diphoton, generically given by:

\begin{equation}
R_{\gamma\gamma}(h_1) = \frac{\Gamma{(h_1\to\,gg)}\times\,BR{(h_1\to\gamma\gamma)}}{\Gamma^{(SM)}{(h_1\to\,gg)}\times\,BR^{(SM)}{(h_1\to\gamma\gamma}} 
\label{mu}
\end{equation}

It is worth to notice that two approximations have been used when identifying this ratio, namely: 
$1)$ The ratio  $R_{\gamma\gamma} $concerns only the leading parton level gluon fusion Higgs production contribution. $2)$ The narrow width approximation is assumed.\\
The ratios relevant for the other decay channels $Z\gamma$, $b\bar{b}$, $\tau^+\tau^-$, $W^+W^-$ and $ZZ$ are defined in a similar way. For the constraints and bounds from their corresponding signal strength measurements, we require agreement with the ATLAS and CMS at least at $1 \sigma$.  \\
 
To conclude this subsection, it should be stressed that, throughout this paper,  we have used the public code HiggsBounds-5.3.2beta to test compatibility of the theoretical Higgs predictions in our model against  various exclusion bounds and limits from LEP, ATLAS, CMS and Tevatron experiments.  HiggsSignals-2.2.3beta is also used to implement the Higgs rate measurements at the LHC (and the Tevatron).  

\begin{figure*}[!h]
	\centering
	\resizebox{0.32\textwidth}{!}{
		\includegraphics{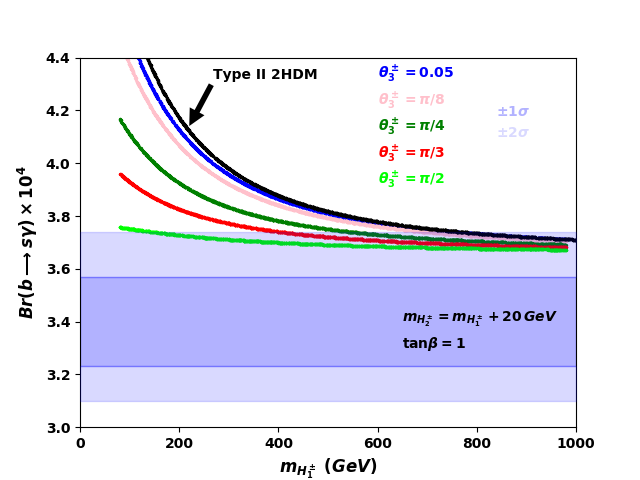}}
	\resizebox{0.32\textwidth}{!}{
		\includegraphics{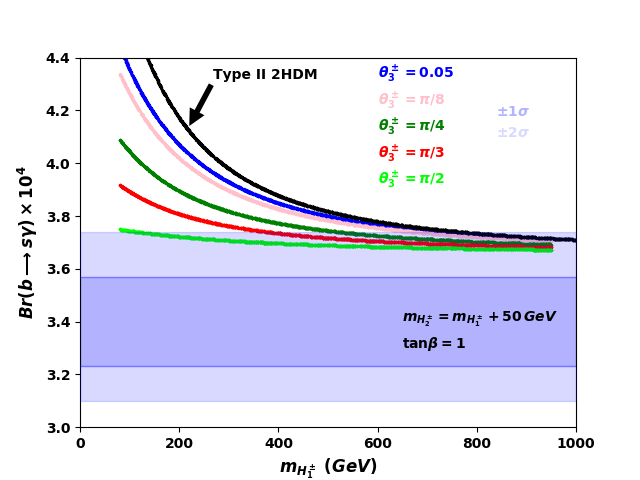}}
	\resizebox{0.32\textwidth}{!}{
		\includegraphics{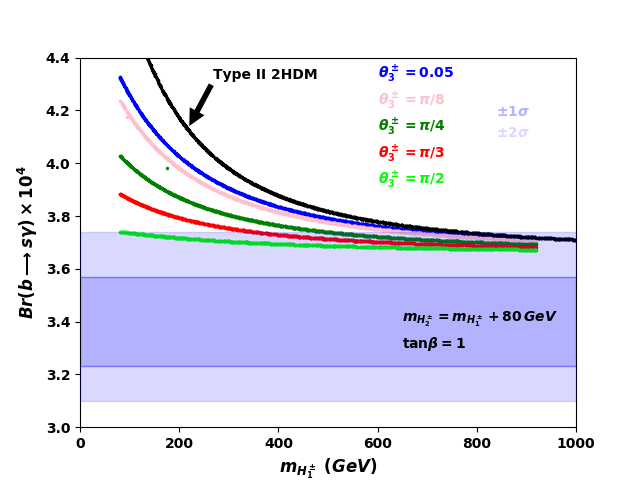}}
	\resizebox{0.32\textwidth}{!}{
		\includegraphics{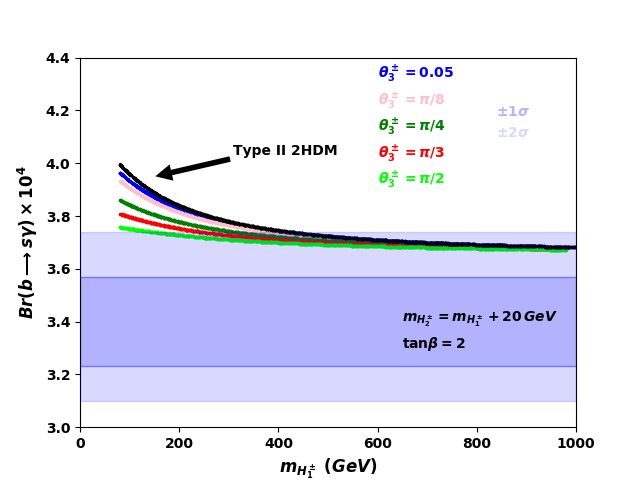}}
	\resizebox{0.32\textwidth}{!}{
		\includegraphics{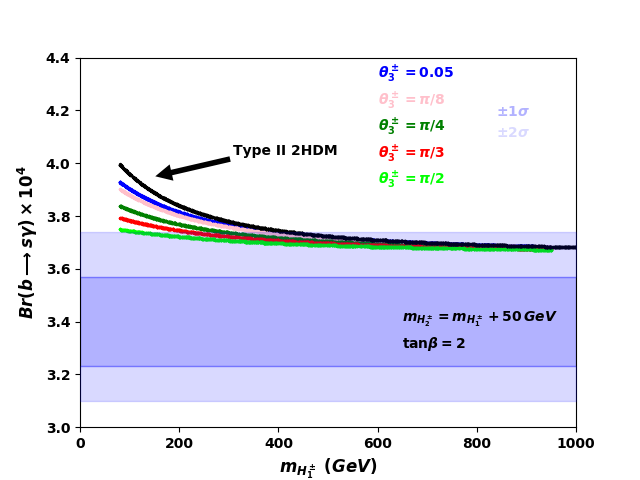}}
	\resizebox{0.32\textwidth}{!}{
		\includegraphics{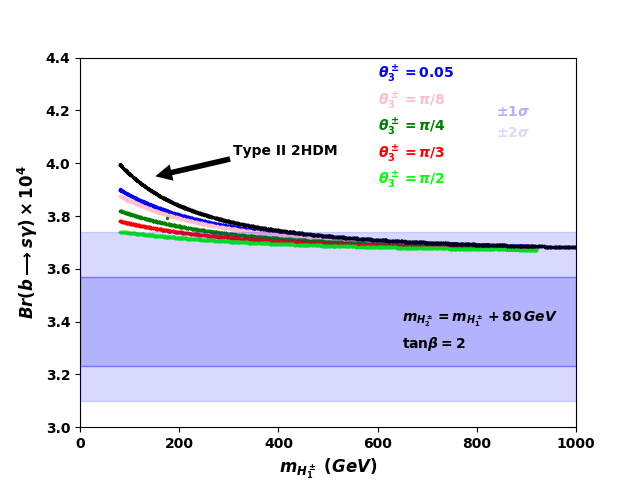}}
	\resizebox{0.32\textwidth}{!}{
		\includegraphics{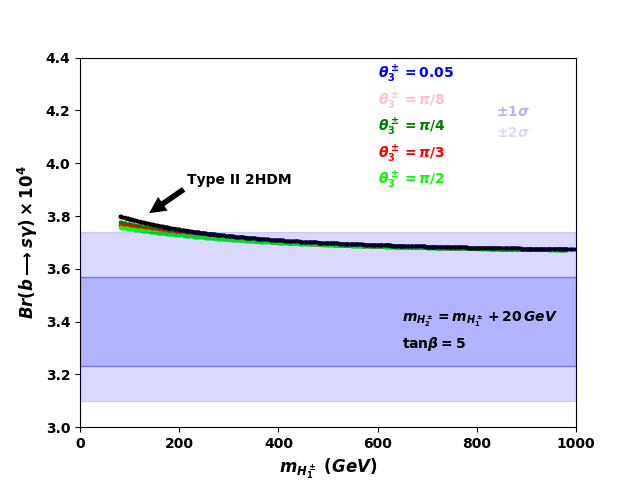}}
	\resizebox{0.32\textwidth}{!}{
		\includegraphics{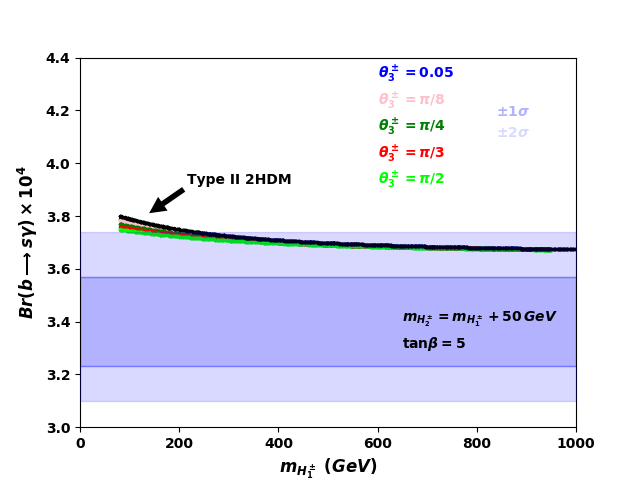}}
	\resizebox{0.32\textwidth}{!}{
		\includegraphics{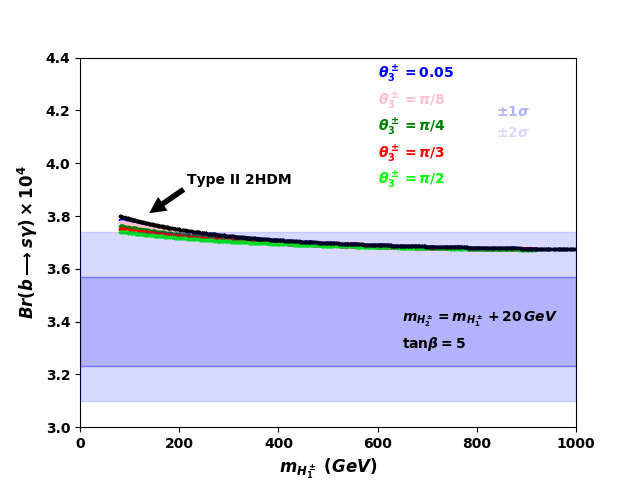}}
	\resizebox{0.32\textwidth}{!}{
		\includegraphics{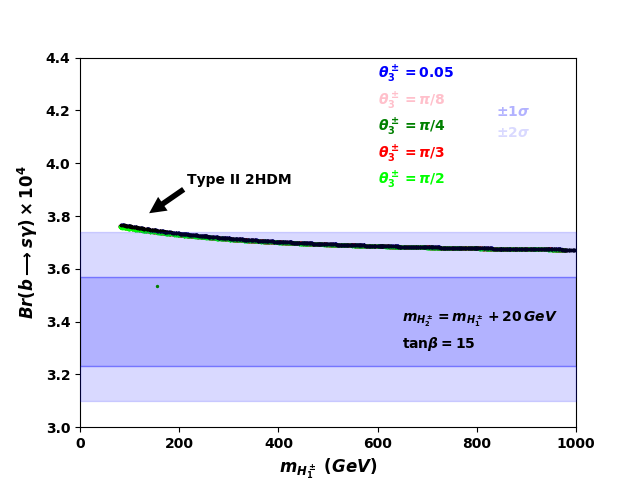}}
	\resizebox{0.32\textwidth}{!}{
		\includegraphics{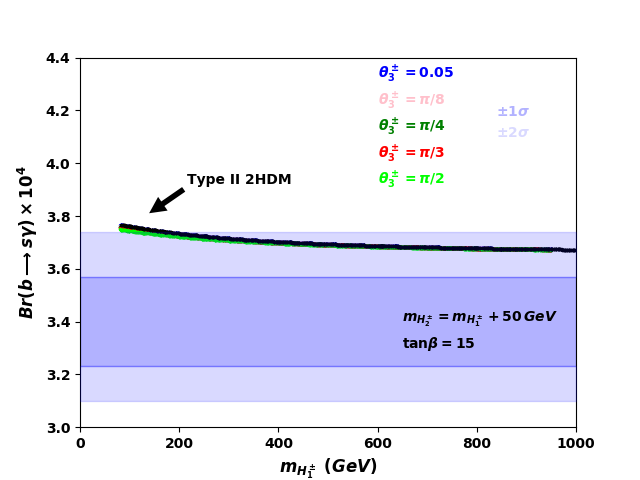}}
	\resizebox{0.32\textwidth}{!}{
		\includegraphics{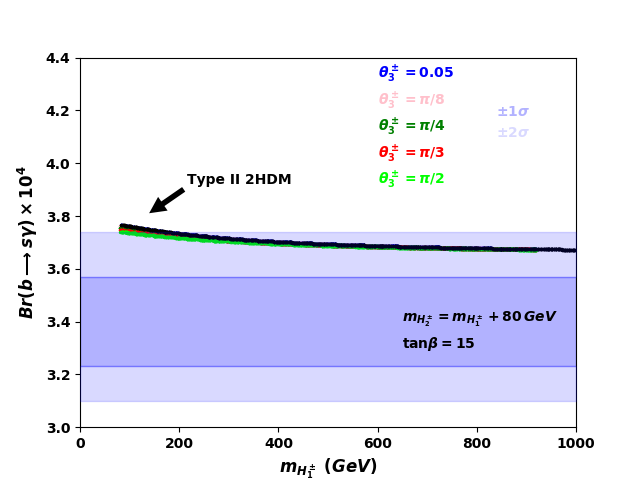}}
	\caption{The branching ratio of $B \rightarrow X_s \gamma$ in $2HDM+T$ with 
		$m_A \in [80,\;1000]$ (GeV), $m_{H^\pm_1} \in [80,\;1000]$ (GeV), $\lambda_a \in[0,\;5]$, $\lambda_c \in[0,\;4]$, $\lambda_4 \in[-7,\;8]$, $\mu_1=20$, $v_t=1$ GeV and $\alpha_{1,2,3}\in[-\pi/2\,,\pi/2]$ as a function of $m_{H^\pm_1}$. The left, centre and right panels correspond respectively to  $m_{H_2^\pm}-m_{H_1^\pm}=20,\;50$ and $80$ GeV, with  $\tan \beta=1,\;2,\;5,\;15$.}
	\label{fig:constraints}
	
\end{figure*}	

\subsection{Constraint from $B \to X_s \gamma$ on charged Higgs}
In this subsection we aim  to see whether  $2\mathcal{HDM+T}$ model can accomodate a light charged Higgs. This feature occured in some models with two charged Higgs bosons. For example, in Three Higgs Double Model ($3HDM$), where it has been shown that the mass of one  the two charged Higgs bosons can be smaller than the top mass without contradicting the $b \to s \gamma$ data \cite{Akeroyd2016_1, Akeroyd2018}. Hence we performed a preliminary study of the $b \to s \gamma$ decay rate at leading order in $2HDM+T$ where both charged Higgs are contributing. For this, we simply extended the analysis of $2HDM$ in \cite{Borzumati1998, Akeroyd2016_2}.  

Hence, we consider the branching ratio for $b \to s \gamma$ defined as:
\begin{eqnarray}
BR(B \rightarrow X_s \gamma) = \frac{\Gamma(B\rightarrow X_s \gamma)}{\Gamma_{SL}}BR_{SL}	
\end{eqnarray}
where $BR_{SL}$ and $\Gamma_{SL}$ are respectively the measured semileptonic branching ratio, and the semileptonic decay
width. The latter reads :
\begin{eqnarray}
\Gamma_{SL} = \frac{G_F^2}{192 \pi^3}|V_{cb}|^2m_b^5 g(z)\left(1-\frac{2\alpha_s(\bar\mu_b)}{3\pi}f(z)+\frac{\delta^{NP}_{SL}}{m_b^2}\right)
\end{eqnarray}
with $z=\frac{m_c^2}{m_b^2}$. The phase space function $g(z)$, the QCD radiation function $f(z)$ and the non–perturbative correction $\delta^{NP}_{SL}$ are given in  \cite{Borzumati1998}.  \\

 The  $B\rightarrow X_s \gamma$ transition proceeds via penguin diagrams, which involve both top quark and $W^\pm$ gauge bosons  in the loop with charged Higgs bosons exchange. It is generally sensitive to the values of $\tan  \beta$ and $M_{H^\pm}$. At leading order its decay width is  given by,
\begin{eqnarray} 
\Gamma(B\rightarrow X_s \gamma) = \frac{G_F^2}{32 \pi^4}|V_{ts}^*V_{tb}|^2\alpha_{em}m_b^5|\bar D|^2 
\end{eqnarray}
where the amplitude $\bar D$ is given by:
\begin{eqnarray}
\bar D = C^{0,eff}_7(\mu_b) + \frac{\alpha_s(\mu_b)}{4\pi}V(\mu_b)
\end{eqnarray}
The effective Wilson coefficient $C^{0,eff}_7(\mu_b)$ \cite{Akeroyd2016_1} read,
\begin{eqnarray}
C^{0,eff}_7(\mu_b)= \eta^{\frac{16}{23}}C^{0,eff}_7(\mu_W)+\frac{8}{3}(\eta^{\frac{14}{23}}-\eta^{\frac{16}{23}})C^{0,eff}_8(\mu_W)    +\sum_{i=1}^{8}h_i\eta^{a_i}
\end{eqnarray}
\\
where $\mu_b$ and $ \mu_W$ are the b$-$quark and $W$boson mass scales respectively.  The parameter $\eta$ is defined as $\eta=\frac{\alpha_s(\mu_W) }{\alpha_s(\mu_b)}$ and $a_i$ and $h_i$ are the leading log QCD corrections in the SM \cite{Borzumati1998}. \\
\\

 The effective Wilson coefficient $C^{0, eff}_7(\mu_W)$ is given by:\\ 
\begin{eqnarray}
C^{0,eff}_7(\mu_W)=C^{0}_{7,SM}+|Y_1|^2C^{0}_{7,Y_1Y_1}+|Y_2|^2C^{0}_{7,Y_2Y_2}+X_1Y_1^*C^{0}_{7,X_1Y_1}+X_2Y_2^*C^{0}_{7,X_2Y_2} 
\end{eqnarray}
The coefficient $C_{7,SM}^0(\mu_W)$ is function of $x=m^2_t/M_W^2$ , while $C^0_{7,j}(\mu_W)$ $(j=YY,\;XY )$ is function of $y=m_t^2/m^2_{H^\pm_a}$ ($H^\pm_a=H^\pm_1,\;H^\pm_2$). In our model, the couplings $Y_1$, $Y_2$, $X_1$ and $X_2$ read:
\begin{eqnarray}
Y_1= \frac{\cos\theta_3^\pm}{\tan \beta},\;\;\; Y_2= -\frac{\sin\theta_3^\pm}{\tan \beta},\;\;\; X_1=\cos\theta_3^\pm\tan \beta\;\;\; and \;\;\; 	X_2= -\sin\theta_3^\pm\tan \beta
\end{eqnarray} 

All the necessary ingredients relevant to examine, at leading order, the radiative $b \to s \gamma$ in $2HDM+T$ are now defined.  We present  the branching ratio of $B \rightarrow X_s \gamma$ in the type II $2THD+T$ as a function of $m_{H^\pm_1}$. The mass difference $\Delta  M_{H^\pm}=m_{H^\pm_2}-m_{H^\pm_1}$ is taken to be $20$ (left panel), $50$ (centre panel) and $80$ GeV (right panel) with $\tan \beta=1,\;2,\;15$. We also used different values of the mixing angle  $\theta_3$ ranging from  $0.05$ ($\pi/2$) for a mostly triplet $H_2^\pm  (H_1^\pm )$, to $ \pi/4$ corresponding a nearly equal doublet-triplet contributions to the charged Higgs bosons. Besides, to compare the predictions in Type II $2HDM$ with those in Type-II $2HDM+T$, we also plot the black curves to present the results in $2HDM$. First it is worth to notice that the difference between the prediction in $2HDM$ and $2HDM+T$ becomes slightly larger as the mass difference $\Delta  M_{H^\pm}$ increases. \\
\\
As illustrated in Fig.~\ref{fig:constraints}, for certain ranges of the model parameters, at $ 1\sigma $,  the resulting value of $b \to s \gamma$  exceeds its current world average measurements  \cite{pdg}. However the consistency with experimental data can be reached within $ 2\sigma $, for small values of $\tan \beta$. In this case,  a part of the $m_{H^\pm_1}$ is excluded, but a light charged Higgs ${H^\pm_1} $ is still viable, essentially when the doublet triplet mixing $\theta_3$ is far from the extremal values $0$ and $\pi/2$, with a mass lower limit in the range between $154 - 174$ GeV. For high $\tan \beta$, the results being almost insensitive to $\theta_3$ and $\Delta M_{H^\pm}$ values, coincide with those of $2HDM$ at leading order. It is clearly seen that $m_{H^\pm_1} < m_t$ is still allowed and do not contradict $b \to s \gamma$ at $ 2\sigma $.\\

\section{ANALYSIS AND RESULTS }
\subsection{Allowed Parameter Space}

In this section, we generate points in parameter space that pass all theoretical constraints previously derived. A  particular emphasis is placed on the effects of three Veltman conditions mVC's.   In the subsequent analysis,  we assume that the deviations on $\delta T$ such as $|\delta T|\le 5 $.\\

We also require that all these points comply with LEP and LHC signal strengths $\mu_f$ for all final states $\gamma\gamma$, $\gamma Z$, $\tau^+\tau^-$, $W^+W^-$, $ZZ$, and $b\bar{b}$. \\

The following inputs are used  in the numerical analysis, 
\begin{widetext}
	\[
	\begin{matrix}
	m_{h_1}=125.09\,\,\text{GeV},\,128\, \text{GeV} \leq m_{h_2}\leq m_{h_3}\leq 1000\,\text{GeV},\,\,\\
	80\,\text{GeV}\leq m_{H^{\pm}_1} \leq m_{H^{\pm}_2}\leq 1000\,\text{GeV},\,\, 80\,\text{GeV}\leq m_{A^{0}}\leq 1000\,\text{GeV},\,\,\,\frac{-\pi}{2}\leq \alpha_{1,2,3}\leq \frac{\pi}{2}\\
	0.5\leq \tan\beta\leq 25,\,\,-10^2\leq \mu_1\leq 10^2,\,\,0\le v_t\leq 3 \,\text{GeV},\,\,\,\frac{-\pi}{2}\leq \theta_{3}^\pm\leq \frac{\pi}{2},\lambda_4\in[-16,\;16]
	\end{matrix}
	\]
\end{widetext}

\begin{figure*}[!h]
	\centering
	\resizebox{0.42\textwidth}{!}{
		\includegraphics{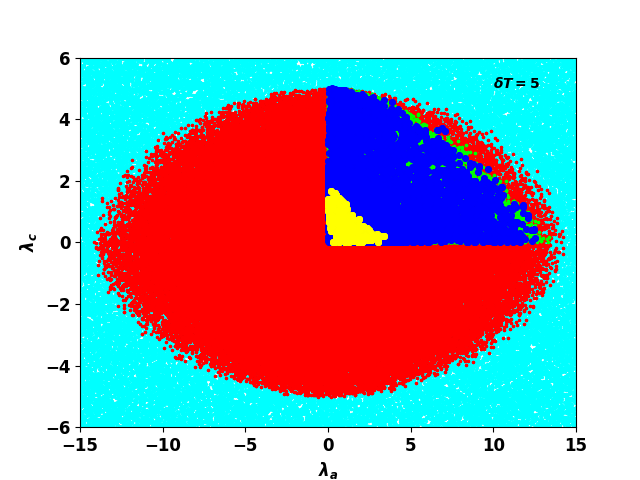}}
	\resizebox{0.42\textwidth}{!}{
		\includegraphics{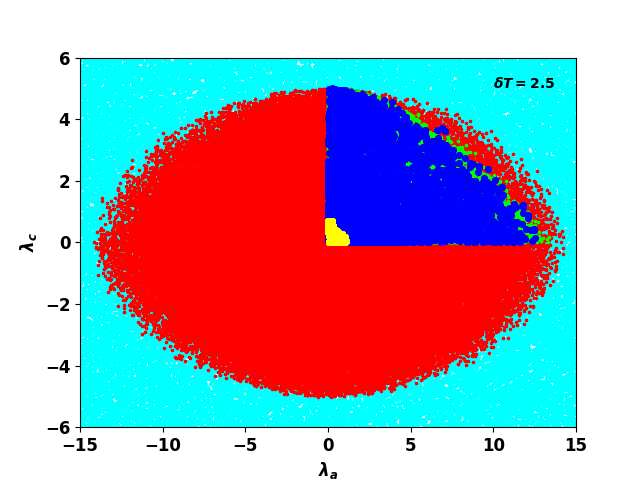}}
	\resizebox{0.42\textwidth}{!}{
		\includegraphics{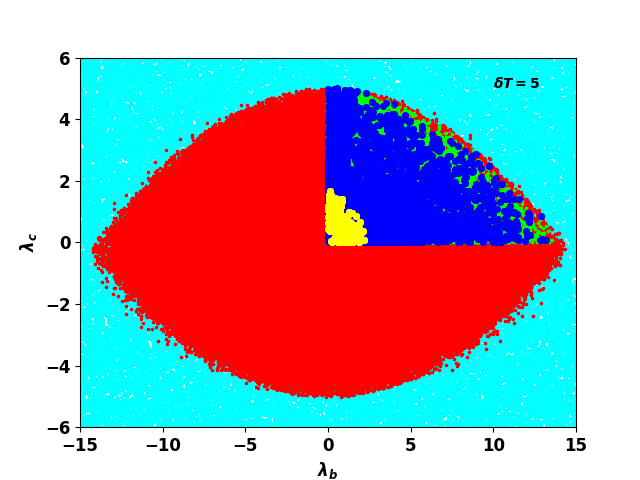}}
	\resizebox{0.42\textwidth}{!}{
		\includegraphics{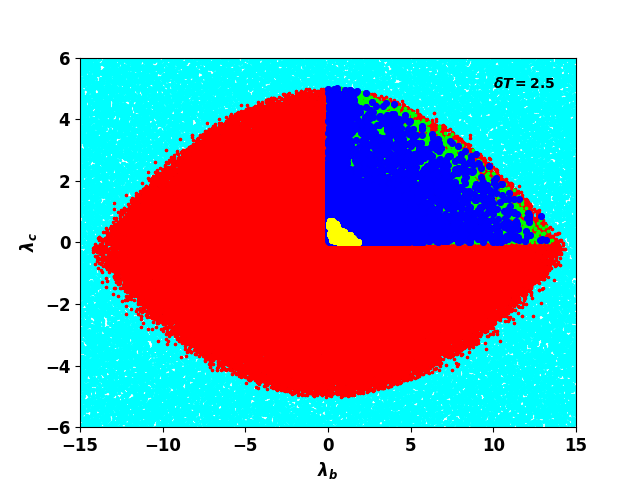}}
	\caption{The allowed regions in ($\lambda_a$, $\lambda_c$) (Top) and ($\lambda_b$, $\lambda_c$) (Bottom) after imposing theoretical and experimental constraints. (\textcolor{cyan}{cyan}): Excluded by Unitarity constraints; (\textcolor{red}{red}):
		Unitarity + BFB constraints; (\textcolor{green}{green}): Excluded
		by Unitarity+BFB+LHC constraints; (\textcolor{blue}{blue}): Excluded by
		Unitarity + BFB +  LHC  $\&$ $T_{d_1}=0$ $\wedge$ $T_{d_2}=0$ $\wedge$ $T_{t}=0$ constraints. Only the
		\textcolor{yellow}{yellow} areas obey all constraints.  Here, the errors
		for $\chi^2$ fit are $95.5\%$ C.L. }
	\label{fig:1}
\end{figure*}

We must stress that the $b  \to s \gamma$ stringent restrictions on the charged Higgs mass bound $m_{H^{\pm}} > 570$ GeV is relaxed in our analysis. Indeed, the extra scalars in a $BSM$ model can absolutely induce a significant alteration of these limits, since the re-interpretation of these  flavor measurements are generally model dependent. \\

\begin{widetext}
	\begin{figure}[!h]
		\centering	
		\resizebox{0.32\textwidth}{!}{
			\includegraphics{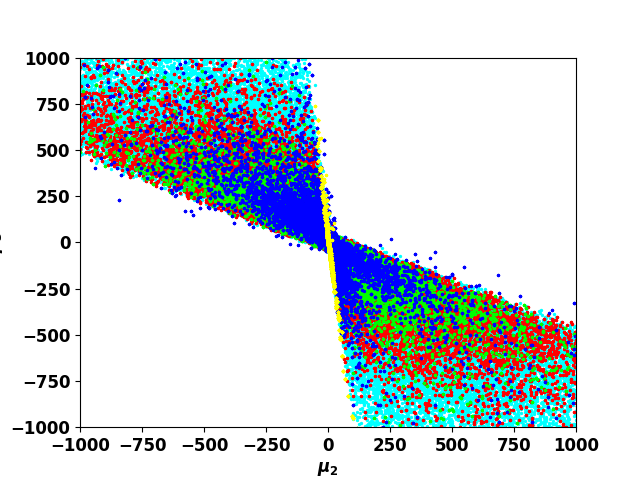}}
		\resizebox{0.32\textwidth}{!}{
			\includegraphics{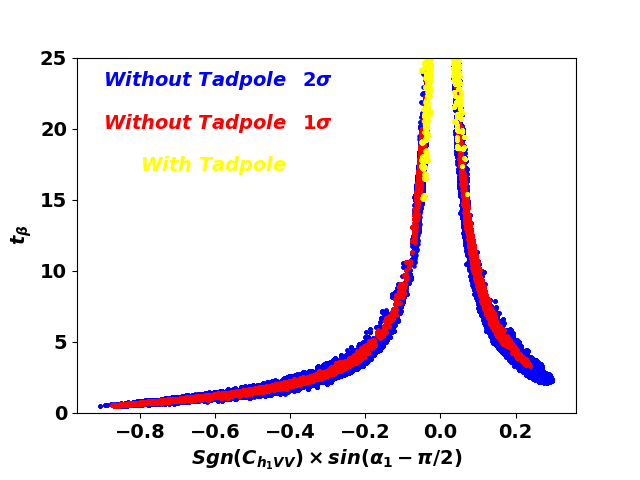}}
		\caption[titre court]{The allowed regions after imposing theoretical and experimental constraints in:  ($\mu_2$, $\mu_3$) plane (left ); ($sgn(C_V^{h_1})\sin(\alpha_1-\pi/2)$, $\tan\beta$) for $\delta T=2.5$, with and without Veltman conditions (right). The color caption in left panel is similar to Fig.~\ref{fig:1}. The errors for $\chi^2$ fit in right panel are $95.5\%$ C.L. (blue), and $68\%$ C.L. (red and yellow).}
		\label{fig:2}
	\end{figure} 
\end{widetext}

Fig.~\ref{fig:1} displays the excluded regions in $\left(\lambda_a\,, \,\lambda_c\right)$ and $\left(\lambda_b \,, \,\lambda_c\right)$ by various theoretical constraints and LHC measurements. The variation of Veltman conditions are fixed to  $\delta T= 5$ in the left panel, and $\delta T= 2.5$ in the right panel. We clearly see that the allowed regions undergo drastic reduction as we add constraints. Once naturalness is invoked, the parameter spaces are sizably shrinked to limited areas, indicated in yellow, with extent depending on $\delta T$ values.  As results, the allowed ranges for these potential parameters are: 
$$\lambda_a\in [0\;,\;3.5], \hspace{0.8cm}  \lambda_b \in [0\;,\,2.25],  \hspace{0.8cm} \lambda_c \in [0\;,\;1.18]  \hspace{0.6cm} when \hspace{0.2cm} \delta T= 5$$
In Fig.~\ref{fig:2}, the left panel presents the points in $(\mu_2,\mu_3)$ plane that pass both theoretical and experimental constraints.  We show the excluded regions of  parameter space by unitarity in cyan, and  by the combined sets of BFB and unitarity in red. When consistency with  combined data from LEP, ATLAS and CMS is imposed as well the blue area is also ruled out.  At last, if naturalness induced conditions take place, the allowed parameter space is reduced even more and only a small strip marked in yellow survives. More precisely, we find that $\mu_2$ and $\mu_3$ parameters are more sensitive to the naturalness conditions, mainly  to $ \delta T_t$,  than to the other theoretical constraints. As a result, $\mu_2$ and $\mu_3$ could be either positive or negative varying within the range  $ [-42\,,\,61]$ and  $[-695\,,\,523]$ respectively. The right panel illustrates the scatter plot in $\tan \beta$ and $sgn(C_V^{h_1})\times\sin(\alpha_1-\pi/2)$ for $\Delta \xi^2 \le 5.99$ and $2.3$ respectively. Without naturalness consideration, the corresponding generated samples are illustrated in red at $1\sigma$ and in blue at $2\sigma$ while the yellow points signal inclusion of Veltman conditions at $1\sigma$.  This plot shows that only $\tan \beta >  15$ and $\sin \alpha_1$ delineated by the interval $[-1.57\,;-1.51]U[1.51\,;1.57]$ comply with all constraints.  At this stage, note that the left branch with $\sin \alpha_1 < 0$, lies close to  $\sin(\beta - \alpha_1) = 1$ corresponding to the SM-alignment limit, that is where the couplings of CP even scalars to gauge bosons are assumed to mimic the SM Higgs coupling.  The right branch corresponding to  $\sin \alpha_1 > 0 $  represents the so-called  wrong sign Yukaya coupling limit.  

\subsection{Implications on Heavy Higgs masses}
In this subsection, the light CP-even Higgs boson $h_1$ being identified to  the SM-like Higgs with the mass of $125$ GeV, we explore to what extent the nonstandard Higgs spectrum of $2\mathcal{HDM+T}$, namely  $ h_2, h_3, A_0, H^{\pm}_1$ and $ H^{\pm}_2$ could be probed via theoretical constraints. A particular emphasis is put on naturalness to show how it has impacted their masses. In addition,  the corresponding parameter spaces have to comply  with LEP and LHC measurements for all Higgs decay channels, though in the subsequent  analysis,  we only show results with the diphoton mode and its correlation with $Z \gamma$ mode. 

That being said, it is also worth to remind that the decay $h_1\to\gamma\gamma, Z\gamma$ are loop processes mediated at one loop level by virtual exchange of SM particles (fermions and gauge bosons) and new charged Higgs states ($H^{\pm}_1$ and $H^{\pm}_2$) predicted by $2\mathcal{HDM+T}$. All tree level Higgs couplings to fermions and bosons in this model  depend essentially on the mixing angles  $\alpha_i$, $\theta_i^\pm$ and $\beta_i$. Thus,  the interference between charged scalar loop contributions and  those of the $W^\pm$ and $f=(t, b, c,\tau)$ loops depends on the sign of $g_{h_1H^{\pm}_iH^{\pm}_j}$ couplings,  which could result either in an enhancement or suppression of the $h_1 \to\gamma\gamma, Z\gamma$ decay modes with respect to SM predictions. 

\begin{figure*}[t]
	\centering	
	\resizebox{0.42\textwidth}{!}{
		\includegraphics{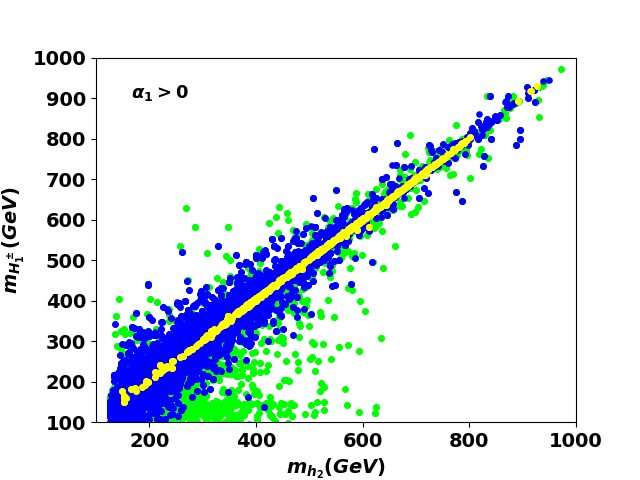}}
	\resizebox{0.42\textwidth}{!}{
		\includegraphics{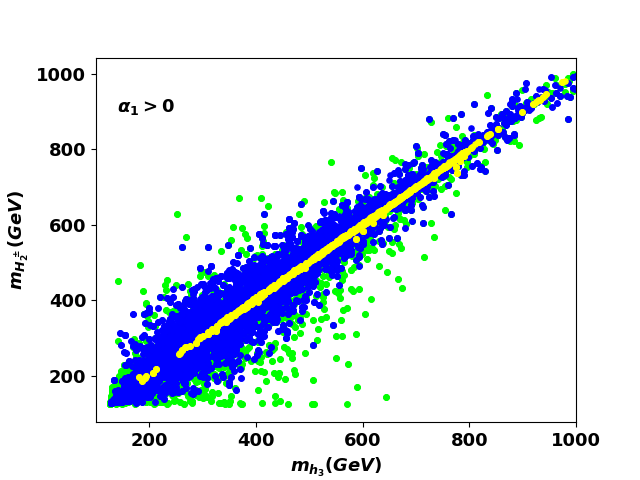}}
	\resizebox{0.42\textwidth}{!}{
		\includegraphics{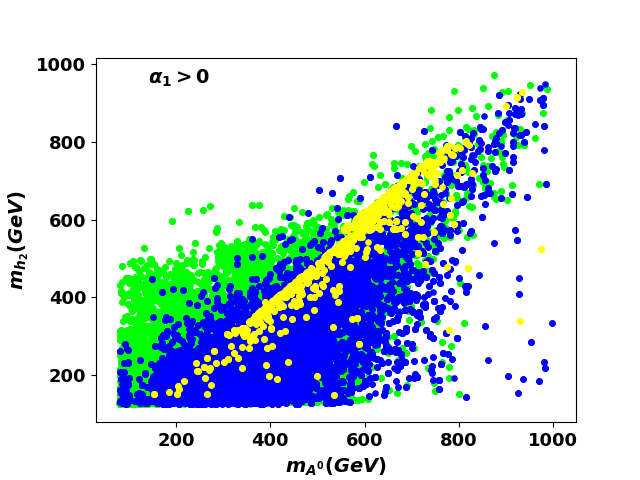}}
	\resizebox{0.42\textwidth}{!}{
		\includegraphics{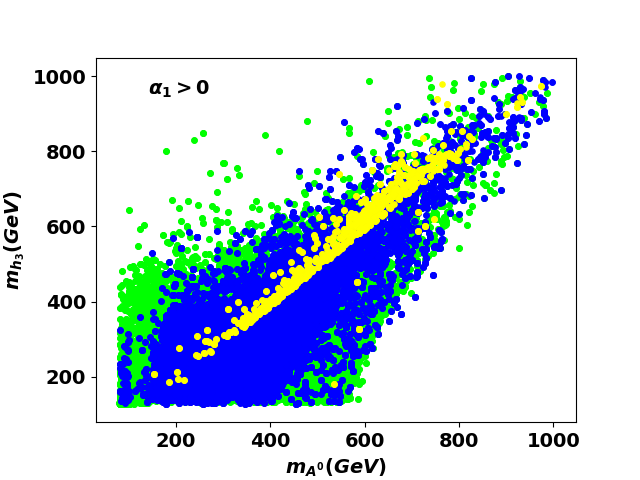}}
	\caption{Allowed Higgs mass ranges  in the planes $m_{\phi_i}$ vs $m_{\phi_j}$  for $\sin \alpha_1 > 0 $ ($\phi_i=h_i,\,A$ and  $\phi_j=H^{\pm}_j$). All theoretical and experimental constraints are taken into account with colors caption similar to Fig.~\ref{fig:1}. The yellow region indicates  surviving regions to Veltman conditions for $\delta T=2.5$. The error for $\chi^2$ fit is $95.5\%$ C.L.}
	\label{fig:3}
\end{figure*}

\begin{figure*}[t]
	\centering	
	\resizebox{0.42\textwidth}{!}{
		\includegraphics{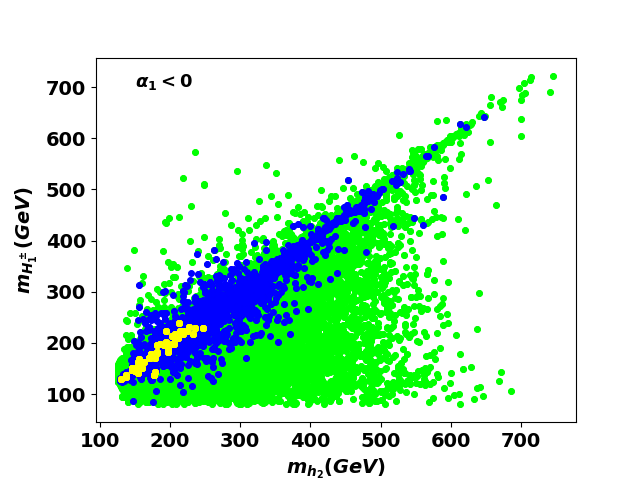}}
	\resizebox{0.42\textwidth}{!}{
		\includegraphics{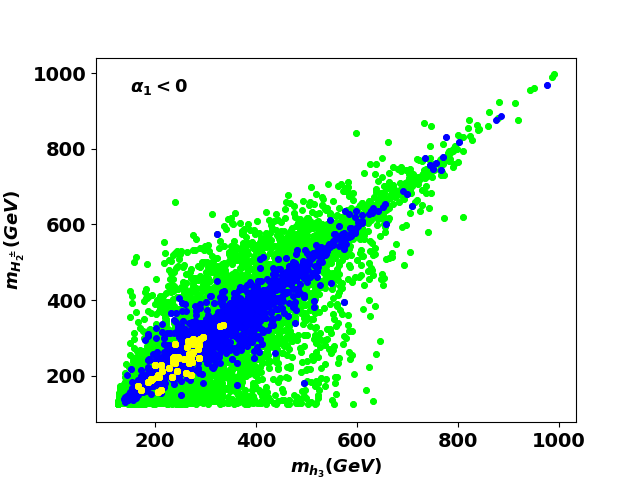}}
	\resizebox{0.42\textwidth}{!}{
		\includegraphics{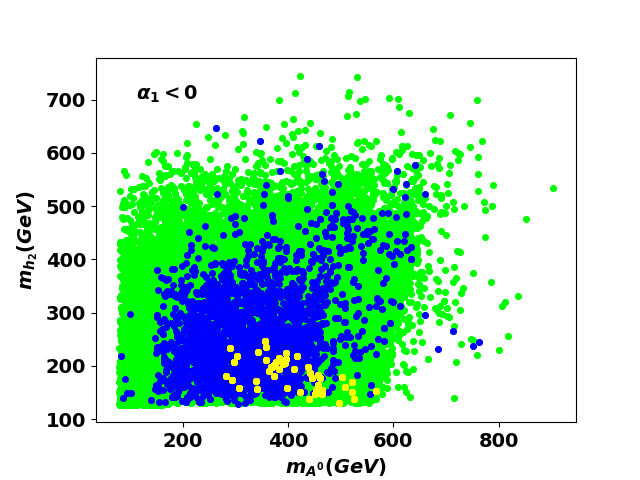}}
	\resizebox{0.42\textwidth}{!}{
		\includegraphics{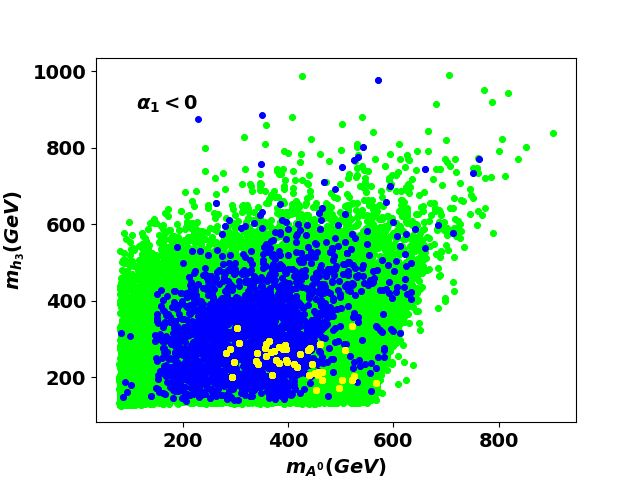}}
	\caption{Same as Fig.~3 for $\sin \alpha_1 < 0 $.}
	\label{fig:4}
\end{figure*}
Hereafter, we analyze the two scenarios corresponding to $\sin \alpha_1 > 0$ and to $\sin \alpha_1 < 0$. Fig.~\ref{fig:3} illustrates the allowed masses ranges plotted in the planes $m_{\phi_i}$ vs $m_{\phi_j}$ ($\phi_i=h_i,\,A$,  $\phi_j=H^{\pm}_j$)  resulting from scans over different values of potential parameters when $\sin \alpha_1 > 0$. The yellow samples indicate surviving regions to Veltman conditions with $\delta T=2.5$ and  $\Delta \chi^2 \le 5.99$. We  can readily see that the Higgs masses are bounded and most of the yellow points lie in ranges of $m_{h_2}$, $m_{h_3}$, $m_{A_0}$, $m_{H^{\pm}_1}$ and $m_{H_2^{\pm}}$ between $151$~GeV and $979$ GeV. The obtained results in this scenario are summarized in Table~\ref{tbl:2} 

In Fig.~\ref{fig:4} we perform a similar analysis for the scenario where  $\sin \alpha_1 < 0$.  Again, all plotted points passed the constraints mentioned above at $2\sigma$. As already noted, the area marked in yellow encodes the cancellation of quadratic divergencies. We see that most of the nonstandard Higgs masses are relatively light and strongly constrained by naturalness. Here, the excluded Higgs mass regions are significantly extended with lower bounds above $130$ GeV and upper bounds not exceeding $335$~GeV, except for the pseudoscalar Higgs ${A_0}$ for which upper mass limit can go up to $566$~GeV, as can be read from Table~\ref{tbl:3}. \\

Remarkably, the above effects of naturalness on the nonstandard Higgs masses can also be probed via the $R_{\gamma\gamma}(h_1)$  and $R_{Z \gamma}(h_1)$ when consistency with ATLAS \cite{atlas2018} and CMS \cite{cms2018} signal strengths measurements is imposed:
 $$\mu_{\gamma \gamma}^{ATLAS} = 0.99 \pm 0.14 \hspace{1cm}  ,  \hspace{1cm} \mu_{\gamma \gamma}^{CMS}= 1.18 _{-0.14}^{+0.17}
\hspace{1cm}  and \hspace{1cm}  \mu_{\gamma Z} \le 6.2$$  

\begin{figure*}[!h]
	\centering
	\resizebox{0.42\textwidth}{!}{
		\includegraphics{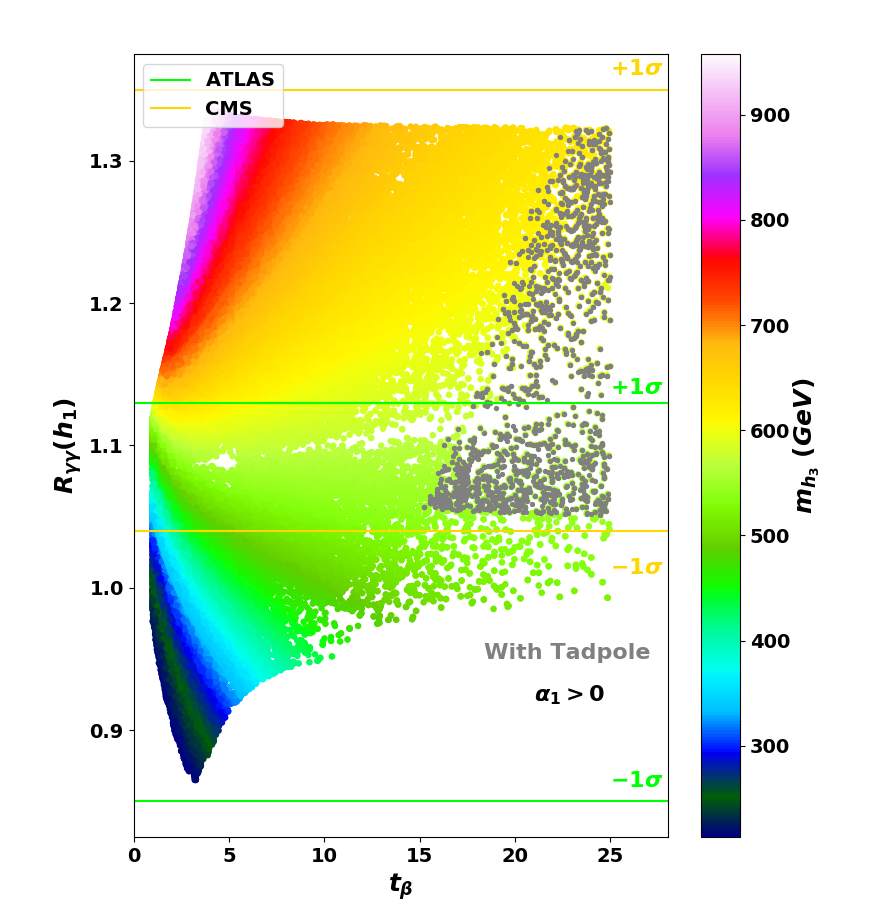}}
	\resizebox{0.42\textwidth}{!}{
		\includegraphics{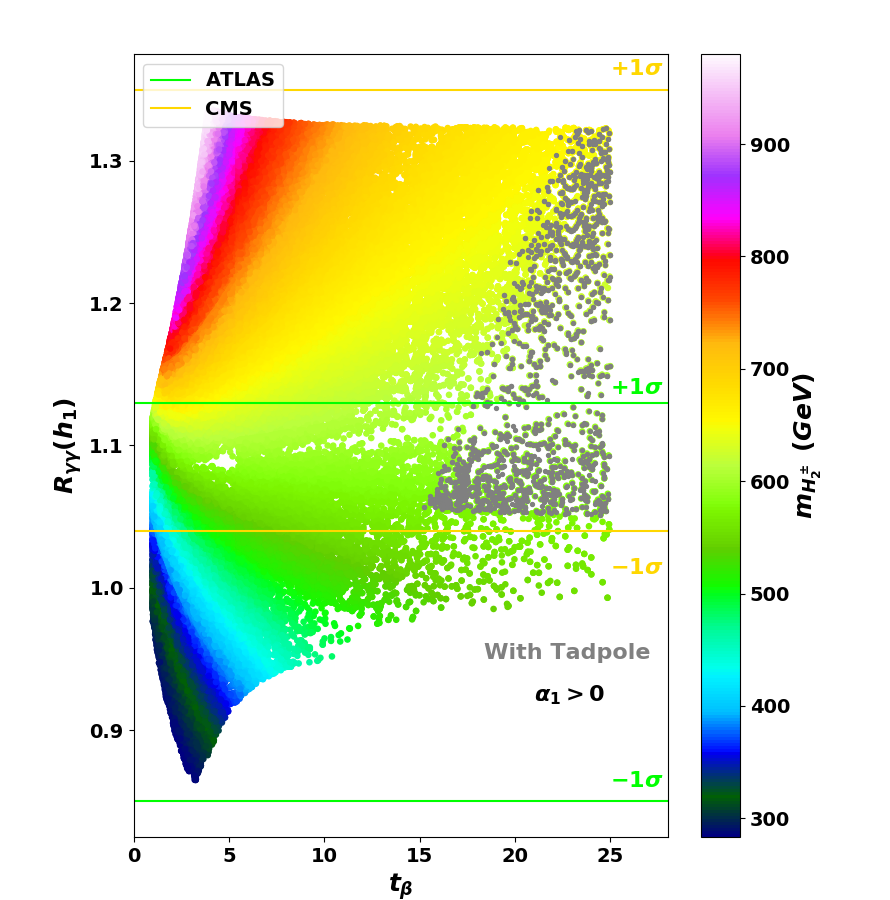}}
	\caption{ $R_{\gamma\gamma}(h_1)$ as a function of $\tan \beta$  and either $m_{h_3}$ (left) or  $m_{H^\pm_2}$ (right) for $\alpha_1 > 0$. The
grey color indicates the surviving regions to Veltman conditions. Our inputs are $\lambda_4=-0.57$, $\lambda_5=-1.18$, $\lambda_a=0.33$, $\lambda_b=1.37$, $\lambda_c=0.1$, $\mu_1=77$, $v_t=0.8$, $\alpha_1\in[0.6\,,1.57]$, $\alpha_2=5.7\times 10^{-3}$, $\alpha_3=-0.35$, $\theta^\pm_3=-1.36$, $\tan\beta\in[0.5\,,25]$, $m_{h_1}=125.09\,GeV$ and $m_{h_2}\in[127,\,1000]\,GeV$. The error for $\chi^2$ fit is $95.5\%$ C.L.}
	\label{fig:5}
\end{figure*}

\begin{figure*}[!h]
	\centering
	\resizebox{0.42\textwidth}{!}{
		\includegraphics{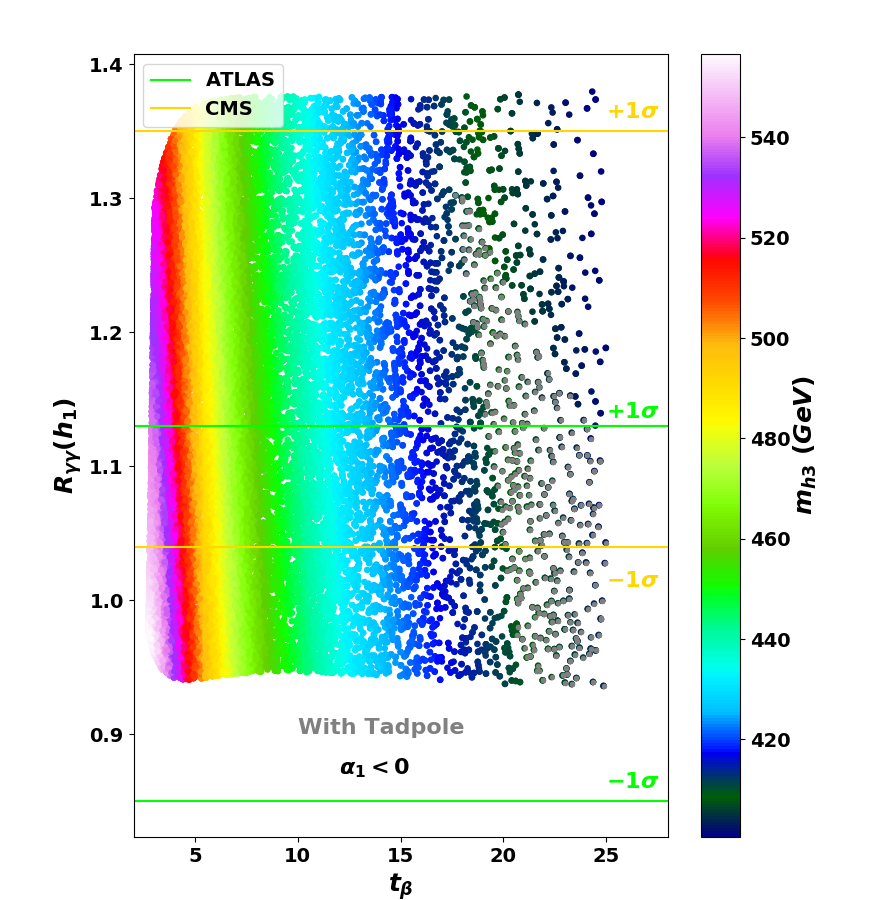}}
	\resizebox{0.42\textwidth}{!}{
		\includegraphics{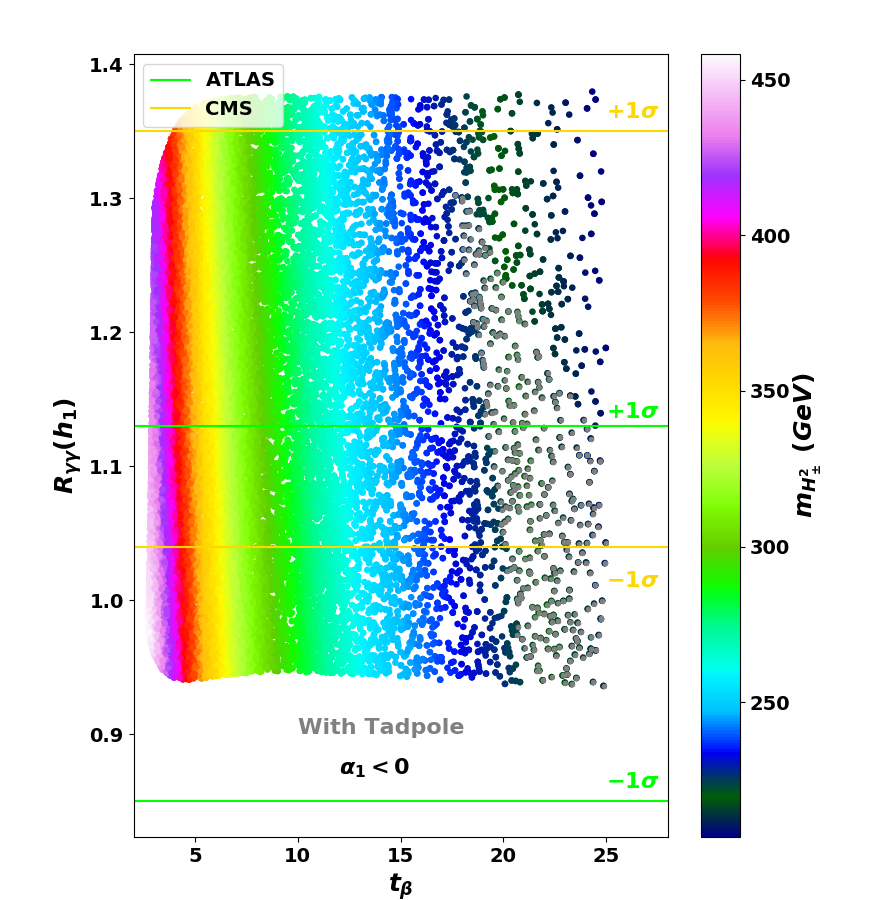}}
	\caption{$R_{\gamma\gamma}(h_1)$ as a function of $\tan \beta$  and either $m_{h_3}$ (left) or  $m_{H^\pm_2}$ (right) for $\alpha_1 < 0$. The grey color indicates the surviving regions to Veltman conditions. Our inputs are $\lambda_4=3.0346$, $\lambda_5=1.33$, $\lambda_a=0.16$, $\lambda_b=0.88$, $\lambda_c=0.31$, $\mu_1=54$, $v_t=1$, $\alpha_1\in[-1.57\,,-1.25]$, $\alpha_2=-4.6\times 10^{-2}$, $\alpha_3=-1.48$, $\theta^\pm_3=0.62$, $\tan\beta\in[0.5\,,25]$, $m_{h_1}=125.09\,GeV$ and $m_{h_2}\in[127,\,1000]\,GeV$. The error for $\chi^2$ fit is $95.5\%$ C.L.}
		\label{fig:6}
	\label{tanb_vs_masses_with_alpn}
\end{figure*}

Indeed, the scatter plots of Figs.~(\ref{fig:5}, \ref{fig:6}) display $R_{\gamma\gamma}(h_1)$ ratio as a function of $\tan \beta$ and either $m_{h_3}$ (left) or $m_{H_2^{\pm}}$ (right) with LHC experimental data taken into account within $1 \sigma$.  As illustrating benchmark scenarios, highlighting how the Higgs masses evolve  with respect to various constraints,  we fix all parameters within the allowed parameter ranges except for the nonstandard scalar masses $m_{A_0}$ and  $m_{H_1^{\pm}}$.  For $\alpha_1 > 0$ (Fig.~\ref{fig:5}),  we see that for $\tan \beta \le 5$, lower bounds on $h_3$ and $H^{\pm}_2$ initially set at $80$~GeV, are raised to about $213$~GeV for $h_3$ and $283$~GeV for $H^{\pm}_2$ while their upper bounds are slightly decreased to $957$~GeV and $980$~GeV respectively. However, once  $\tan \beta$ gets larger values, the upper bounds on $m_{h_3}$ and $m_{H_2^{\pm}}$ decreased significantly from $1000$ GeV to less than $860$~GeV for $h_3$ and $883$~GeV for $H_2^\pm$. If, in addition, Veltman conditions are activated (grey area), we show that only $\tan \beta$ values within $[15\,,\,25] $ are relevant which constrain $m_{h_3}$ and $m_{H_2^{\pm}}$ to vary within relatively tightened ranges $[542\,,\,631]$~GeV and $[570\,,\,653]$~GeV respectively. In this scenario, the Higgs masses for $A_0$ and $H_1^{\pm}$ are predicted as:   $587 \le m_{A_0}  \le 670$~GeV and  $538 \le m_{H_1^{\pm}}  \le 627$~GeV. Similar analysis is performed for $\alpha_1 < 0$ as  illustrated in Fig.~\ref{fig:6}. However in this case, the masses upper limits behave quite the opposite of the previous scenario: upper bounds dropped sharply to $580$~GeV for $m_{h_3}$ and $480$~GeV for $m_{H_2^{\pm}}$ whatever the value given to $\tan \beta$. Furthermore,  when the Veltman conditions are considered, we notice two salient features: 1) $\tan \beta$ is compelled to vary within the reduced interval $[18\,,\,25]$;   2) Deeply affected lower mass limits which are pushed up to almost reach the lower bounds. On the other hand,  our results also show that $m_{A_0}$ and  $m_{H_1^{\pm}}$ are predicted as: $283 \le m_{A_0} \le 296$~GeV and  $145 \le m_{H_1^{\pm}}  \le 170$~GeV. To conclude, we have clearly seen the leading role played by naturalness comparatively to the other constraints and how Veltman conditions deeply affect the analysis excluding substantial mass regions of non standards scalars. The overall resulting ranges of $2\mathcal{HDM+T}$ spectrum are summarized in Tables \ref{tbl:2}  and \ref{tbl:3} .

\begin{figure*}[!h]
	\centering
	\resizebox{0.42\textwidth}{!}{
		\includegraphics{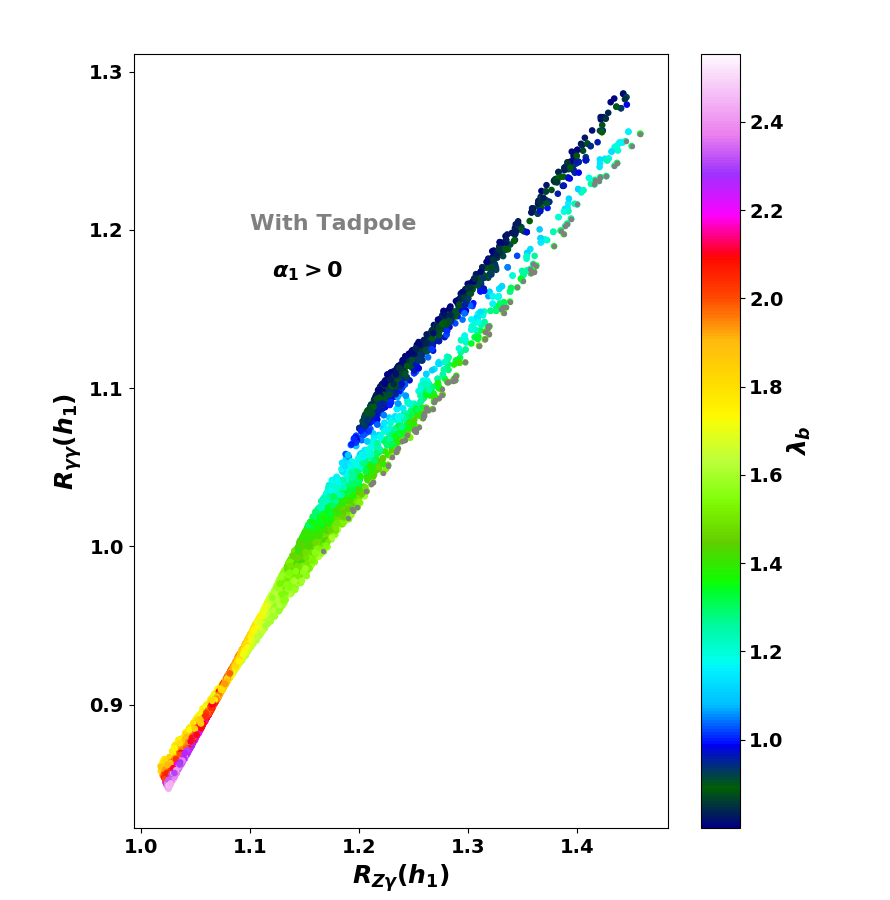}}
	\resizebox{0.42\textwidth}{!}{
		\includegraphics{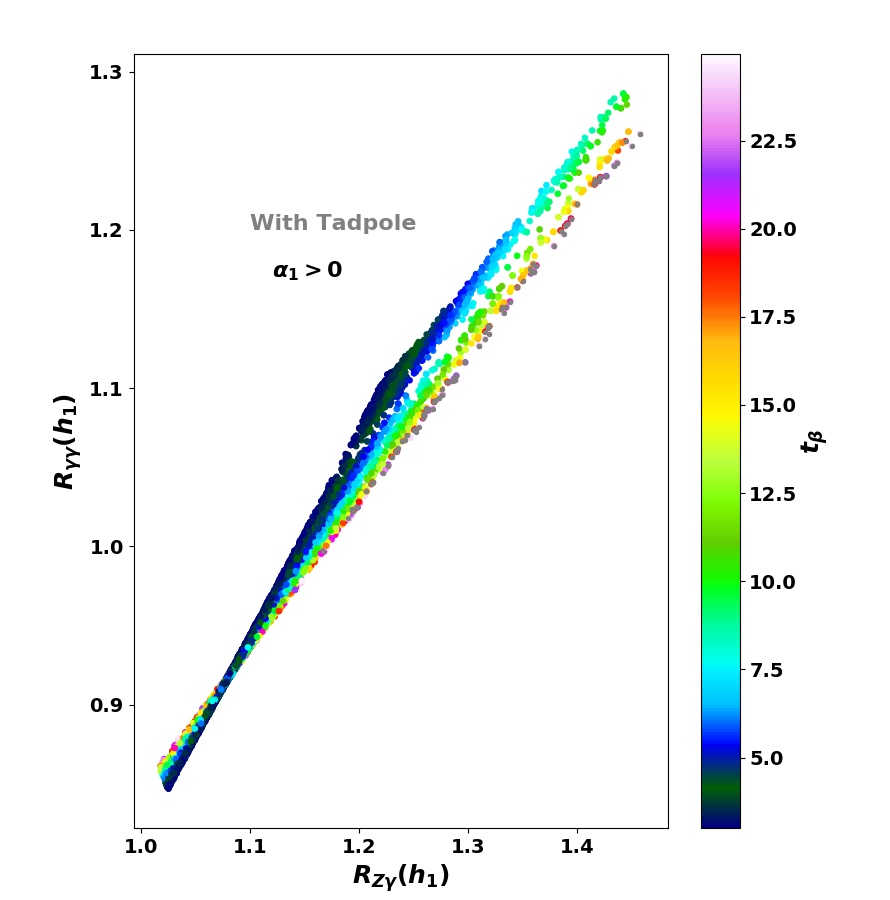}}
	\caption{$R_{\gamma \gamma}(h_1)$ and $R_{Z\gamma}(h_1)$ correlation versus either $\lambda_b$ (left) or  $tan \beta$ (right) for $\alpha_1 > 0$. The  grey color indicates surviving regions to Veltman conditions.  Our inputs are $\lambda_4=-0.57$, $\lambda_5=-1.18$, $\lambda_a=0.33$, $\lambda_b\in [0.5\,,4]$, $\lambda_c=0.1$, $\mu_1=77$, $v_t=0.8$, $\alpha_1\in[0.6\,,1.57]$, $\alpha_2=0.73\times 10^{-3}$, $\alpha_3=-0.35$, $\theta^\pm_3=-1.36$, $\tan\beta\in[0.5\,,25]$, $m_{h_1}=125.09\,GeV$ and $m_{h_2}\in[180,\;187]\,GeV$. The error for $\chi^2$ fit is $95.5\%$ C.L.}
	\label{fig:7}
\end{figure*}

\begin{figure*}[!h]
	\centering
	\resizebox{0.42\textwidth}{!}{
		\includegraphics{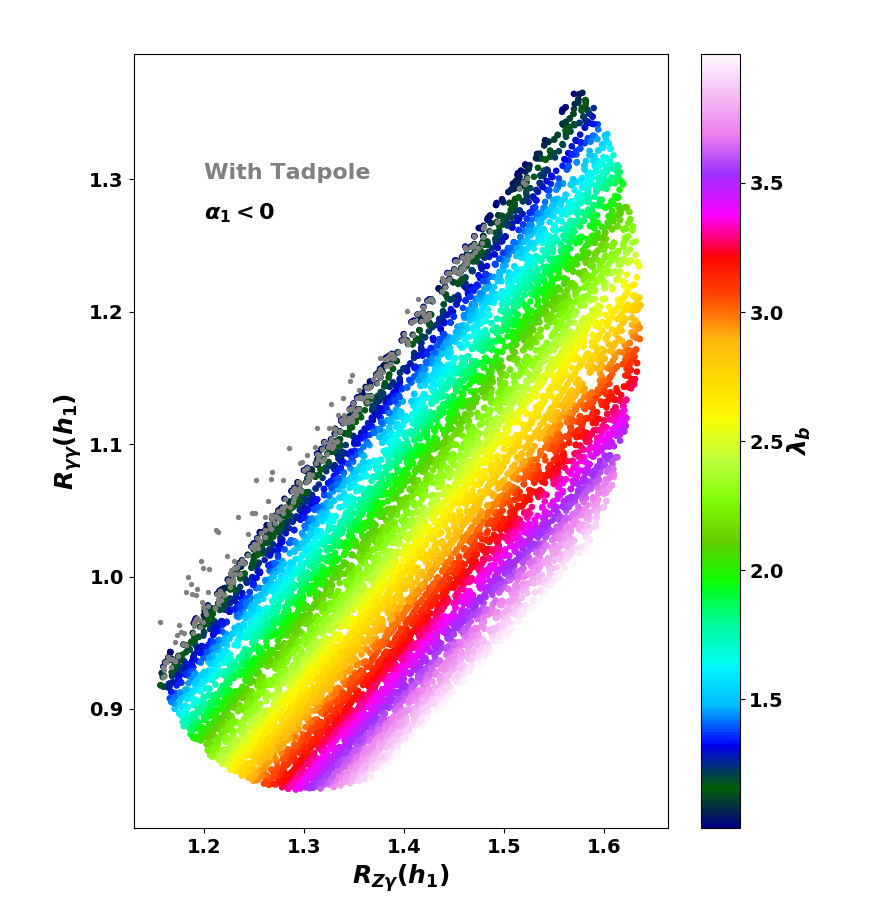}}
	\resizebox{0.42\textwidth}{!}{
		\includegraphics{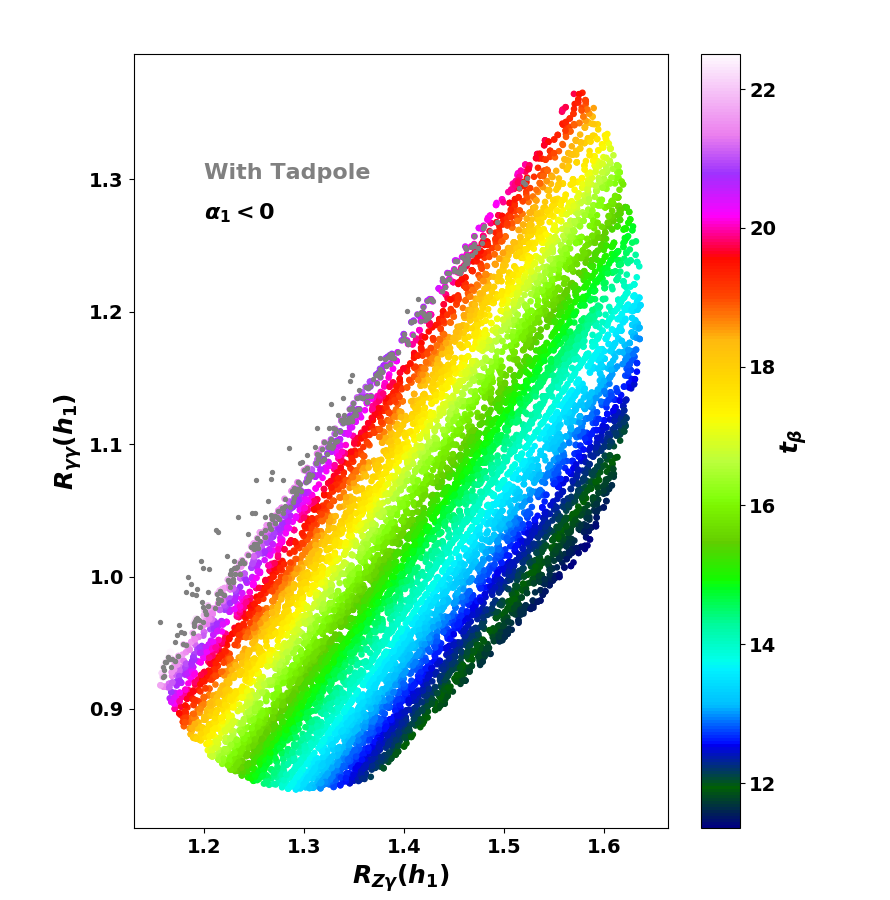}}
		\caption{$R_{\gamma \gamma}(h_1)$ and $R_{Z\gamma}(h_1)$ correlation versus either $\lambda_b$ (left) or  $tan \beta$ (right) for $\alpha_1 < 0$. The  grey color indicates surviving regions to Veltman conditions. Our inputs are $\lambda_4=3.03$, $\lambda_5=1.33$, $\lambda_a=0.16$, $\lambda_b\in [0\,,4]$, $\lambda_c=0.16$, $\mu_1=54$, $v_t=1$, $\alpha_1\in[-1.57\,,-0.6]$, $\alpha_2=-0.04$, $\alpha_3=-1.48$, $\theta^\pm_3=0.62$, $\tan\beta\in[0.5\,,25]$, $m_{h_1}=125.09\,GeV$ and $m_{h_2}=201\,GeV$. The error for $\chi^2$ fit is $95.5\%$ C.L.}
	\label{fig:8}
\end{figure*}

\paragraph{}
Finally, we study  the correlation between $R_{\gamma\gamma}(h_1)$ and $R_{\gamma\,Z}(h_1)$. Again Higgs masses $m_{A_0}$ and  $m_{H_1^{\pm}}$ are considered as output parameters in the analysis. At first sight, from Figs.~(\ref{fig:7},\ref{fig:8}), we see that  $R_{\gamma\,Z}(h_1)$ deviates slightly with respect to its standard value, with a maximum below $1.6$ in both scenarios.   We also find that $R_{\gamma\gamma}(h_1)$ and $R_{\gamma\,Z}(h_1)$ are always correlated regardless of the sign of $\sin \alpha_1$. Yet this correlation only happens when $\tan \beta$ lies within $[17\,,25]$  ($[19\,,25]$) with $\lambda_b$ ranging from $1.25$ to $1.4$ ($1$ to $1.14$) in $\sin \alpha_1 > 0$ ($\sin \alpha_1 < 0$) scenario.\\

		\begin{widetext}
			
				\begin{table}[ht]
				
			 \begin{tabular}{l c c c c}
	\hline \hline
	\textbf{parameters} & \textbf{U} & \textbf{U+BFB} & \textbf{U+BFB+LEP+LHC}& \textbf{All Constraints} \\ \hline\hline
	
	$\lambda_1$&$[-8.19\,,\,8.31]$&$[0\,,\,8.29]$&$[0\,,\,8]$&$[0\,,\,0.72]$ \\
	$\lambda_2$&$[-8.15\,,\,8.26]$&$[0\,,\,8.24]$&$[0\,,\,8]$&$[0.25\,,\,0.26]$ \\
	$\lambda_3$&$[-12.13\,,\,15.39]$&$[-3.7\,,\,14.55]$&$[-2.5\,,\,12]$&$[0.22\,,\,2.47]$ \\
	$\lambda_4$&$[-15.68\,,\,13.46]$&$[-14.72\,,\,8.]$&$[-7.03\,,\,5.55]$&$[-0.59\,,\,4.93]$ \\
	$\lambda_5$&$[-8.1\,,\,8.24]$&$[-7.74\,,\,5.97]$&$[-7.10\,,\,5.88]$&$[-3.5\,,\,2.8]$ \\
	$\lambda_a$&$[-13.36\,,\,13.73]$&$[0\,,\,12.38]$&$[0\,,\,12.60]$&$[0\,,\,1.4]$ \\
	$\lambda_b$&$[-13.23\,,\,13.35]$&$[0\,,\,13.57]$&$[0\,,\,13.51]$& $[0.1\,,\,1.62]$ \\
	$\lambda_c$&$[-5.02\,,\,4.98]$&$[0\,,\,5.01]$&$[0\,,\,4.98]$&$[0\,,\,0.8]$ \\
	$\mu_1$&$[-10^2\,,\,10^2]$&$[-10^2\,,\,10^2]$&$[-10^2\,,\,10^2]$&$[-10^2\,,\,10^2]$ \\
	$\mu_2$&$[-10\,,\,12.2]\times 10^3$&$[-8.93\,,\,6.3]\times 10^3$&$[-7\,,\,6.2]\times 10^3$&$[-42\,,\,61]$ \\
	$\mu_3$&$[-3.6\,,\,3.4]\times 10^3$&$[-3.84\,,\,2.88]\times 10^3$&$[-3.2\,,\,2.7]\times 10^3$&$[-695\,,\,523]$ \\
	$m_{12}^2$&$[-8.1\,,\,43.9]\times 10^4$&$[-15\,,\,38]\times 10^4$&$[-10.1\,,\,37.3]\times 10^4$&$[0.1150\,,\,9.8]\times 10^4$ \\
	$\alpha_1$&$[-\frac{\pi}{2}\,,\,\frac{\pi}{2}]$&$[-\frac{\pi}{2}\,,\,\frac{\pi}{2}]$&$[-1.57\,,-1.25]$&$[-1.57\,,-1.51]$ \\
	&&&$U[0.42\,,1.57]$&$U[1.51\,,1.57]$\\
	$\alpha_2$&$[-\frac{\pi}{2}\,,\,\frac{\pi}{2}]$&$[-\frac{\pi}{2}\,,\,\frac{\pi}{2}]$&$[-0.43\,,\,0.52]$&$[-0.14\,,\,0.13]$ \\
	$\alpha_3$&$[-\frac{\pi}{2}\,,\,\frac{\pi}{2}]$&$[-\frac{\pi}{2}\,,\,\frac{\pi}{2}]$&$[-\frac{\pi}{2}\,,\,\frac{\pi}{2}]$&$[-\frac{\pi}{2}\,,\,\frac{\pi}{2}]$ \\
	$\theta^\pm$&$[-\frac{\pi}{2}\,,\,\frac{\pi}{2}]$&$[-\frac{\pi}{2}\,,\,\frac{\pi}{2}]$&$[-\frac{\pi}{2}\,,\,\frac{\pi}{2}]$&$[-\frac{\pi}{2}\,,\,\frac{\pi}{2}]$ \\
	$v_t$
	(GeV)&$[0\,,\,3]$&$[0\,,\,3]$&$[0\,,\,3]$&$[0\,,\,3]$ \\
	$tan\beta$&$[0.5\,,\,25]$&$[0.5\,,\,25]$&$[0.52\,,\,25]$&$[15\,,\,25]$ \\ \hline\hline
\end{tabular}
				\centering	
				\caption{Allowed ranges for the potential parameters from various constraints. The last column includes effects due to Veltman conditions with $\delta T=2.5$.}
				\centering 
				\label{tbl:1}
			\end{table}
	\end{widetext}

\begin{widetext}
 \begin{table}[h!]
	
	\begin{tabular}{l c c c c}
		\hline \hline
		\textbf{$m_{\phi_i}$} & \textbf{U} & \textbf{U+BFB} & \textbf{U+BFB+LEP+LHC}& \textbf{All Constraints} \\ \hline\hline
		
		$m_{h_2}$ (GeV)&$[126\,,\,980]$&$[126\,,\,971]$&$[126\,,\,948]$&$[151\,,\,928]$ \\
		$m_{h_3}$ (GeV)&$[126\,,\,1000]$&$[127\,,\,999]$&$[127\,,\,999]$& $[186\,,\,979]$ \\
		$m_{A^0}$ (GeV)&$[80\,,\,999]$&$[80\,,\,998]$&$[82\,,\,997.9]$&$[153\,,\,973]$ \\
		$m_{H^\pm_1}$ (GeV)&$[80\,,\,1000]$&$[81\,,\,970]$&$[81\,,\,944]$&$[151\,,\,929]$ \\
		$m_{H^\pm_2}$ (GeV)&$[80\,,\,1000]$&$[127\,,\,994]$&$[127\,,\,995]$&$[186\,,\,979]$  \\ \hline\hline
	\end{tabular}
	\centering	
	\caption{Allowed Higgs bosons masses from various constraints for $\alpha_1 > 0$ scenario. The last column includes effects due to Veltman conditions with $\delta T=2.5$.}
	\centering 
	\label{tbl:2}
\end{table}
	\end{widetext}

\begin{widetext}
	\begin{table}[h!]
		
		\begin{tabular}{l c c c c}
			\hline \hline
			\textbf{$m_{\phi_i}$} & \textbf{U} & \textbf{U+BFB} & \textbf{U+BFB+LEP+LHC}& \textbf{All Constraints} \\ \hline\hline
			
			$m_{h_2}$ (GeV)&$[126\,,\,684]$&$[125\,,\,667]$&$[128\,,\,656]$&$[130\,,\,246]$ \\
			$m_{h_3}$ (GeV)&$[126\,,\,1000]$&$[126\,,\,990]$&$[137\,,\,976]$& $[163\,,\,335]$ \\
			$m_{A^0}$ (GeV)&$[80\,,\,950]$&$[80\,,\,772]$&$[83\,,\,762]$&$[282\,,\,566]$ \\
			$m_{H^\pm_1}$ (GeV)&$[80\,,\,670]$&$[80\,,\,650]$&$[85\,,\,641]$&$[131\,,\,243]$ \\
			$m_{H^\pm_2}$ (GeV)&$[80\,,\,1000]$&$[126\,,\,997]$&$[130\,,\,762]$& $[160\,,\,334]$ \\ \hline\hline
		\end{tabular}
		\centering	
		\caption{Like in Table.~\ref{tbl:2} for $\alpha_1 < 0$ scenario.}
		\centering 
		\label{tbl:3}
	\end{table}
\end{widetext}

\section{Conclusions}
This work arose as a continuation of activities around Beyond Standard Models extended with a Triplet scalar. In this paper we have performed a comprehensive study of the Higgs potential of $2\mathcal{HDM}$ model augmented by a real triplet scalar (dubbed $2\mathcal{HDM+T}$).  First, we have presented the salient features of the Higgs sector, with the lighter CP even scalar $h_1$ identified to the 125 observed Higgs, then  derived constraints originating from perturbative unitarity, vacuum stability and naturalness problem. We have checked that, the theoretical constraints and the Higgs spectrum of $\mathcal{HTM}$ and $2\mathcal{HDM}$ are recovered when the extra couplings parameters to these models are removed. Then by imposing theoretical constraints and incorporating  limits from combined LEP and LHC results we obtained the  potential parameters ranges of variation, and the allowed parameter space of $2\mathcal{HDM+T}$.    \\

The second aim of our analysis is to gain more insight on the masses of heavy Higgs bosons and probe the effect of naturalness on their range of variations. Depending on the mixing angles, essentially the sign of $\alpha_1$ we have shown how the nonstandard Higgs masses evolve and the established bounds for both scenarios,  $\sin \alpha_1 > 0$ and  $\sin \alpha_1 <0 $. Given the theoretical constraints on potential parameters, we have also investigated how important the contributions from new heavy scalars to $h_1 \to \gamma \gamma$ decay  while remaining compatible with the Higgs signal measurements within $1 \sigma$. Such analysis has placed stringent limits on the masses variation and confirmed the crucial role of naturalness constraints in controlling the parameter space and phenomenological studies of $2\mathcal{HDM+T}$. Lastly we have found that Higgs decays to the diphoton and to $Z \gamma$ are generally correlated in both scenarios when $\tan \beta$ varies within $ 17 \le \tan \beta \le 25$ and $\lambda_b$ is almost constant lying about $1.3$ for $\sin \alpha_1 > 0$,  and $1.1$  for $\sin \alpha_1 < 0$.   \\

\newpage
\section*{Acknowledgments}
This work is partially supported by the Moroccan Ministry of Higher Education and Scientific Research MESRSFC and CNRST: Projet PPR/2015/6.

\section*{Appendices}
\appendix

\setcounter{equation}{0}
\renewcommand{\theequation}{A\arabic{equation}}
\section{Hybrid parameterisation}
\label{sec-para}
Using Eqs.~(\ref{rota-matrix-cp-even}, \ref{masseod}) and $\mathcal{C} \mathcal{M}^2_{{charge}} \mathcal{C}^T=diag(0, m^2_{H_1^\pm}, m^2_{H_2^\pm})$, one can easily express non physical parameters in terms of the physical Higgs masses, mixing angle, $\lambda_4$, $\mu_1$ and $v_t$: 

\begin{eqnarray}
	\begin{matrix}
	\lambda _c&=& -\frac{v_1^2 \left(\mathcal{C}_{21}^2 m_{H_1^\pm}^2+\mathcal{C}_{31}^2 m_{H_2^\pm}^2\right)+2 v_2 v_1 \left(\mathcal{C}_{21} \mathcal{C}_{22} m_{H_1^\pm}^2+\mathcal{C}_{31} \mathcal{C}_{32} m_{H_2^\pm}^2\right)+v_2^2 \left(\mathcal{C}_{22}^2 m_{H_1^\pm}^2+\mathcal{C}_{32}^2 m_{H_2^\pm}^2\right)-4 v_t^2 \left(\mathcal{E}_{13}^2 m_{h_1}^2+\mathcal{E}_{23}^2 m_{h_2}^2+\mathcal{E}_{33}^2 m_{h_3}^2\right)}{8 v_t^4}\\
	\lambda _b&=& \frac{\mathcal{C}_{21} \mathcal{C}_{22} v_1 m_{H_1^\pm}^2+\mathcal{C}_{22}^2 v_2 m_{H_1^\pm}^2+\mathcal{C}_{32} \left(\mathcal{C}_{31} v_1+\mathcal{C}_{32} v_2\right) m_{H_2^\pm}^2+2 v_t \left(\mathcal{E}_{12} \mathcal{E}_{13} m_{h_1}^2+\mathcal{E}_{22} \mathcal{E}_{23} m_{h_2}^2+\mathcal{E}_{32} \mathcal{E}_{33} m_{h_3}^2\right)}{2 v_2 v_t^2}\hspace*{3cm}\\
	\lambda _a&=& \frac{\mathcal{C}_{21}^2 v_1 m_{H_1^\pm}^2+\mathcal{C}_{21} \mathcal{C}_{22} v_2 m_{H_1^\pm}^2+\mathcal{C}_{31} \left(\mathcal{C}_{31} v_1+\mathcal{C}_{32} v_2\right) m_{H_2^\pm}^2+2 v_t \left(\mathcal{E}_{11} \mathcal{E}_{13} m_{h_1}^2+\mathcal{E}_{21} \mathcal{E}_{23} m_{h_2}^2+\mathcal{E}_{31} \mathcal{E}_{33} m_{h_3}^2\right)}{2 v_1 v_t^2}\hspace*{3cm}\\
	m_{12}^2&=& \frac{1}{2} \left(\frac{v_1 \left(\frac{2 v_2^2 m_A^2}{v_d^2}-\mathcal{C}_{31}^2 m_{H_2}^2+2 \lambda_5 v_2^2+\mu _1 v_t\right)}{v_2}-\frac{\mathcal{C}_{21}^2 v_1 m_{H_1^\pm}^2}{v_2}-\mathcal{C}_{21} \mathcal{C}_{22} m_{H_1^\pm}^2-\mathcal{C}_{31} \mathcal{C}_{32} m_{H_2^\pm}^2\right)\hspace*{3.5cm}\\
	\lambda _3&=& -\frac{m_A^2}{v_d^2}+\frac{2 v_1 \left(\mathcal{C}_{21}^2 m_{H_1^\pm}^2+\mathcal{C}_{31}^2 m_{H_2^\pm}^2-\mu _1 v_t\right)+v_2 \left(\mathcal{E}_{11} \mathcal{E}_{12} m_{h_1}^2+\mathcal{E}_{21} \mathcal{E}_{22} m_{h_2}^2+\mathcal{E}_{31} \mathcal{E}_{32} m_{h_3}^2\right)}{v_1 v_2^2}-\lambda_5\hspace*{4cm}\\
	\mu_3&=& \frac{\mathcal{C}_{21}^2 v_1 m_{H_1^\pm}^2+\mathcal{C}_{21} \mathcal{C}_{22} v_2 m_{H_1^\pm}^2+\mathcal{C}_{31} \left(\mathcal{C}_{31} v_1+\mathcal{C}_{32} v_2\right) m_{H_2^\pm}^2-v_1 \mu _1 v_t}{v_2 v_t}\hspace*{8cm}\\
	\mu _2&=& \frac{-\mathcal{C}_{21}^2 v_1^2 m_{H_1^\pm}^2-\mathcal{C}_{31}^2 v_1^2 m_{H_2^\pm}^2+v_2^2 \left(\mathcal{C}_{22}^2 m_{H_1}^2+\mathcal{C}_{32}^2 m_{H_2^\pm}^2\right)+v_1^2 \mu _1 v_t}{v_2^2 v_t}\hspace*{8cm}\\
	\lambda _4&=& \frac{2 m_A^2}{v_d^2}-\frac{2 \left(\mathcal{C}_{21}^2 m_{H_1^\pm}^2+\mathcal{C}_{31}^2 m_{H_2^\pm}^2-\mu _1 v_t\right)}{v_2^2}+\lambda_5\hspace*{10cm}\\
	\lambda _2&=& \frac{v_d^2 \left(\mathcal{E}_{12}^2 m_{h_1}^2+\mathcal{E}_{22}^2 m_{h_2}^2+\mathcal{E}_{32}^2 m_{h_3}^2\right)-v_1^2 \left(m_A^2+\lambda_5 v_d^2\right)}{v_d^2 v_2^2}\hspace*{9.5cm}\\
	\lambda _1&=& \frac{v_d^2 \left(\mathcal{E}_{11}^2 m_{h_1}^2+\mathcal{E}_{21}^2 m_{h_2}^2+\mathcal{E}_{31}^2 m_{h_3}^2\right)-v_2^2 \left(m_A^2+\lambda_5 v_d^2\right)}{v_d^2 v_1^2}\hspace*{9.5cm}
    \end{matrix}
    \end{eqnarray}
    
    The $\theta^\pm_3$ mixing angle is given by:
    
  \begin{eqnarray}
   \cos \theta_3^\pm = -\frac{v_1 v_d \left(v_d^2 (m_{H_1^\pm}^2 m^2_{G^+H^-} v v_2+2 m_{H_2^\pm}^2 m^2_{G^+\delta^-} v_1 v_t)+\sqrt{2} m^2_{G^+\delta^-} v_t \left(m_{H_1^\pm}^2 v v_2^2-4 m_{H_2^\pm}^2 v_1 v_t^2\right)\right)}{m^4_{G^+H^-} v^2 v_d^4+2 \sqrt{2} m^2_{G^+H^-} m^2_{G^+\delta^-} v^2 v_2 v_d^2 v_t+m^4_{G^+\delta^-} \left(2 v^2 v_2^2 v_t^2+v_1^2 \left(v_d^4-4 \sqrt{2} v_d^2 v_t^2+8 v_t^4\right)\right)}   
   \end{eqnarray}

   \begin{eqnarray}
   \sin \theta_3^\pm = \frac{v_1 v_d \left(-2 \sqrt{2} m^2_{G^+\delta^-} v_2 v_t^2 (m_{H_1^\pm}^2 v_1+m_{H_2^\pm}^2 v)+m_{H_1^\pm}^2 m^2_{G^+\delta^-} v_1 v_2 v_d^2-2 m_{H_2^\pm}^2 m^2_{G^+H^-} v v_d^2 v_t\right)}{m^4_{G^+H^-} v^2 v_d^4+2 \sqrt{2} m^2_{G^+H^-} m^2_{G^+\delta^-} v^2 v_2 v_d^2 v_t+m^4_{G^+\delta^-} \left(2 v^2 v_2^2 v_t^2+v_1^2 \left(v_d^4-4 \sqrt{2} v_d^2 v_t^2+8 v_t^4\right)\right)}   
   \end{eqnarray}
\label{sec-diagonal}

\setcounter{equation}{0}
\renewcommand{\theequation}{B\arabic{equation}}
\section{Unitarity constraints}
\label{sec-U}

The first submatrix $\mathcal{M}_1$ corresponds to scattering whose initial and final states are one of the following: $\phi_1^+\delta^-$, $\phi_1^-\delta^+$, $\phi_2^+\delta^-$, $\phi_2^-\delta^+$, $\phi_1^+\phi_2^-$, $\phi_1^-\phi_2^+$, $\rho_0\eta_1$, $\rho_0\eta_2$, $\rho_1\eta_2$, $\rho_2\eta_1$, $\eta_1\eta_2$, $\rho_0\rho_1$, $\rho_0\rho_2$, $\rho_1\rho_2$. by Wolfram Mathematica we found:

\begin{widetext}
	\begin{equation}
	\mathcal{M}_1=\left(
	\begin{array}{cccccccccccccc}
	\lambda _a & 0 & 0 & 0 & 0 & 0 & 0 & 0 & 0 & 0 & 0 & 0 & 0 & 0 \\
	0 & \lambda _a & 0 & 0 & 0 & 0 & 0 & 0 & 0 & 0 & 0 & 0 & 0 & 0 \\
	0 & 0 & \lambda _b & 0 & 0 & 0 & 0 & 0 & 0 & 0 & 0 & 0 & 0 & 0 \\
	0 & 0 & 0 & \lambda _b & 0 & 0 & 0 & 0 & 0 & 0 & 0 & 0 & 0 & 0 \\
	0 & 0 & 0 & 0 & \lambda _{34} & 2 \lambda _5 & -\frac{1}{2} i \bar{\lambda}_{45} & 0 & 0 & \frac{1}{2} i \bar{\lambda}_{45} & \frac{\lambda _{45}}{2} & 0 & 0 & \frac{\lambda _{45}}{2} \\
	0 & 0 & 0 & 0 & 2 \lambda _5 & \lambda _{34} & \frac{1}{2} i \bar{\lambda}_{45} & 0 & 0 & -\frac{1}{2} i \bar{\lambda}_{45} & \frac{\lambda _{45}}{2} & 0 & 0 & \frac{\lambda _{45}}{2} \\
	0 & 0 & 0 & 0 & \frac{1}{2} i \bar{\lambda}_{45} & -\frac{1}{2} i \bar{\lambda}_{45} & \bar{\lambda}_{345} & 0 & 0 & \lambda _5 & 0 & 0 & 0 & 0 \\
	0 & 0 & 0 & 0 & 0 & 0 & 0 & \lambda _a & 0 & 0 & 0 & 0 & 0 & 0 \\
	0 & 0 & 0 & 0 & 0 & 0 & 0 & 0 & \lambda _b & 0 & 0 & 0 & 0 & 0 \\
	0 & 0 & 0 & 0 & -\frac{1}{2} i \bar{\lambda}_{45} & \frac{1}{2} i \bar{\lambda}_{45} & \lambda _5 & 0 & 0 & \bar{\lambda}_{345} & 0 & 0 & 0 & 0 \\
	0 & 0 & 0 & 0 & \frac{\lambda _{45}}{2} & \frac{\lambda _{45}}{2} & 0 & 0 & 0 & 0 & \lambda _{345} & 0 & 0 & \lambda _5 \\
	0 & 0 & 0 & 0 & 0 & 0 & 0 & 0 & 0 & 0 & 0 & \lambda _a & 0 & 0 \\
	0 & 0 & 0 & 0 & 0 & 0 & 0 & 0 & 0 & 0 & 0 & 0 & \lambda _b & 0 \\
	0 & 0 & 0 & 0 & \frac{\lambda _{45}}{2} & \frac{\lambda _{45}}{2} & 0 & 0 & 0 & 0 & \lambda _5 & 0 & 0 & \lambda _{345} \\
	\end{array}
	\right)
	\end{equation}
\end{widetext}
where $\lambda _{ij}=\lambda_i+\lambda_j$, $\bar{\lambda} _{ij}=\lambda_i-\lambda_j$,  $\lambda _{ijk}=\lambda_i+\lambda_j+\lambda_k$ and $\bar{\lambda} _{ijk}=\lambda_i+\lambda_j-\lambda_k$. We find that $\mathcal{M}_1$ has the following eigenvalues:
\begin{equation}
\begin{aligned}
a_1=\lambda _3+\lambda _4
\end{aligned}
\end{equation}
\begin{equation}
\begin{aligned}
a_2^\pm=\lambda _3+2 \lambda _4\pm 3 \lambda _5
\end{aligned}
\end{equation}
\begin{equation}
\begin{aligned}
a_3^\pm=\lambda _3\pm\lambda _5
\end{aligned}
\end{equation}
\begin{equation}
\begin{aligned}
a_4=\lambda _a
\end{aligned}
\end{equation}
\begin{equation}
\begin{aligned}
a_5=\lambda _b
\end{aligned}
\end{equation}

The submatrix $\mathcal{M}_2$ corresponds to scattering processes with initial and final states within the following set: $\phi_1^-\phi_1^+$, $\phi_2^-\phi_2^+$, $\delta^+\delta^-$, $\frac{\rho_1\rho_1}{\sqrt 2}$, $\frac{\rho_2\rho_2}{\sqrt 2}$, $\frac{\rho_0\rho_0}{\sqrt 2}$, $\frac{\eta_1\eta_1}{\sqrt 2}$, $\frac{\eta_2\eta_2}{\sqrt 2}$, where the $\sqrt 2$ accounts for identical particle statistics. This submatrix reads that:

\begin{widetext}
	\begin{equation}
\mathcal{M}_2=
\left(
\begin{array}{cccccccc}
2 \lambda _1 & \lambda _3+\lambda _4 & \lambda _{a} & \frac{\lambda _1}{\sqrt{2}} & \frac{\lambda _3}{\sqrt{2}} & \frac{\lambda _1}{\sqrt{2}} & \frac{\lambda _3}{\sqrt{2}} & \frac{\lambda _{a}}{\sqrt{2}} \\
\lambda _3+\lambda _4 & 2 \lambda _2 & \lambda _{b} & \frac{\lambda _3}{\sqrt{2}} & \frac{\lambda _2}{\sqrt{2}} & \frac{\lambda _3}{\sqrt{2}} & \frac{\lambda _2}{\sqrt{2}} & \frac{\lambda _{b}}{\sqrt{2}} \\
\lambda _{a} & \lambda _{b} & 4 \lambda _{c} & \frac{\lambda _{a}}{\sqrt{2}} & \frac{\lambda _{b}}{\sqrt{2}} & \frac{\lambda _{a}}{\sqrt{2}} & \frac{\lambda _{b}}{\sqrt{2}} & \sqrt{2} \lambda _{c} \\
\frac{\lambda _1}{\sqrt{2}} & \frac{\lambda _3}{\sqrt{2}} & \frac{\lambda _{a}}{\sqrt{2}} & \frac{3 \lambda _1}{2} & \frac{1}{2} \lambda_{345} & \frac{\lambda _1}{2} & \frac{1}{2} \bar{\lambda_{345}} & \frac{\lambda _{a}}{2} \\
\frac{\lambda _3}{\sqrt{2}} & \frac{\lambda _2}{\sqrt{2}} & \frac{\lambda _{b}}{\sqrt{2}} & \frac{1}{2} \lambda_{345} & \frac{3 \lambda _2}{2} & \frac{1}{2} \bar{\lambda_{345}} & \frac{\lambda _2}{2} & \frac{\lambda _{b}}{2} \\
\frac{\lambda _1}{\sqrt{2}} & \frac{\lambda _3}{\sqrt{2}} & \frac{\lambda _{a}}{\sqrt{2}} & \frac{\lambda _1}{2} & \frac{1}{2} \bar{\lambda_{345}} & \frac{3 \lambda _1}{2} & \frac{1}{2} \lambda_{345} & \frac{\lambda _{a}}{2} \\
\frac{\lambda _3}{\sqrt{2}} & \frac{\lambda _2}{\sqrt{2}} & \frac{\lambda _{b}}{\sqrt{2}} & \frac{1}{2} \bar{\lambda_{345}} & \frac{\lambda _2}{2} & \frac{1}{2} \lambda_{345} & \frac{3 \lambda _2}{2} & \frac{\lambda _{b}}{2} \\
\frac{\lambda _{a}}{\sqrt{2}} & \frac{\lambda _{b}}{\sqrt{2}} & \sqrt{2} \lambda _{c} & \frac{\lambda _{a}}{2} & \frac{\lambda _{b}}{2} & \frac{\lambda _{a}}{2} & \frac{\lambda _{b}}{2} & 3 \lambda _{c} \\
\end{array}
\right)
	\end{equation}
\end{widetext}

where $\lambda_{345}=\lambda_3+\lambda_4+\lambda_5$ and $\bar{\lambda_{345}}=\lambda_3+\lambda_4-\lambda_5$. The corresponding eigenvalues are:

\begin{equation}
\begin{aligned}
b_1^\pm=\frac{1}{2} \left(\pm\sqrt{\lambda _1^2-2 \lambda _2 \lambda _1+\lambda _2^2+ 4 \lambda _4^2}+\lambda _1+\lambda _2\right)
\end{aligned}
\end{equation}
\begin{equation}
\begin{aligned}
b_2^\pm=\frac{1}{2} \left(\pm\sqrt{\lambda _1^2-2 \lambda _2 \lambda _1+\lambda _2^2+ 4 \lambda _5^2}+\lambda _1+\lambda _2\right)
\end{aligned}
\end{equation}
\begin{equation}
\begin{aligned}
b_3=2 \lambda _{c}
\end{aligned}
\end{equation}

The three other eigenvalues, $b_{4; 5; 6}$,  are located as roots of the cubic
polynomial equation given in Eq.~(\ref{eq:cubic-polynom}).

The third submatrix $\mathcal{M}_3$ encodes the scattering with initial and final states being either $\rho_1\eta_1$ state, or $\rho_2\eta_2$ state. It reads,

\begin{equation}
\mathcal{M}_3=
\left(
\begin{array}{cc}
\lambda _1 & \lambda _5 \\
\lambda _5 & \lambda _2 \\
\end{array}
\right)
\end{equation}
Its 2 eigenvalues read as follows:
\begin{equation}
\begin{aligned}
c^\pm=b_2^\pm
\end{aligned}
\end{equation}

The fourth submatrix $\mathcal{M}_4$ corresponds to scattering with initial and final states being one of the $12$ following states: ($\rho_0\phi_1^+$, $\rho_1\phi_1^+$, $\rho_2\phi_1^+$,  $\eta_1\phi_1^+$, $\eta_2\phi_1^+$, $\rho_0\phi_2^+$, $\rho_1\phi_2^+$, $\rho_2\phi_2^+$,  $\eta_1\phi_2^+$, $\eta_2\phi_2^+$, $\rho_0\delta^+$, $\rho_1\delta^+$, $\rho_2\delta^+$,  $\eta_1\delta^+$, $\eta_2\delta^+$). $\mathcal{M}_4$ is given by:

\begin{widetext}
	\begin{equation}
\mathcal{M}_4=
\left(
\begin{array}{ccccccccccccccc}
\lambda _{a} & 0 & 0 & 0 & 0 & 0 & 0 & 0 & 0 & 0 & 0 & 0 & 0 & 0 & 0 \\
0 & \lambda _1 & 0 & 0 & 0 & 0 & 0 & \frac{\lambda _{45}}{2} & 0 & -\frac{i \bar{\lambda} _{45}}{2} & 0 & 0 & 0 & 0 & 0 \\
0 & 0 & \lambda _3 & 0 & 0 & 0 & \frac{\lambda _{45}}{2} & 0 & \frac{i \bar{\lambda} _{45}}{2} & 0 & 0 & 0 & 0 & 0 & 0 \\
0 & 0 & 0 & \lambda _1 & 0 & 0 & 0 & \frac{i \bar{\lambda} _{45}}{2} & 0 & \frac{\lambda _{45}}{2} & 0 & 0 & 0 & 0 & 0 \\
0 & 0 & 0 & 0 & \lambda _3 & 0 & -\frac{i \bar{\lambda} _{45}}{2} & 0 & \frac{\lambda _{45}}{2} & 0 & 0 & 0 & 0 & 0 & 0 \\
0 & 0 & 0 & 0 & 0 & \lambda _{b} & 0 & 0 & 0 & 0 & 0 & 0 & 0 & 0 & 0 \\
0 & 0 & \frac{\lambda _{45}}{2} & 0 & \frac{i \bar{\lambda} _{45}}{2} & 0 & \lambda _3 & 0 & 0 & 0 & 0 & 0 & 0 & 0 & 0 \\
0 & \frac{\lambda _{45}}{2} & 0 & -\frac{i \bar{\lambda} _{45}}{2} & 0 & 0 & 0 & \lambda _2 & 0 & 0 & 0 & 0 & 0 & 0 & 0 \\
0 & 0 & -\frac{i \bar{\lambda} _{45}}{2} & 0 & \frac{\lambda _{45}}{2} & 0 & 0 & 0 & \lambda _3 & 0 & 0 & 0 & 0 & 0 & 0 \\
0 & \frac{i \bar{\lambda} _{45}}{2} & 0 & \frac{\lambda _{45}}{2} & 0 & 0 & 0 & 0 & 0 & \lambda _2 & 0 & 0 & 0 & 0 & 0 \\
0 & 0 & 0 & 0 & 0 & 0 & 0 & 0 & 0 & 0 & 2 \lambda _{c} & 0 & 0 & 0 & 0 \\
0 & 0 & 0 & 0 & 0 & 0 & 0 & 0 & 0 & 0 & 0 & \lambda _{a} & 0 & 0 & 0 \\
0 & 0 & 0 & 0 & 0 & 0 & 0 & 0 & 0 & 0 & 0 & 0 & \lambda _{b} & 0 & 0 \\
0 & 0 & 0 & 0 & 0 & 0 & 0 & 0 & 0 & 0 & 0 & 0 & 0 & \lambda _{a} & 0 \\
0 & 0 & 0 & 0 & 0 & 0 & 0 & 0 & 0 & 0 & 0 & 0 & 0 & 0 & \lambda _{b} \\
\end{array}
\right)
	\end{equation}
\end{widetext}

with its eigenvalues reading as,:

\begin{equation}
\begin{aligned}
f_1^\pm=\lambda_3\pm\lambda_4
\end{aligned}
\end{equation}
\begin{equation}
\begin{aligned}
f_2^\pm=b_1^\pm
\end{aligned}
\end{equation}
\begin{equation}
\begin{aligned}
f_3^\pm=a_3^\pm
\end{aligned}
\end{equation}
\begin{equation}
\begin{aligned}
f_4^\pm=b_2^\pm
\end{aligned}
\end{equation}
\begin{equation}
\begin{aligned}
f_5=a_4
\end{aligned}
\end{equation}
\begin{equation}
\begin{aligned}
f_6=a_5
\end{aligned}
\end{equation}
\begin{equation}
\begin{aligned}
f_7=b_3
\end{aligned}
\end{equation}

The fifth submatrix $\mathcal{M}_5$ corresponds to scattering with
initial and final states being one of the following 6 sates: ($\frac{\phi_1^+\phi_1^+}{\sqrt{2}}$, $\frac{\phi_2^+\phi_2^+}{\sqrt{2}}$, $\frac{\delta^+\delta^+}{\sqrt{2}}$, $\phi_1^+\phi_2^+$, $\phi_1^+\delta^+$, $\phi_2^+\delta^+$). It is represented by,

\begin{equation}
\mathcal{M}_5=\left(
\begin{array}{cccccc}
\lambda _1 & \lambda _5 & 0 & 0 & 0 & 0 \\
\lambda _5 & \lambda _2 & 0 & 0 & 0 & 0 \\
0 & 0 & 2 \lambda _{c} & 0 & 0 & 0 \\
0 & 0 & 0 & \lambda _3+\lambda _4 & 0 & 0 \\
0 & 0 & 0 & 0 & \lambda _{a} & 0 \\
0 & 0 & 0 & 0 & 0 & \lambda _{b} \\
\end{array}
\right)
\end{equation}

with its six eigenvalues reading as,

\begin{equation}
\begin{aligned}
e_1=a_1
\end{aligned}
\end{equation}
\begin{equation}
\begin{aligned}
e_2^\pm=b_2^\pm
\end{aligned}
\end{equation}
\begin{equation}
\begin{aligned}
e_3=a_4
\end{aligned}
\end{equation}
\begin{equation}
\begin{aligned}
e_4=a_5
\end{aligned}
\end{equation}
\begin{equation}
\begin{aligned}
e_5=b_3
\end{aligned}
\end{equation}
%

\setcounter{equation}{0}
\renewcommand{\theequation}{C\arabic{equation}}
\section{Boundedness from below Constraints}
\label{bfb-appendix}
To proceed to the most general case, we adopt a different parameterization of the fields that will turn out to be
particularly convenient to entirely solve the problem. For that we combine both parameterizations used in \cite{aa2011, Bonilla2015} and define:
\begin{eqnarray}
r &\equiv& \sqrt{H_1^\dagger{H_1} + H_2^\dagger{H_2} + Tr\Delta^{\dagger}{\Delta}} \label{eq:para1}\\
H_1^\dagger{H_1} &\equiv& r^2 \cos^2 \theta \sin^2 \phi  \label{eq:para2}\\
H_2^\dagger{H_2} &\equiv& r^2 \sin^2 \theta \sin^2 \phi  \label{eq:para3}\\
Tr\Delta^{\dagger}{\Delta} &\equiv& r^2 \cos^2 \phi  \label{eq:para4}\\
Tr(\Delta^{\dagger}{\Delta})^2/(Tr\Delta^{\dagger}{\Delta})^2 &\equiv& \epsilon  \label{eq:para5}\\
(H_1^\dagger{\Delta}{\Delta}^{\dagger}H_1)/(H_1^\dagger{H_1}Tr\Delta^{\dagger}{\Delta}) &\equiv&  \eta  \label{eq:para6}\\
(H_2^\dagger{\Delta}{\Delta}^{\dagger}H_2)/(H_2^\dagger{H_2}Tr\Delta^{\dagger}{\Delta}) &\equiv&  \zeta \label{eq:para7}
\end{eqnarray}
\noindent
Obviously, when $H_1$, $H_2$ and $\Delta$ scan all the field space, the radius $r$ scans the domain $[0, \infty[$, the angle $\theta \in [0, 2 \pi]$ and the angle $\phi \in [0, \frac{\pi}{2}]$. Moreover, as $\frac{H_1^\dagger\cdot H_2}{|H_1||H_2|}$ is a product of unit spinor, it is a complex number $\alpha + i \beta$ such that $|\alpha + i \beta| \le\,1$. We can rewrite it in polar coordinates as $\alpha + i \beta = \xi e^{i\psi}$ with $\xi \in (0,1)$. We can also show that $\epsilon = \eta = \zeta = \frac{1}{2}$. \\

\noindent
With this parameterization, one can cast $V^{(4)}(H_1,H_2,\Delta)$ into the following simple form,
\begin{eqnarray}
V^{(4)}(r,c^2_\theta,s^2_\phi,c_{2\psi},\xi,\epsilon,\eta,\zeta) &=& r^4\Big\{ \lambda_1c^4_\theta s^4_\phi + \lambda_2s^4_\theta s^4_\phi + \lambda_3 c^2_\theta s^2_\theta s^4_\phi + \lambda_4 c^2_\theta s^2_\theta s^4_\phi \xi^2 + \lambda_5c^2_\theta s^2_\theta s^4_\phi \xi^2 \cos2\psi \nonumber\\
&& + c^4_\phi (\bar{\lambda}_8 + \epsilon\bar{\lambda}_9) +  c^2_\theta c^2_\phi s^2_\phi (\lambda_6 + \eta\lambda_8) + s^2_\theta c^2_\phi s^2_\phi (\lambda_7 + \zeta\lambda_9)\Big\}
\label{eq:V4general2}
\end{eqnarray}
\noindent
By using the notation:

\begin{eqnarray}
x &\equiv& \cos^2\theta \\
y &\equiv& \sin^2\phi  \\
z &\equiv& \cos2\psi \in (-1,1)
\end{eqnarray}
we can transform the potential to more a convenient form :
\begin{eqnarray}
V^{(4)}/r^4 &=& \big\{\frac{\lambda_1}{2}\,x^2 + \frac{\lambda_2}{2}\,(1-x)^2 + \lambda_3\,x(1-x) + \lambda_4\,x(1-x)\xi^2 + \lambda_5\,x(1-x)\xi^2\,z \big\}\,y^2\nonumber\\
&+&  \big\{\bar{\lambda}_8 + \epsilon\bar{\lambda}_9\big\}\,(1-y)^2\nonumber\\
&+&  \big\{ (\lambda_6 + \eta\lambda_8)\,x + (\lambda_7 + \zeta\lambda_9)\,(1-x)\big\}\,y(1-y)
\label{eq:V4general3}
\end{eqnarray}
One can derive the BFB condition  by studying $V^{(4)}(x,y,z,\xi,\epsilon,\eta,\zeta) $ as a quadratic function using the fact that :
\begin{eqnarray}
f(y) =  a\,y^2 + b\,(1-y)^2 + c\,y\,(1 - y),\quad y \in (0,1)\quad \Leftrightarrow \quad a > 0,\, b > 0\,\,{\rm and}\,\,c +2\sqrt{ab} > 0\nonumber\\
\label{lemme}
\end{eqnarray}
Then the following set of constraints is readily deduced:
\begin{eqnarray}
F_{I}(\xi,z) &\equiv& \frac{\lambda_1}{2}\,x^2 + \frac{\lambda_2}{2}\,(1-x)^2 + \lambda_3\,x(1-x) + \lambda_4\,x(1-x)\xi^2 + \lambda_5\,x(1-x)\xi^2\,z  > 0 \label{condA}\\
F_{II}(\epsilon) &\equiv& \bar{\lambda}_8 + \epsilon\bar{\lambda}_9 > 0 \label{condB}  \\
F_{III}(\eta,\zeta) &\equiv& (\lambda_6 + \eta\lambda_8)\,x + (\lambda_7 + \zeta\lambda_9)\,(1-x) > -2\sqrt{F_{I}(\xi,z)\,F_{II}(\epsilon)} \label{condC}
\end{eqnarray}
For $F_{I}(\xi,z) > 0$, using again Eq.~\ref{lemme}, we recover the usual BFB constraints of $2\mathcal{HDM}$ if  $\xi = {0;1}$ and $z={-1;1}$:
\begin{eqnarray}
\lambda_1\,,\,\lambda_2 &>& 0 \label{2hdm_1}\\
\lambda_3 + \sqrt{\lambda_1\lambda_2} &>& 0 \label{2hdm_2}  \\
\lambda_3 + \lambda_4 - |\lambda_5| + \sqrt{\lambda_1\lambda_2} &>& 0 \label{2hdm_3} 
\end{eqnarray}
Since $F_{II}(\epsilon)$ is a monotonic function, the condition $0 < F_{II}(\epsilon)$ is equivalent to $0 < F_{II}(\frac{1}{2})$. So Eq.~\ref{condB} becomes,
\begin{eqnarray}
\qquad \bar{\lambda}_8 + \frac{1}{2}\bar{\lambda}_9 > 0 \label{generic_condB}
\end{eqnarray}
As to Eq.~\ref{condC}, one can re-write it as:
\begin{eqnarray}
F_{III}(\eta,\zeta) + 2\sqrt{F_{I}(\xi,z)\,F_{II}(\epsilon)} > 0 \, \Leftrightarrow \, \left\{
\begin{aligned}
F_{III}(\eta,\zeta)\,&> 0&\text{and} && F_{I}(\xi,z) F_{II}(\epsilon) > 0 &&\text{(i)}\\
&&\text{or}\\
F_{III}(\eta,\zeta)\,&\leqslant 0&\text{and} && 4\,F_{I}(\xi,z) F_{II}(\epsilon) > F^2_{III}(\eta,\zeta) &&\text{(ii)}\\
\end{aligned}
\right.
\end{eqnarray}

\begin{itemize}
	\item scenario (i) : starting with the fact that $x = \cos^2\theta > 0$ and $1-x =\sin^2\theta > 0$, thus $F_{III}(\eta,\zeta) > 0  \Rightarrow $ generic relations :
	\begin{eqnarray} 
	&&\lambda_6 + \frac{1}{2}\lambda_8 > 0 \label{generic1_condC}\\
	&&\lambda_7 + \frac{1}{2}\lambda_9 > 0 \label{generic2_condC} 
	\end{eqnarray}
	\item scenario (ii) : This scenario implies that $(\lambda_6 + \eta\lambda_8)\, {\rm and}\, (\lambda_7 + \zeta\lambda_9) \le 0$,  and leads to:
	\begin{eqnarray}
	&&\Big\{2\lambda_1(\bar{\lambda}_8 + \epsilon\bar{\lambda}_9) - (\lambda_6 + \eta\lambda_8)^2\Big\}\,x^2 + 
	\Big\{2\lambda_2(\bar{\lambda}_8 + \epsilon\bar{\lambda}_9) - (\lambda_7 + \zeta\lambda_9)^2\Big\}\,(1-x)^2 \nonumber\\
	&&+ \Big\{ 4(\lambda_3 + \lambda_4 \xi^2 + \lambda_5 \xi^2 z)(\bar{\lambda}_8 + \epsilon\bar{\lambda}_9) - 2(\lambda_6 + \eta\lambda_8)\,(\lambda_7 + \zeta\lambda_9) \Big\}x\,(1 - x) > 0\nonumber\\
	\end{eqnarray}
	Applying the {\it lemma} given by Eq.~\ref{lemme}, we obtain the generic new constraints,
	\begin{eqnarray}
	&& \lambda_6 + \eta\lambda_8 > -\sqrt{2\lambda_1(\bar{\lambda}_8 + \epsilon\bar{\lambda}_9)} \label{generic3_condC}\\
	&& \lambda_7 + \zeta\lambda_9 > -\sqrt{2\lambda_2(\bar{\lambda}_8 + \epsilon\bar{\lambda}_9)} \label{generic4_condC}\\
	&& 4(\lambda_3 + \lambda_4 \xi^2 + \lambda_5 \xi^2 z)(\bar{\lambda}_8 + \epsilon\bar{\lambda}_9) - 2(\lambda_6 + \eta\lambda_8)\,(\lambda_7 + \zeta\lambda_9) > \nonumber\\
	&&\hspace{1cm}-2\sqrt{\bigg(2\lambda_1(\bar{\lambda}_8 + \epsilon\bar{\lambda}_9) - (\lambda_6 + \eta\lambda_8)^2\bigg)\bigg(2\lambda_2(\bar{\lambda}_8 +\epsilon\bar{\lambda}_9) - (\lambda_7 + \zeta\lambda_9)^2\bigg)}
	\label{generic5_condC} 
	\end{eqnarray}
\end{itemize}
Then from Eqs.~(\ref{generic3_condC},~\ref{generic4_condC}), we deduce:

\begin{eqnarray}
&&\lambda_6 + \frac{1}{2}\lambda_8 > -\sqrt{2\lambda_1(\bar{\lambda}_8 + \frac{1}{2}\bar{\lambda}_9)}\\
&& \lambda_7 + \frac{1}{2}\lambda_9 > -\sqrt{2\lambda_2(\bar{\lambda}_8 + \frac{1}{2}\bar{\lambda}_9)}
\end{eqnarray}

Lastly, by considering $\xi\in[0;1]$ and $z\in[-1;1]$, we can see that Eq.~\ref{generic5_condC} leads to the constraints:
\begin{equation}
\begin{aligned}
4\left(\lambda_3+\lambda_4-|\lambda_5|\right)\lambda_c-2\lambda_a\lambda_b>\\-2\sqrt{\left(2\lambda_1\lambda_c-\lambda_a^2\right)\left(2\lambda_2\lambda_c-\lambda_b^2\right)}
\end{aligned}
\end{equation}
\begin{equation}
\begin{aligned}
4\lambda_3\lambda_c-2\lambda_a\lambda_b>-2\sqrt{\left(2\lambda_1\lambda_c-\lambda_a^2\right)\left(2\lambda_2\lambda_c-\lambda_b^2\right)}
\end{aligned}
\end{equation}
\setcounter{equation}{0}
\renewcommand{\theequation}{C\arabic{equation}}


\section{Scalar couplings}
\label{scalarcoup}

\setcounter{equation}{0}
\renewcommand{\theequation}{D\arabic{equation}}
\label{sec-higgs couplings}
\paragraph*{}
In this appendix, we present hereafter the triple scalar couplings needed for our study. More precisely, we present the couplings used to calculate the tadpoles of  two neutral $CP$-even Higgs $h_1$, $h_2$ and $h_3$. Here only three-leg couplings will be considered since we are interested in one-loop contributions. Further, within this restricted class, we look for vertices such as $h_1F_iF_i$, $h_2F_iF_i$ or $h_3F_iF_i$ , where
$F_i$ stands for any quantum field of our model: scalar and vectorial bosons, fermions, Goldstone fields $G_i$ , and Faddeev-Popov ghost fields $\eta_i$.
\paragraph*{}
We note $C^{h_1}_{F_iF_i}$, $C^{h_2}_{F_iF_i}$ and $C^{h_3}_{F_iF_i}$, the couplings to the Higgs $h_1$, $h_2$ and $h_3$.
Since the field $F_i$ fixes the propagator, we also give the values $t^{h_1}_i$, $t^{h_2}_i$ and $t^{h_3}_i$ of the loop due to the propagator of the $F_i$ particle which gain a factor of 2 in the case of charged fields
and the symmetry factor $s_i$:
\begin{scriptsize}
	
	\begin{align}
	C_1^{h_1}\equiv C^{h_1}_{h_1h_1}=&-\frac{3}{2}\left(2 v_1 \mathcal{E}_{11} \left(\mathcal{E}_{13}^2 \lambda _a+\lambda _1 \mathcal{E}_{11}^2+\lambda _{345} \mathcal{E}_{12}^2\right)+2 v_2 \mathcal{E}_{12} \left(\mathcal{E}_{13}^2 \lambda _b+\lambda _{345} \mathcal{E}_{11}^2+\lambda _2 \mathcal{E}_{12}^2\right)\right.\nonumber\\
	&\left.+\mathcal{E}_{13} \left(2 v_t \left(\mathcal{E}_{11}^2 \lambda _a+\mathcal{E}_{12}^2 \lambda _b+2 \mathcal{E}_{13}^2 \lambda _c\right)-\mu _2 \mathcal{E}_{12}^2-\mathcal{E}_{11} \left(\mu _1 \mathcal{E}_{11}+2 \mu _3 \mathcal{E}_{12}\right)\right)\right)
	\end{align}
	\begin{align}
	C_1^{h_2}\equiv C^{h_2}_{h_1h_1}&=-\frac{1}{2}\left(2 \mathcal{E}_{23} \mathcal{E}_{11}^2 \lambda _a v_t+4 \mathcal{E}_{13} \mathcal{E}_{21} \mathcal{E}_{11} \lambda _a v_{t}+2 v_1 \left(2 \mathcal{E}_{11} \left(\mathcal{E}_{13} \mathcal{E}_{23} \lambda _a+\lambda _{345} \mathcal{E}_{12} \mathcal{E}_{22}\right)+\mathcal{E}_{21} \left(\mathcal{E}_{13}^2 \lambda _a+\lambda _{345} \mathcal{E}_{12}^2\right)+3 \lambda _1 \mathcal{E}_{21} \mathcal{E}_{11}^2\right)\right.\nonumber\\
	&+6 \lambda _2 v_2 \mathcal{E}_{12}^2 \mathcal{E}_{22}+2 \mathcal{E}_{23} \mathcal{E}_{12}^2 \lambda _b v_{t}+4 \mathcal{E}_{13} \mathcal{E}_{22} \mathcal{E}_{12} \lambda _b v_{t}+2 v_2 \left(\mathcal{E}_{13} \left(\mathcal{E}_{13} \mathcal{E}_{22}+2 \mathcal{E}_{12} \mathcal{E}_{23}\right) \lambda _b+\lambda _{345} \mathcal{E}_{11} \left(2 \mathcal{E}_{12} \mathcal{E}_{21}+\mathcal{E}_{11} \mathcal{E}_{22}\right)\right)+\nonumber\\
	&\left.12 \mathcal{E}_{13}^2 \mathcal{E}_{23} \lambda _c v_{t}+\mu _1 \left(-\mathcal{E}_{23}\right) \mathcal{E}_{11}^2-2 \mu _1 \mathcal{E}_{13} \mathcal{E}_{21} \mathcal{E}_{11}-2 \mu _2 \mathcal{E}_{12} \mathcal{E}_{13} \mathcal{E}_{22}-\mu _2 \mathcal{E}_{12}^2 \mathcal{E}_{23}-2 \mu _3 \left(\mathcal{E}_{11} \mathcal{E}_{13} \mathcal{E}_{22}+\mathcal{E}_{12} \left(\mathcal{E}_{13} \mathcal{E}_{21}+\mathcal{E}_{11} \mathcal{E}_{23}\right)\right)\right)
	\end{align}
	\begin{align}
	C_1^{h_3}\equiv C^{h_3}_{h_1h_1}&-\frac{1}{2}\left(2 \mathcal{E}_{33} \mathcal{E}_{11}^2 \lambda _a v_{t}+4 \mathcal{E}_{13} \mathcal{E}_{31} \mathcal{E}_{11} \lambda _a v_{t}+2 v_1 \left(2 \mathcal{E}_{11} \left(\mathcal{E}_{13} \mathcal{E}_{33} \lambda _a+\lambda _{345} \mathcal{E}_{12} \mathcal{E}_{32}\right)+\mathcal{E}_{31} \left(\mathcal{E}_{13}^2 \lambda _a+\lambda _{345} \mathcal{E}_{12}^2\right)+3 \lambda _1 \mathcal{E}_{31} \mathcal{E}_{11}^2\right)\right.\nonumber\\
	&+2\mathcal{E}_{33} \mathcal{E}_{12}^2 \lambda _b v_{t}+4 \mathcal{E}_{13} \mathcal{E}_{32} \mathcal{E}_{12} \lambda _b v_{t}+2 v_2 \left(\mathcal{E}_{13} \left(\mathcal{E}_{13} \mathcal{E}_{32}+2 \mathcal{E}_{12} \mathcal{E}_{33}\right) \lambda _b+\lambda _{345} \mathcal{E}_{11} \left(2 \mathcal{E}_{12} \mathcal{E}_{31}+\mathcal{E}_{11} \mathcal{E}_{32}\right)\right)+12 \mathcal{E}_{13}^2 \mathcal{E}_{33} \lambda _c v_{t}\nonumber\\
	&\left.+6 \lambda _2 v_2 \mathcal{E}_{12}^2 \mathcal{E}_{32}-\mu _1 \mathcal{E}_{33} \mathcal{E}_{11}^2-2 \mu _1 \mathcal{E}_{13} \mathcal{E}_{31} \mathcal{E}_{11}-2 \mu _2 \mathcal{E}_{12} \mathcal{E}_{13} \mathcal{E}_{32}-\mu _2 \mathcal{E}_{12}^2 \mathcal{E}_{33}-2 \mu _3 \left(\mathcal{E}_{11} \mathcal{E}_{13} \mathcal{E}_{32}+\mathcal{E}_{12} \left(\mathcal{E}_{13} \mathcal{E}_{31}+\mathcal{E}_{11} \mathcal{E}_{33}\right)\right)\right)\nonumber\\
	\end{align}
	\begin{align}
	t_1^{h_1}&=t_1^{h_2}=t_1^{h_3}=iA_0\left(m^2_{h_1}\right)
	\end{align}
	\begin{align}
	C_2^{h_1}\equiv C^{h_1}_{h_2h_2}&=-i\frac{1}{2}\left(2 \mathcal{E}_{13} \mathcal{E}_{21}^2 \lambda _a v_{t}+4 \mathcal{E}_{11} \mathcal{E}_{23} \mathcal{E}_{21} \lambda _a v_{t}+2 v_1 \left(2 \mathcal{E}_{21} \left(\mathcal{E}_{13} \mathcal{E}_{23} \lambda _a+\lambda _{345} \mathcal{E}_{12} \mathcal{E}_{22}\right)+\mathcal{E}_{11} \left(\mathcal{E}_{23}^2 \lambda _a+3 \lambda _1 \mathcal{E}_{21}^2+\lambda _{345} \mathcal{E}_{22}^2\right)\right)\right.\nonumber\\
	&+2 \mathcal{E}_{13} \mathcal{E}_{22}^2 \lambda _b v_{t}+4 \mathcal{E}_{12} \mathcal{E}_{23} \mathcal{E}_{22} \lambda _b v_{t}+2 v_2 \left(\mathcal{E}_{23} \left(2 \mathcal{E}_{13} \mathcal{E}_{22}+\mathcal{E}_{12} \mathcal{E}_{23}\right) \lambda _b+\lambda _{345} \mathcal{E}_{21} \left(\mathcal{E}_{12} \mathcal{E}_{21}+2 \mathcal{E}_{11} \mathcal{E}_{22}\right)\right)+12 \mathcal{E}_{13} \mathcal{E}_{23}^2 \lambda _c v_{t}\nonumber\\
	&\left.+6 \lambda _2 v_2 \mathcal{E}_{12} \mathcal{E}_{22}^2-\mu _1\mathcal{E}_{13} \mathcal{E}_{21}^2-2 \mu _1 \mathcal{E}_{11} \mathcal{E}_{23} \mathcal{E}_{21}-\mu _2 \mathcal{E}_{13} \mathcal{E}_{22}^2-2 \mu _2 \mathcal{E}_{12} \mathcal{E}_{22} \mathcal{E}_{23}-2 \mu _3 \left(\mathcal{E}_{13} \mathcal{E}_{21} \mathcal{E}_{22}+\left(\mathcal{E}_{12} \mathcal{E}_{21}+\mathcal{E}_{11} \mathcal{E}_{22}\right) \mathcal{E}_{23}\right)\right)\nonumber\\
	\end{align}
	\begin{align}
	C_2^{h_2}\equiv C^{h_2}_{h_2h_2}&=-i\frac{3}{2}\left(2 v_1 \mathcal{E}_{21} \left(\mathcal{E}_{23}^2 \lambda _a+\lambda _1 \mathcal{E}_{21}^2+\lambda _{345} \mathcal{E}_{22}^2\right)+2 v_2 \mathcal{E}_{22} \left(\mathcal{E}_{23}^2 \lambda _b+\lambda _{345} \mathcal{E}_{21}^2+\lambda _2 \mathcal{E}_{22}^2\right)\right.\nonumber\\
	&\left.+\mathcal{E}_{23} \left(2 v_{t} \left(\mathcal{E}_{21}^2 \lambda _a+\mathcal{E}_{22}^2 \lambda _b+2 \mathcal{E}_{23}^2 \lambda _c\right)-\mu _2 \mathcal{E}_{22}^2-\mathcal{E}_{21} \left(\mu _1 \mathcal{E}_{21}+2 \mu _3 \mathcal{E}_{22}\right)\right)\right)\nonumber\\
	\end{align}
	\begin{align}
	C_2^{h_3}\equiv C^{h_3}_{h_2h_2}&=-i\frac{1}{2}\left(2 \mathcal{E}_{33} \mathcal{E}_{21}^2 \lambda _a v_{t}+4 \mathcal{E}_{23} \mathcal{E}_{31} \mathcal{E}_{21} \lambda _a v_{t}+2 v_1 \left(2 \mathcal{E}_{21} \left(\mathcal{E}_{23} \mathcal{E}_{33} \lambda _a+\lambda _{345} \mathcal{E}_{22} \mathcal{E}_{32}\right)+\mathcal{E}_{31} \left(\mathcal{E}_{23}^2 \lambda _a+\lambda _{345} \mathcal{E}_{22}^2\right)+3 \lambda _1 \mathcal{E}_{31} \mathcal{E}_{21}^2\right)\right.\nonumber\\
	&+6 \lambda _2 v_2 \mathcal{E}_{22}^2 \mathcal{E}_{32}+2 \mathcal{E}_{33} \mathcal{E}_{22}^2 \lambda _b v_{t}+4 \mathcal{E}_{23} \mathcal{E}_{32} \mathcal{E}_{22} \lambda _b v_{t}+2 v_2 \left(\mathcal{E}_{23} \left(\mathcal{E}_{23} \mathcal{E}_{32}+2 \mathcal{E}_{22} \mathcal{E}_{33}\right) \lambda _b+\lambda _{345} \mathcal{E}_{21} \left(2 \mathcal{E}_{22} \mathcal{E}_{31}+\mathcal{E}_{21} \mathcal{E}_{32}\right)\right)\nonumber\\
	&\left.+12 \mathcal{E}_{23}^2 \mathcal{E}_{33} \lambda _c v_{t}+\mu _1 \left(-\mathcal{E}_{33}\right) \mathcal{E}_{21}^2-2 \mu _1 \mathcal{E}_{23} \mathcal{E}_{31} \mathcal{E}_{21}-2 \mu _2 \mathcal{E}_{22} \mathcal{E}_{23} \mathcal{E}_{32}-\mu _2 \mathcal{E}_{22}^2 \mathcal{E}_{33}-2 \mu _3 \left(\mathcal{E}_{21} \mathcal{E}_{23} \mathcal{E}_{32}+\mathcal{E}_{22} \left(\mathcal{E}_{23} \mathcal{E}_{31}+\mathcal{E}_{21} \mathcal{E}_{33}\right)\right)\right)\nonumber\\
	\end{align}
	\begin{align}
	t_2^{h_1}&=t_2^{h_2}=t_2^{h_3}=iA_0\left(m^2_{h_2}\right)\nonumber\\
	\end{align}
	
	\begin{align}
	C_3^{h_1}\equiv C^{h_1}_{h_3h_3}&=-i\frac{1}{2}\left(2 \mathcal{E}_{13} \mathcal{E}_{31}^2 \lambda _a v_{t}+4 \mathcal{E}_{11} \mathcal{E}_{33} \mathcal{E}_{31} \lambda _a v_{t}+2 v_1 \left(2 \mathcal{E}_{31} \left(\mathcal{E}_{13} \mathcal{E}_{33} \lambda _a+\lambda _{345} \mathcal{E}_{12} \mathcal{E}_{32}\right)+\mathcal{E}_{11} \left(\mathcal{E}_{33}^2 \lambda _a+3 \lambda _1 \mathcal{E}_{31}^2+\lambda _{345} \mathcal{E}_{32}^2\right)\right)\right.\nonumber\\
	&+2 \mathcal{E}_{13} \mathcal{E}_{32}^2 \lambda _b v_{t}+4 \mathcal{E}_{12} \mathcal{E}_{33} \mathcal{E}_{32} \lambda _b v_{t}+2 v_2 \left(\mathcal{E}_{33} \left(2 \mathcal{E}_{13} \mathcal{E}_{32}+\mathcal{E}_{12} \mathcal{E}_{33}\right) \lambda _b+\lambda _{345} \mathcal{E}_{31} \left(\mathcal{E}_{12} \mathcal{E}_{31}+2 \mathcal{E}_{11} \mathcal{E}_{32}\right)\right)+12 \mathcal{E}_{13} \mathcal{E}_{33}^2 \lambda _c v_{t}\nonumber\\
	&\left.+6 \lambda _2 v_2 \mathcal{E}_{12} \mathcal{E}_{32}^2-\mu _1 \mathcal{E}_{13} \mathcal{E}_{31}^2-2 \mu _1 \mathcal{E}_{11} \mathcal{E}_{33} \mathcal{E}_{31}-\mu _2 \mathcal{E}_{13} \mathcal{E}_{32}^2-2 \mu _2 \mathcal{E}_{12} \mathcal{E}_{32} \mathcal{E}_{33}-2 \mu _3 \left(\mathcal{E}_{13} \mathcal{E}_{31} \mathcal{E}_{32}+\left(\mathcal{E}_{12} \mathcal{E}_{31}+\mathcal{E}_{11} \mathcal{E}_{32}\right) \mathcal{E}_{33}\right)\right)\nonumber\\
	\end{align}
	\begin{align}
	C_3^{h_2}\equiv C^{h_2}_{h_3h_3}&=-i\frac{1}{2}\left(2 \mathcal{E}_{23} \mathcal{E}_{31}^2 \lambda _a v_{t}+4 \mathcal{E}_{21} \mathcal{E}_{33} \mathcal{E}_{31} \lambda _a v_{t}+2 v_1 \left(2 \mathcal{E}_{31} \left(\mathcal{E}_{23} \mathcal{E}_{33} \lambda _a+\lambda _{345} \mathcal{E}_{22} \mathcal{E}_{32}\right)+\mathcal{E}_{21} \left(\mathcal{E}_{33}^2 \lambda _a+3 \lambda _1 \mathcal{E}_{31}^2+\lambda _{345} \mathcal{E}_{32}^2\right)\right)\right.\nonumber\\
	&+6 \lambda _2 v_2 \mathcal{E}_{22} \mathcal{E}_{32}^2+2 \mathcal{E}_{23} \mathcal{E}_{32}^2 \lambda _b v_{t}+4 \mathcal{E}_{22} \mathcal{E}_{33} \mathcal{E}_{32} \lambda _b v_{t}+2 v_2 \left(\mathcal{E}_{33} \left(2 \mathcal{E}_{23} \mathcal{E}_{32}+\mathcal{E}_{22} \mathcal{E}_{33}\right) \lambda _b+\lambda _{345} \mathcal{E}_{31} \left(\mathcal{E}_{22} \mathcal{E}_{31}+2 \mathcal{E}_{21} \mathcal{E}_{32}\right)\right)\nonumber\\
	&\left.+12 \mathcal{E}_{23} \mathcal{E}_{33}^2 \lambda _c v_{t}+\mu _1 \left(-\mathcal{E}_{23}\right) \mathcal{E}_{31}^2-2 \mu _1 \mathcal{E}_{21} \mathcal{E}_{33} \mathcal{E}_{31}-\mu _2 \mathcal{E}_{23} \mathcal{E}_{32}^2-2 \mu _2 \mathcal{E}_{22} \mathcal{E}_{32} \mathcal{E}_{33}-2 \mu _3 \left(\mathcal{E}_{23} \mathcal{E}_{31} \mathcal{E}_{32}+\left(\mathcal{E}_{22} \mathcal{E}_{31}+\mathcal{E}_{21} \mathcal{E}_{32}\right) \mathcal{E}_{33}\right)\right)\nonumber\\
	\end{align}
	\begin{align}
	C_3^{h_3}\equiv C^{h_3}_{h_3h_3}&=-i\frac{3}{2}\left(2 v_1 \mathcal{E}_{31} \left(\mathcal{E}_{33}^2 \lambda _a+\lambda _1 \mathcal{E}_{31}^2+\lambda _{345} \mathcal{E}_{32}^2\right)+2 v_2 \mathcal{E}_{32} \left(\mathcal{E}_{33}^2 \lambda _b+\lambda _{345} \mathcal{E}_{31}^2+\lambda _2 \mathcal{E}_{32}^2\right)+\mathcal{E}_{33}\left(2 v_{t} \left(\mathcal{E}_{31}^2 \lambda _a+\mathcal{E}_{32}^2 \lambda _b+2 \mathcal{E}_{33}^2 \lambda _c\right)\right.\right.\nonumber\\
	&\left.\left.-\mu _2\mathcal{E}_{32}^2-\mathcal{E}_{31} \left(\mu _1 \mathcal{E}_{31}+2 \mu _3 \mathcal{E}_{32}\right)\right)\right)\nonumber\\
	\end{align}
	\begin{align}
	t_3^{h_1}&=t_3^{h_2}=t_3^{h_3}=iA_0\left(m^2_{h_3}\right)\nonumber\\
	\end{align}
	
	\begin{align}
	C_4^{h_1}\equiv C^{h_1}_{G_0G_0}&=-i\frac{v_1^2 \left(\mathcal{E}_{13} \left(2 \lambda _a v_{t}-\mu _1\right)+2 \lambda _{345} v_2 \mathcal{E}_{12}\right)+v_2^2 \left(\mathcal{E}_{13} \left(2 \lambda _b v_{t}-\mu _2\right)+2 \lambda _2 v_2 \mathcal{E}_{12}\right)+2 v_2 v_1 \left(\lambda _{345} v_2 \mathcal{E}_{11}-\mu _3 \mathcal{E}_{13}\right)+2 \lambda _1 v_1^3 \mathcal{E}_{11}}{2 v_d^2}\nonumber\\
	\end{align}
	\begin{align}
	C_4^{h_2}\equiv C^{h_2}_{G_0G_0}&=-\frac{i \left(v_1^2 \left(\mathcal{E}_{23} \left(2 \lambda _a v_{t}-\mu _1\right)+2 \lambda _{345} v_2 \mathcal{E}_{22}\right)+v_2^2 \left(\mathcal{E}_{23} \left(2 \lambda _b v_{t}-\mu _2\right)+2 \lambda _2 v_2 \mathcal{E}_{22}\right)+2 v_2 v_1 \left(\lambda _{345} v_2 \mathcal{E}_{21}-\mu _3 \mathcal{E}_{23}\right)+2 \lambda _1 v_1^3 \mathcal{E}_{21}\right)}{2 v_d^2}\nonumber\\
	\end{align}
	\begin{align}
	C_4^{h_3}\equiv C^{h_3}_{G_0G_0}&=-\frac{i \left(v_1^2 \left(\mathcal{E}_{33} \left(2 \lambda _a v_{t}-\mu _1\right)+2 \lambda _{345} v_2 \mathcal{E}_{32}\right)+v_2^2 \left(\mathcal{E}_{33} \left(2 \lambda _b v_{t}-\mu _2\right)+2 \lambda _2 v_2 \mathcal{E}_{32}\right)+2 v_2 v_1 \left(\lambda _{345} v_2 \mathcal{E}_{31}-\mu _3 \mathcal{E}_{33}\right)+2 \lambda _1 v_1^3 \mathcal{E}_{31}\right)}{2 v_d^2}\nonumber\\
	\end{align}
	\begin{align}
	t_4^{h_1}&=t_4^{h_2}=t_4^{h_3}=iA_0\left(\xi_Z m^2_{Z}\right)\nonumber\\
	\end{align}
	
	\begin{align}
	C_5^{h_1}\equiv C^{h_1}_{A_1A_1}&=-i\frac{1}{2}\left(s_{\beta }^2 \left(\mathcal{E}_{13} \left(2 \lambda _a v_{t}-\mu _1\right)+2 \lambda _1 v_1 \mathcal{E}_{11}+2 \left(\lambda _3+\lambda _4-\lambda _5\right) v_2 \mathcal{E}_{12}\right)+c_{\beta }^2 \left(\mathcal{E}_{13} \left(2 \lambda _b v_{t}-\mu _2\right)+2 \lambda _2 v_2 \mathcal{E}_{12}\right)\right.\nonumber\\
	&\left.+2 \left(\lambda _3+\lambda _4-\lambda _5\right) v_1 \mathcal{E}_{11}+2 c_{\beta } s_{\beta } \left(\mu _3 \mathcal{E}_{13}-2 \lambda _5 \left(v_2 \mathcal{E}_{11}+v_1 \mathcal{E}_{12}\right)\right)\right)\nonumber\\
	\end{align}
	\begin{align}
	C_5^{h_2}\equiv C^{h_2}_{A_1A_1}&=-i\frac{1}{2}\left(s_{\beta }^2 \left(\mathcal{E}_{23} \left(2 \lambda _a v_{t}-\mu _1\right)+2 \lambda _1 v_1 \mathcal{E}_{21}+2 \left(\lambda _3+\lambda _4-\lambda _5\right) v_2 \mathcal{E}_{22}\right)+c_{\beta }^2 \left(\mathcal{E}_{23} \left(2 \lambda _b v_{t}-\mu _2\right)+2 \lambda _2 v_2 \mathcal{E}_{22}\right)\right.\nonumber\\
	&\left.+2 \left(\lambda _3+\lambda _4-\lambda _5\right) v_1 \mathcal{E}_{21}+2 c_{\beta } s_{\beta } \left(\mu _3 \mathcal{E}_{23}-2 \lambda _5 \left(v_2 \mathcal{E}_{21}+v_1 \mathcal{E}_{22}\right)\right)\right)\nonumber\\
	\end{align}
	\begin{align}
	C_5^{h_3}\equiv C^{h_3}_{A_1A_1}&=-i\frac{1}{2}\left(s_{\beta }^2 \left(\mathcal{E}_{33} \left(2 \lambda _a v_{t}-\mu _1\right)+2 \lambda _1 v_1 \mathcal{E}_{31}+2 \left(\lambda _3+\lambda _4-\lambda _5\right) v_2 \mathcal{E}_{32}\right)+c_{\beta }^2 \left(\mathcal{E}_{33} \left(2 \lambda _b v_{t}-\mu _2\right)+2 \lambda _2 v_2 \mathcal{E}_{32}\right)\right.\nonumber\\
	&\left.+2 \left(\lambda _3+\lambda _4-\lambda _5\right) v_1 \mathcal{E}_{31}+2 c_{\beta } s_{\beta } \left(\mu _3 \mathcal{E}_{33}-2 \lambda _5 \left(v_2 \mathcal{E}_{31}+v_1 \mathcal{E}_{32}\right)\right)\right)\nonumber\\
	\end{align}
	\begin{align}
	t_5^{h_1}&=t_5^{h_2}=t_5^{h_3}=iA_0\left(m^2_{A_1}\right)\nonumber\\
	\end{align}
	\begin{align}
	C_6^{h_1}\equiv C^{h_1}_{G^\pm G^\pm}&=-\frac{1}{2}i\left(2 \mathcal{C}_{11}^2 \mathcal{E}_{13} \lambda _a v_{t}+2 v_1 \left(\mathcal{E}_{11} \left(\mathcal{C}_{13}^2 \lambda _a+\mathcal{C}_{12}^2 \lambda _3\right)+\mathcal{C}_{11}^2 \lambda _1 \mathcal{E}_{11}+\mathcal{C}_{12} \mathcal{C}_{11} \left(\lambda _4+\lambda _5\right) \mathcal{E}_{12}\right)+2 \mathcal{C}_{12}^2 \mathcal{E}_{13} \lambda _b v_{t}+\right.\nonumber\\
	&2 v_2 \left(\mathcal{E}_{12} \left(\mathcal{C}_{13}^2 \lambda _b+\mathcal{C}_{11}^2 \lambda _3\right)+\mathcal{C}_{12}^2 \lambda _2 \mathcal{E}_{12}+\mathcal{C}_{11} \mathcal{C}_{12} \left(\lambda _4+\lambda _5\right) \mathcal{E}_{11}\right)+4 \mathcal{C}_{13}^2 \mathcal{E}_{13} \lambda _c v_{t}+2 \mathcal{C}_{11} \mathcal{C}_{13} \mu _1 \mathcal{E}_{11}\nonumber\\
	&\left.+\mathcal{C}_{12}^2 \mu _2 \mathcal{E}_{13}+2 \mathcal{C}_{13} \mathcal{C}_{12} \mu _2 \mathcal{E}_{12}+\mathcal{C}_{11}^2 \mu _1 \mathcal{E}_{13}+2 \mu _3 \left(\mathcal{C}_{11} \mathcal{C}_{13} \mathcal{E}_{12}+\mathcal{C}_{12} \left(\mathcal{C}_{13} \mathcal{E}_{11}+\mathcal{C}_{11} \mathcal{E}_{13}\right)\right)\right)\nonumber\\
	\end{align}
	\begin{align}
	C_6^{h_2}\equiv C^{h_2}_{G^\pm G^\pm}&=-\frac{1}{2}i\left(2 \mathcal{C}_{11}^2 \mathcal{E}_{23} \lambda _a v_{t}+2 v_1 \left(\mathcal{E}_{21} \left(\mathcal{C}_{13}^2 \lambda _a+\mathcal{C}_{12}^2 \lambda _3\right)+\mathcal{C}_{11}^2 \lambda _1 \mathcal{E}_{21}+\mathcal{C}_{12} \mathcal{C}_{11} \left(\lambda _4+\lambda _5\right) \mathcal{E}_{22}\right)+2 \mathcal{C}_{12}^2 \mathcal{E}_{23} \lambda _b v_{t}+\right.\nonumber\\
	&2 v_2 \left(\mathcal{E}_{22} \left(\mathcal{C}_{13}^2 \lambda _b+\mathcal{C}_{11}^2 \lambda _3\right)+\mathcal{C}_{12}^2 \lambda _2 \mathcal{E}_{22}+\mathcal{C}_{11} \mathcal{C}_{12} \left(\lambda _4+\lambda _5\right) \mathcal{E}_{21}\right)+4 \mathcal{C}_{13}^2 \mathcal{E}_{23} \lambda _c v_{t}+2 \mathcal{C}_{11} \mathcal{C}_{13} \mu _1 \mathcal{E}_{21}+\nonumber\\
	&\left.\mathcal{C}_{11}^2 \mu _1 \mathcal{E}_{23}+2 \mathcal{C}_{12} \mathcal{C}_{13} \mu _2 \mathcal{E}_{22}+\mathcal{C}_{12}^2 \mu _2 \mathcal{E}_{23}+2 \mu _3 \left(\mathcal{C}_{11} \mathcal{C}_{13} \mathcal{E}_{22}+\mathcal{C}_{12} \left(\mathcal{C}_{13} \mathcal{E}_{21}+\mathcal{C}_{11} \mathcal{E}_{23}\right)\right)\right)\nonumber\\
	\end{align}
	\begin{align}
	C_6^{h_3}\equiv C^{h_3}_{G^\pm G^\pm}&=-\frac{1}{2}i\left(2 \mathcal{C}_{11}^2 \mathcal{E}_{33} \lambda _a v_{t}+2 v_1 \left(\mathcal{E}_{31} \left(\mathcal{C}_{13}^2 \lambda _a+\mathcal{C}_{12}^2 \lambda _3\right)+\mathcal{C}_{11}^2 \lambda _1 \mathcal{E}_{31}+\mathcal{C}_{12} \mathcal{C}_{11} \left(\lambda _4+\lambda _5\right) \mathcal{E}_{32}\right)+2 \mathcal{C}_{12}^2 \mathcal{E}_{33} \lambda _b v_{t}+\right.\nonumber\\
	&2 v_2 \left(\mathcal{E}_{32} \left(\mathcal{C}_{13}^2 \lambda _b+\mathcal{C}_{11}^2 \lambda _3\right)+\mathcal{C}_{12}^2 \lambda _2 \mathcal{E}_{32}+\mathcal{C}_{11} \mathcal{C}_{12} \left(\lambda _4+\lambda _5\right) \mathcal{E}_{31}\right)+4 \mathcal{C}_{13}^2 \mathcal{E}_{33} \lambda _c v_{t}+2 \mathcal{C}_{11} \mathcal{C}_{13} \mu _1 \mathcal{E}_{31}+\nonumber\\
	&\left.\mathcal{C}_{11}^2 \mu _1 \mathcal{E}_{33}+2 \mathcal{C}_{12} \mathcal{C}_{13} \mu _2 \mathcal{E}_{32}+\mathcal{C}_{12}^2 \mu _2 \mathcal{E}_{33}+2 \mu _3 \left(\mathcal{C}_{11} \mathcal{C}_{13} \mathcal{E}_{32}+\mathcal{C}_{12} \left(\mathcal{C}_{13} \mathcal{E}_{31}+\mathcal{C}_{11} \mathcal{E}_{33}\right)\right)\right)\nonumber\\
	\end{align}
	\begin{align}
	t_6^{h_1}&=t_7^{h_2}=t_7^{h_3}=i2A_0\left(\xi_W m^2_{W}\right)\nonumber\\
	\end{align}
	
	\begin{align}
	C_7^{h_1}\equiv C^{h_1}_{H^\pm_1H^\pm_1}&=-\frac{1}{2}i\left(2 \mathcal{C}_{21}^2 \mathcal{E}_{13} \lambda _a v_{t}+2 v_1 \left(\mathcal{E}_{11} \left(\mathcal{C}_{23}^2 \lambda _a+\mathcal{C}_{22}^2 \lambda _3\right)+\mathcal{C}_{21}^2 \lambda _1 \mathcal{E}_{11}+\mathcal{C}_{22} \mathcal{C}_{21} \left(\lambda _4+\lambda _5\right) \mathcal{E}_{12}\right)+2 \mathcal{C}_{22}^2 \mathcal{E}_{13} \lambda _b v_{t}+\right.\nonumber\\
	&2 v_2 \left(\mathcal{E}_{12} \left(\mathcal{C}_{23}^2 \lambda _b+\mathcal{C}_{21}^2 \lambda _3\right)+\mathcal{C}_{22}^2 \lambda _2 \mathcal{E}_{12}+\mathcal{C}_{21} \mathcal{C}_{22} \left(\lambda _4+\lambda _5\right) \mathcal{E}_{11}\right)+4 \mathcal{C}_{23}^2 \mathcal{E}_{13} \lambda _c v_{t}+2 \mathcal{C}_{21} \mathcal{C}_{23} \mu _1 \mathcal{E}_{11}+\nonumber\\
	&\left.\mathcal{C}_{21}^2 \mu _1 \mathcal{E}_{13}+2 \mathcal{C}_{22} \mathcal{C}_{23} \mu _2 \mathcal{E}_{12}+\mathcal{C}_{22}^2 \mu _2 \mathcal{E}_{13}+2 \mu _3 \left(\mathcal{C}_{21} \mathcal{C}_{23} \mathcal{E}_{12}+\mathcal{C}_{22} \left(\mathcal{C}_{23} \mathcal{E}_{11}+\mathcal{C}_{21} \mathcal{E}_{13}\right)\right)\right)\nonumber\\
	\end{align}
	\begin{align}
	C_7^{h_2}\equiv C^{h_2}_{H^\pm_1H^\pm_1}&=-\frac{1}{2}i\left(2 \mathcal{C}_{21}^2 \mathcal{E}_{23} \lambda _a v_{t}+2 v_1 \left(\mathcal{E}_{21} \left(\mathcal{C}_{23}^2 \lambda _a+\mathcal{C}_{22}^2 \lambda _3\right)+\mathcal{C}_{21}^2 \lambda _1 \mathcal{E}_{21}+\mathcal{C}_{22} \mathcal{C}_{21} \left(\lambda _4+\lambda _5\right) \mathcal{E}_{22}\right)+2 \mathcal{C}_{22}^2 \mathcal{E}_{23} \lambda _b v_{t}+\right.\nonumber\\
	&2 v_2 \left(\mathcal{E}_{22} \left(\mathcal{C}_{23}^2 \lambda _b+\mathcal{C}_{21}^2 \lambda _3\right)+\mathcal{C}_{22}^2 \lambda _2 \mathcal{E}_{22}+\mathcal{C}_{21} \mathcal{C}_{22} \left(\lambda _4+\lambda _5\right) \mathcal{E}_{21}\right)+4 \mathcal{C}_{23}^2 \mathcal{E}_{23} \lambda _c v_{t}+2 \mathcal{C}_{21} \mathcal{C}_{23} \mu _1 \mathcal{E}_{21}+\nonumber\\
	&\left.\mathcal{C}_{21}^2 \mu _1 \mathcal{E}_{23}+2 \mathcal{C}_{22} \mathcal{C}_{23} \mu _2 \mathcal{E}_{22}+\mathcal{C}_{22}^2 \mu _2 \mathcal{E}_{23}+2 \mu _3 \left(\mathcal{C}_{21} \mathcal{C}_{23} \mathcal{E}_{22}+\mathcal{C}_{22} \left(\mathcal{C}_{23} \mathcal{E}_{21}+\mathcal{C}_{21} \mathcal{E}_{23}\right)\right)\right)\nonumber\\
	\end{align}
	\begin{align}
	C_7^{h_3}\equiv C^{h_3}_{H^\pm_1H^\pm_1}&=-\frac{1}{2}i\left(2 \mathcal{C}_{21}^2 \mathcal{E}_{33} \lambda _a v_{t}+2 v_1 \left(\mathcal{E}_{31} \left(\mathcal{C}_{23}^2 \lambda _a+\mathcal{C}_{22}^2 \lambda _3\right)+\mathcal{C}_{21}^2 \lambda _1 \mathcal{E}_{31}+\mathcal{C}_{22} \mathcal{C}_{21} \left(\lambda _4+\lambda _5\right) \mathcal{E}_{32}\right)+2 \mathcal{C}_{22}^2 \mathcal{E}_{33} \lambda _b v_{t}+\right.\nonumber\\
	&2 v_2 \left(\mathcal{E}_{32} \left(\mathcal{C}_{23}^2 \lambda _b+\mathcal{C}_{21}^2 \lambda _3\right)+\mathcal{C}_{22}^2 \lambda _2 \mathcal{E}_{32}+\mathcal{C}_{21} \mathcal{C}_{22} \left(\lambda _4+\lambda _5\right) \mathcal{E}_{31}\right)+4 \mathcal{C}_{23}^2 \mathcal{E}_{33} \lambda _c v_{t}+2 \mathcal{C}_{21} \mathcal{C}_{23} \mu _1 \mathcal{E}_{31}+\nonumber\\
	&\left.\mathcal{C}_{21}^2 \mu _1 \mathcal{E}_{33}+2 \mathcal{C}_{22} \mathcal{C}_{23} \mu _2 \mathcal{E}_{32}+\mathcal{C}_{22}^2 \mu _2 \mathcal{E}_{33}+2 \mu _3 \left(\mathcal{C}_{21} \mathcal{C}_{23} \mathcal{E}_{32}+\mathcal{C}_{22} \left(\mathcal{C}_{23} \mathcal{E}_{31}+\mathcal{C}_{21} \mathcal{E}_{33}\right)\right)\right)\nonumber\\
	\end{align}
	\begin{align}
	t_7^{h_1}&=t_7^{h_3}=t_7^{h_3}=i2A_0\left(m^2_{H_1^\pm}\right)\nonumber\\
	\end{align}

	\begin{align}
	C_8^{h_1}\equiv C^{h_1}_{H^\pm_2H^\pm_2}&=-\frac{1}{2}i\left(2 \mathcal{C}_{31}^2 \mathcal{E}_{13} \lambda _a v_{t}+2 v_1 \left(\mathcal{E}_{11} \left(\mathcal{C}_{33}^2 \lambda _a+\mathcal{C}_{32}^2 \lambda _3\right)+\mathcal{C}_{31}^2 \lambda _1 \mathcal{E}_{11}+\mathcal{C}_{32} \mathcal{C}_{31} \left(\lambda _4+\lambda _5\right) \mathcal{E}_{12}\right)+2 \mathcal{C}_{32}^2 \mathcal{E}_{13} \lambda _b v_{t}+\right.\nonumber\\
	&2 v_2 \left(\mathcal{E}_{12} \left(\mathcal{C}_{33}^2 \lambda _b+\mathcal{C}_{31}^2 \lambda _3\right)+\mathcal{C}_{32}^2 \lambda _2 \mathcal{E}_{12}+\mathcal{C}_{31} \mathcal{C}_{32} \left(\lambda _4+\lambda _5\right) \mathcal{E}_{11}\right)+4 \mathcal{C}_{33}^2 \mathcal{E}_{13} \lambda _c v_{t}+2 \mathcal{C}_{31} \mathcal{C}_{33} \mu _1 \mathcal{E}_{11}+\nonumber\\
	&\left.\mathcal{C}_{31}^2 \mu _1 \mathcal{E}_{13}+2 \mathcal{C}_{32} \mathcal{C}_{33} \mu _2 \mathcal{E}_{12}+\mathcal{C}_{32}^2 \mu _2 \mathcal{E}_{13}+2 \mu _3 \left(\mathcal{C}_{31} \mathcal{C}_{33} \mathcal{E}_{12}+\mathcal{C}_{32} \left(\mathcal{C}_{33} \mathcal{E}_{11}+\mathcal{C}_{31} \mathcal{E}_{13}\right)\right)\right)\nonumber\\
	\end{align}
	\begin{align}
	C_8^{h_2}\equiv C^{h_2}_{H^\pm_2H^\pm_2}&=-\frac{1}{2}i\left(2 \mathcal{C}_{31}^2 \mathcal{E}_{23} \lambda _a v_{t}+2 v_1 \left(\mathcal{E}_{21} \left(\mathcal{C}_{33}^2 \lambda _a+\mathcal{C}_{32}^2 \lambda _3\right)+\mathcal{C}_{31}^2 \lambda _1 \mathcal{E}_{21}+\mathcal{C}_{32} \mathcal{C}_{31} \left(\lambda _4+\lambda _5\right) \mathcal{E}_{22}\right)+2 \mathcal{C}_{32}^2 \mathcal{E}_{23} \lambda _b v_{t}+\right.\nonumber\\
	&2 v_2 \left(\mathcal{E}_{22} \left(\mathcal{C}_{33}^2 \lambda _b+\mathcal{C}_{31}^2 \lambda _3\right)+\mathcal{C}_{32}^2 \lambda _2 \mathcal{E}_{22}+\mathcal{C}_{31} \mathcal{C}_{32} \left(\lambda _4+\lambda _5\right) \mathcal{E}_{21}\right)+4 \mathcal{C}_{33}^2 \mathcal{E}_{23} \lambda _c v_{t}+2 \mathcal{C}_{31} \mathcal{C}_{33} \mu _1 \mathcal{E}_{21}+\nonumber\\
	&\left.\mathcal{C}_{31}^2 \mu _1 \mathcal{E}_{23}+2 \mathcal{C}_{32} \mathcal{C}_{33} \mu _2 \mathcal{E}_{22}+\mathcal{C}_{32}^2 \mu _2 \mathcal{E}_{23}+2 \mu _3 \left(\mathcal{C}_{31} \mathcal{C}_{33} \mathcal{E}_{22}+\mathcal{C}_{32} \left(\mathcal{C}_{33} \mathcal{E}_{21}+\mathcal{C}_{31} \mathcal{E}_{23}\right)\right)\right)\nonumber\\
	\end{align}
	\begin{align}
	C_8^{h_3}\equiv C^{h_3}_{H^\pm_2H^\pm_2}&=-\frac{1}{2}i\left(2 \mathcal{C}_{31}^2 \mathcal{E}_{33} \lambda _a v_{t}+2 v_1 \left(\mathcal{E}_{31} \left(\mathcal{C}_{33}^2 \lambda _a+\mathcal{C}_{32}^2 \lambda _3\right)+\mathcal{C}_{31}^2 \lambda _1 \mathcal{E}_{31}+\mathcal{C}_{32} \mathcal{C}_{31} \left(\lambda _4+\lambda _5\right) \mathcal{E}_{32}\right)+2 \mathcal{C}_{32}^2 \mathcal{E}_{33} \lambda _b v_{t}+\right.\nonumber\\
	&2 v_2 \left(\mathcal{E}_{32} \left(\mathcal{C}_{33}^2 \lambda _b+\mathcal{C}_{31}^2 \lambda _3\right)+\mathcal{C}_{32}^2 \lambda _2 \mathcal{E}_{32}+\mathcal{C}_{31} \mathcal{C}_{32} \left(\lambda _4+\lambda _5\right) \mathcal{E}_{31}\right)+4 \mathcal{C}_{33}^2 \mathcal{E}_{33} \lambda _c v_{t}+2 \mathcal{C}_{31} \mathcal{C}_{33} \mu _1 \mathcal{E}_{31}\nonumber\\
	&\left.\mathcal{C}_{32}^2 \mu _2 \mathcal{E}_{33}+2 \mathcal{C}_{33} \mathcal{C}_{32} \mu _2 \mathcal{E}_{32}++\mathcal{C}_{31}^2 \mu _1 \mathcal{E}_{33}+2 \mu _3 \left(\mathcal{C}_{31} \mathcal{C}_{33} \mathcal{E}_{32}+\mathcal{C}_{32} \left(\mathcal{C}_{33} \mathcal{E}_{31}+\mathcal{C}_{31} \mathcal{E}_{33}\right)\right)\right)\nonumber\\
	\end{align}
	\begin{align}
	t_8^{h_1}&=t_8^{h_2}=t_8^{h_3}=i2A_0\left(m^2_{H_2^\pm}\right)\nonumber\\
	\end{align}

	\begin{eqnarray}
	C_{9}^{h_1}\equiv C^{h_1}_{ZZ}&\hspace*{-0.4cm}=&\hspace*{-0.2cm}\frac{i e m_W \left(v_1 \mathcal{E}_{11}+v_2 \mathcal{E}_{12}\right)}{v c_W^2 s_W}\nonumber\\
	\end{eqnarray}
	\begin{eqnarray}
	C_{9}^{h_2}\equiv C^{h_2}_{ZZ}&\hspace*{-0.4cm}=&\hspace*{-0.2cm}\frac{i e m_W \left(v_1 \mathcal{E}_{21}+v_2 \mathcal{E}_{22}\right)}{v c_W^2 s_W}\nonumber\\
	C_{9}^{h_3}\equiv C^{h_3}_{ZZ}&\hspace*{-0.4cm}=&\hspace*{-0.2cm}\frac{i e m_W \left(v_1 \mathcal{E}_{31}+v_2 \mathcal{E}_{32}\right)}{v c_W^2 s_W}\nonumber\\
	\end{eqnarray}
	\begin{eqnarray}
	t_{9}^{h_1}&=&t_{9}^{h_2}=t_{9}^{h_3}=-i\left(\left(n-1\right)A_0\left(m_Z^2\right)\right.\nonumber\\
	&&\left.+\xi_ZA_0\left(\xi_Zm_Z^2\right)\right)\nonumber\\
	\end{eqnarray}
	
	\begin{eqnarray}
	C_{10}^{h_1}\equiv C^{h_1}_{WW}&=&\frac{i e m_W \left(4 \mathcal{E}_{13} v_{t}+v_1 \mathcal{E}_{11}+v_2 \mathcal{E}_{12}\right)}{v s_W}\nonumber\\
	\end{eqnarray}
	\begin{eqnarray}
	C_{10}^{h_2}\equiv C^{h_2}_{WW}&=&\frac{i e m_W \left(4 \mathcal{E}_{23} v_{t}+v_1 \mathcal{E}_{21}+v_2 \mathcal{E}_{22}\right)}{v s_W}\nonumber\\
	\end{eqnarray}
	\begin{eqnarray}
	C_{10}^{h_3}\equiv C^{h_3}_{WW}&=&\frac{i e m_W \left(4 \mathcal{E}_{33} v_{t}+v_1 \mathcal{E}_{31}+v_2 \mathcal{E}_{32}\right)}{v s_W}\nonumber\\
	\end{eqnarray}
	\begin{eqnarray}
	t_{10}^{h_1}&=&t_{10}^{h_2}=t_{10}^{h_3}=2\left(-i\left(\left(n-1\right)A_0\left(m_W^2\right)\right.\right.\nonumber\\
	&\hspace*{-0.4cm}&\hspace*{1.4cm}\left.\left.+\xi_WA_0\left(\xi_Wm_Z^2\right)\right)\right)\nonumber\\
	\end{eqnarray}
	
	\begin{eqnarray}
	C_{11}^{h_1}\equiv C^{h_1}_{\eta_Z\bar{\eta}_Z}&\hspace*{-0.4cm}=&\hspace*{-0.2cm}-\frac{i e m_W \xi _Z \left(v_1 \mathcal{E}_{11}+v_2 \mathcal{E}_{12}\right)}{2 v c_W^2 s_W}\nonumber\\
	\end{eqnarray}
	\begin{eqnarray}
	C_{11}^{h_2}\equiv C^{h_2}_{\eta_Z\bar{\eta}_Z}&\hspace*{-0.4cm}=&\hspace*{-0.2cm}-\frac{i e m_W \xi _Z \left(v_1 \mathcal{E}_{21}+v_2 \mathcal{E}_{22}\right)}{2 v c_W^2 s_W}\nonumber\\
	\end{eqnarray}
	\begin{eqnarray}
	C_{11}^{h_3}\equiv C^{h_3}_{\eta_Z\bar{\eta}_Z}&\hspace*{-0.4cm}=&\hspace*{-0.2cm}-\frac{i e m_W \xi _Z \left(v_1 \mathcal{E}_{31}+v_2 \mathcal{E}_{32}\right)}{2 v c_W^2 s_W}\nonumber\\
	\end{eqnarray}
	\begin{eqnarray}
	t_{11}^{h_1}&=&t_{13}^{h_2}=t_{13}^{h_3}=iA_0\left(\xi_Z m^2_{Z}\right)\nonumber\\
	\end{eqnarray}
	
	\begin{eqnarray}
	C_{12}^{h_1}\equiv C^{h_1}_{\eta_\pm\bar{\eta}_\pm}&\hspace*{-0.4cm}=&\hspace*{-0.2cm}-\frac{i e m_W \xi _W \left(4 \mathcal{E}_{13} v_{t}+v_1 \mathcal{E}_{11}+v_2 \mathcal{E}_{12}\right)}{2 v s_W}\nonumber\\
	\end{eqnarray}
	\begin{eqnarray}
	C_{12}^{h_2}\equiv C^{h_2}_{\eta_\pm\bar{\eta}_\pm}&\hspace*{-0.4cm}=&\hspace*{-0.2cm}-\frac{i e m_W \xi _W \left(4 \mathcal{E}_{23} v_{t}+v_1 \mathcal{E}_{21}+v_2 \mathcal{E}_{22}\right)}{2 v s_W}\nonumber\\
	\end{eqnarray}
	\begin{eqnarray}
	C_{12}^{h_3}\equiv C^{h_3}_{\eta_\pm\bar{\eta}_\pm}&\hspace*{-0.4cm}=&\hspace*{-0.2cm}-\frac{i e m_W \xi _W \left(4 \mathcal{E}_{33} v_{t}+v_1 \mathcal{E}_{31}+v_2 \mathcal{E}_{32}\right)}{2 v s_W}\nonumber\\
	t_{12}^{h_1}&=&t_{12}^{h_2}=t_{12}^{h_3}=i2A_0\left(\xi_W m^2_{W}\right)\nonumber\\
	\end{eqnarray}

	\begin{eqnarray}
	C_{13}^{h_1}\equiv C^{h_1}_{f_Df_D}&\hspace*{-0.4cm}=&\hspace*{-0.2cm}-\frac{i e m_{f_{D}} v \mathcal{E}_{11}}{2 v_1 m_W s_W}\nonumber\\
	\end{eqnarray}
	\begin{eqnarray}
	C_{13}^{h_2}\equiv C^{h_2}_{f_Df_D}&\hspace*{-0.4cm}=&\hspace*{-0.2cm}-\frac{i e m_{f_{D}} v \mathcal{E}_{21}}{2 v_1 m_W s_W}\nonumber\\
	C_{13}^{h_3}\equiv C^{h_2}_{f_Df_D}&\hspace*{-0.4cm}=&\hspace*{-0.2cm}-\frac{i e m_{f_{D}} v \mathcal{E}_{31}}{2 v_1 m_W s_W}\nonumber\\
	\end{eqnarray}
	\begin{eqnarray}
	t_{13}^{h_1}&=&t_{13}^{h_2}=t_{13}^{h_3}=im_{f_D}A_0\left(m^2_{f_{D}}\right)Tr\left(I_n\right)\nonumber\\
	\end{eqnarray}

	\begin{eqnarray}
	C_{14}^{h_1}\equiv C^{h_1}_{f_Uf_U}&\hspace*{-0.4cm}=&\hspace*{-0.2cm}-\frac{i e m_{f_{U}} v \mathcal{E}_{12}}{2 v_2 m_W s_W}\nonumber\\
	\end{eqnarray}
	\begin{eqnarray}
	C_{14}^{h_2}\equiv C^{h_2}_{f_Uf_U}&\hspace*{-0.4cm}=&\hspace*{-0.2cm}-\frac{i e m_{f_{U}} v \mathcal{E}_{22}}{2 v_2 m_W s_W}\nonumber\\
	\end{eqnarray}
	\begin{eqnarray}
	C_{14}^{h_3}\equiv C^{h_3}_{f_Uf_U}&\hspace*{-0.4cm}=&\hspace*{-0.2cm}-\frac{i e m_{f_{U}} v \mathcal{E}_{32}}{2 v_2 m_W s_W}\nonumber\\
	\end{eqnarray}
	\begin{eqnarray}
	t_{14}^{h_1}&=&t_{14}^{h_2}=t_{14}^{h_3}=im_{f_U}A_0\left(m^2_{f_{U}}\right)Tr\left(I_n\right)\nonumber\\
	\end{eqnarray}
\end{scriptsize}

\setcounter{equation}{0}
\renewcommand{\theequation}{C\arabic{equation}}

\end{document}